\documentclass[structabstract]{aa}  

\usepackage{graphicx}
\usepackage{float}
\usepackage[varg]{txfonts}
\usepackage{longtable}
\usepackage{longtable,lscape}
\usepackage{lscape}
\usepackage{epsfig}

\newcommand\be{\begin{equation}}
\newcommand\en{\end{equation}}

\usepackage{txfonts}
\begin{document} 

\title{The accretion dynamics of EX Lupi in quiescence:}
 
\subtitle{The star, the spot, and the accretion column}

\author{Aurora Sicilia-Aguilar\inst{1,2}, Min Fang\inst{2}, Veronica Roccatagliata\inst{3}, Andrew Collier Cameron\inst{1}, \\ \'{A}gnes K\'{o}sp\'{a}l\inst{4},  Thomas Henning\inst{5}, Peter \'{A}brah\'{a}m\inst{4}, Nikoletta Sipos\inst{6}}

\institute{\inst{1} School of Physics and Astronomy, University of St Andrews, North Haugh, St Andrews KY16 9SS, UK\\
	\email{asa5@st-andrews.ac.uk}\\
	\inst{2}Departamento de F\'{\i}sica Te\'{o}rica, Facultad de Ciencias, Universidad Aut\'{o}noma de Madrid, 28049 Cantoblanco, Madrid, Spain \\
	\inst{3}Universit\"ats-Sternwarte M\"unchen, Ludwig-Maximilians-Universit\"at,Scheinerstr.~1, 81679 M\"unchen, Germany \\
	\inst{4} Konkoly Observatory, Research Center for Astronomy and Earth Sciences, Hungarian Academy of Sciences, PO Box 67, 1525 Budapest, Hungary\\
 	\inst{5} Max-Planck-Institut f\"{u}r Astronomie, K\"{o}nigstuhl 17, 69117 Heidelberg, Germany\\    
	\inst{6} Institute for Astronomy, ETH Z\"{u}rich, Wolfgang-Pauli-Strasse 27, 8093 Z\"{u}rich, Switzerland\\
}
	
   \date{Submitted February 26, 2015, accepted May 26, 2015}

\abstract
{EX Lupi is a young, accreting M0 star, prototype of EXor variable stars. 
Its spectrum is very rich in emission lines, including many metallic lines with
narrow and broad components.
It has been also proposed to have a close companion.}
{We use the metallic emission lines to study the
accretion structures and to test the companion hypothesis.}
{We analyse 54 spectra obtained during 5 years of quiescence time. 
We study the line profile variability and the
radial velocity of the metallic emission lines. We use 
the velocity signatures of different species with various excitation conditions
and their time dependency to track the dynamics associated to accretion.}
{We observe periodic velocity variations in the broad and the narrow line components,
consistent with rotational modulation. The modulation is stronger for lines with higher
excitation potentials, which are likely produced in a confined area, 
very close to the accretion shock.}
{We propose that the narrow line components are produced in the post-shock region, while the broad
components originate in the more extended, pre-shock material. All the emission lines
suffer velocity modulation due to the rotation of the star. The broad
components are responsible for the line-dependent veiling observed in EX Lupi.  
We demonstrate that a rotationally-modulated line-dependent veiling can explain the 
radial velocity signature of the photospheric absorption lines, making the close-in companion hypothesis
unnecessary. The accretion structure is locked to the star and very stable during the 5 years 
of observations. 
Not all stars with similar spectral types
and accretion rates show the same metallic emission lines, which could be related to differences in 
temperature and density in their accretion
structure(s). The contamination of photospheric signatures by accretion-related processes can be
turned into a very useful tool to determine the innermost details of the accretion 
channels in the proximities of the star. The presence of emission lines from very
stable accretion columns will nevertheless be a very strong limitation for the detection 
of companions by radial velocity
in young stars, given the similarity of the accretion-related signatures with those produced by a
companion.}


\authorrunning{Sicilia-Aguilar et al.}

\titlerunning{EX Lupi: Accretion dynamics}

\maketitle


\section{Introduction \label{intro}}

EX Lupi is the prototype of EXor variables and a remarkable object.
Its strong variability episodes are known for more than 50 years
(McLaughlin 1946; Herbig 1950). The photospheric emission observed during the quiescent phases
is compatible with a young star with spectral type M0 (Herbig 1977; Gras-Vel\'{a}zquez \& Ray 2005). 
The star is known to have short-timescale, moderate (1-2 mags) variability
 (Herbig 1977; Lehman et al. 1995) caused by small variations in the accretion
rate, together with more rare extreme variability episodes (increasing by 4-5 mags in the optical),
of which only two have been documented to date: the first one in the 50's  (Herbig 1977), and a second one in 2008 
(Jones 2008). In these extreme variability episodes, the accretion rate increases by more
than two orders of magnitude during a few months time, before the star goes back to 
quiescence (\'{A}brah\'{a}m et al. 2009).

The main characteristic of the EX Lupi spectra is the large
number of emission lines observed, both in quiescence and in outburst (Patten et al. 1994; 
Herbig et al. 2001; Sipos et al. 2009; K\'{o}sp\'{a}l et al. 2008, 2011; Sicilia-Aguilar et al. 2012, from now on SA12). 
During  its strong 2008 outburst, the whole spectrum was dominated by emission lines and the stellar photosphere
was completely veiled, with the Li I 6708\AA\ line being the only one that still showed a recognisable
absorption (SA12). Such rich emission-line spectra are very rare among classical T Tauri stars (CTTS) and
even other EXors. To our knowledge,
the only other object that shows a similar outburst spectrum is the M5-type EXor ASASN13db (Holoien et al. 2014). 
The number of identified emission lines in the optical
is of the order of a thousand, although there are many more lines that are hard to identify due to blends (SA12).
Most of the emission lines show broad and narrow components (BC and NC), with the remarkable 
exception of the hydrogen lines, which have only broad, very complex, profiles. The
BC shows a strong day-to-day modulation, consistent with bulk motions of clumpy, non-axisymmetric material
rotating and accreting onto the star (SA12). In quiescence, the 
number of emission lines is reduced to about 200, although some photospheric lines 
show cores consistent with weak, NC emission (SA12). Most of the
metallic lines in quiescence are narrow, with very weak or absent BC, although the typical CTTS emission
lines  are broad (e.g. the hydrogen Balmer series)
or show narrow and broad components (e.g. Ca II, some of the strongest Fe II lines, the He I 5875\AA\ line). 
Except for the Ca II IR triplet, most of the BC of metallic and He I lines are weak during
quiescence. All the observed emission lines correspond to permitted transitions,
with no evidence for forbidden emission lines (in both outburst and quiescence). 

The photospheric absorption lines
during quiescence show a strong radial velocity (RV) modulation, which has been interpreted as caused by
a brown dwarf (BD) companion, although hot-spot-induced RV signatures could not be excluded 
(K\'{o}sp\'{a}l et al. 2014; from now on K14). There are several precedents of objects where the RV suggested initially
the presence of a companion, which was later on ruled out as a signal caused by accretion and/or strong
activity, such as RW Auriga (Gahm et al. 1999; Petrov et al. 2001; Dodin et al. 2012) and RU Lupi
(Stempels et al. 2007; Gahm et al. 2013). 
In fact, using radial velocity techniques to detect companions to young
(active, variable, accreting) stars has well-known problems (Crockett et al. 2012; 
Jeffers et al. 2014; Dumusque et al. 2014). 

The emission lines observed in EX Lupi are a key to disentangle accretion and companion effects and
to understand the way accretion proceeds in the EXor prototype.
Emission lines have been used for more than 30 years to study the circumstellar environment,
accretion and winds in CTTS, specially considering the strong
H$\alpha$, He I, and Na I D lines, with the work started by Hartmann (1982), Appenzeller et al. (1983), Edwards et al. (1994),
Hamann \& Persson (1992), and Hartmann et al. (1994) and continued by many others.
In contrast with the more complex strong lines, the weak metallic lines (Fe I/II, Ti II),
have less complex profiles, not affected by self-absorption. 
The metallic lines also 
cover a broad range of excitation conditions, with 
excitation potentials ranging from few eV to several tens of eV and different transition
probabilities. This gives us a unique chance to probe regions of the accreting system
that show different values of temperature and density, constructing a 3-dimensional picture
of the EX Lupi environment. The day-to-day modulation observed in the broad emission lines
during the 2008 outburst showed clear velocity
differences depending on the species, e.g. comparing
 ionised vs neutral lines, or lines with
high excitation potentials such as He I vs lines with weaker excitation potentials such as Ca II 
or Fe I (SA12). The differences between lines were consistent with the
differences in density and temperature expected along an extended ($\sim$0.1 AU) accretion column(s)
or non-axisymmetric rotating/infalling structure(s)
(SA12). A similar approach, but using absorption lines, has been used to track cold matter
moving around intermediate-mass stars (Mora et al. 2002, 2004), and 
to trace the physical conditions along the accretion
column in S CrA SE (Petrov et al. 2014). With the time coverage we now have for EX Lupi in
quiescence, we can not only explore the physical properties, but also use the time variable to
explore the dynamics of the accretion columns. 

In this paper, we analyse the emission lines of the quiescent EX Lupi,
turning the classical problem of accretion-related signatures affecting the
radial velocity measurements into
a tool to investigate the structure of its accretion columns. 
Section \ref{data} introduces the EX Lupi spectroscopy data. 
In Section \ref{analysis}, we explore and quantify the dynamics observed
in narrow and broad emission lines. Simple dynamical and physical models
for the accreting structure are presented in Section \ref{models}. Section \ref{results}
 brings together the 
observations within a unified picture of rotational modulation.
Finally, Section \ref{conclu} summarises our results.

\section{Observations, data reduction, and disk and stellar properties \label{data}}

\subsection{Observations and data reduction}

EX Lupi was monitored using the "Fiber-fed Extended Range Optical Spectrograph" (FEROS; Kaufer et al. 1999)
mounted on the 2.2m telescope 
and the "High Accuracy Radial velocity Planet Searcher" (HARPS; Mayor et al. 2003) 
spectrograph on the 3.6m telescope, both in La Silla, Chile. The observations 
were part of various observing programs (see K14 for details on the quiescence data). 
The spectroscopic followup of EX Lupi consist of 54 spectra (see Table \ref{spectra-table} 
in Appendix \ref{support-appendix}) taken during the quiescence
phase of EX Lupi. These spectra correspond to the high S/N observations listed by K14. The observations cover about 5 years of quiescence data taken between 2007 and 2012 (excluding
the data taken during the 2008 EX Lupi outburst; discussed in SA12). The data were taken at irregular intervals, ranging from
two spectra per night to few days/months intervals, depending on telescope availability.
The exposure times ranged between 10-50 minutes, depending on the brightness of the star and
the atmospheric conditions. FEROS has a resolution of R=48000, and a wavelength coverage between 3500-9200\AA,
containing in total 39 echelle orders with some small gaps. HARPS data have a higher resolution (R=115000)
but a more reduced spectral coverage (3780-6910\AA). For the line comparison in this study, the HARPS spectra
have been resampled to the FEROS resolution, which also allows us to reach a more uniform 
signal-to-noise ratio (S/N).

The data were reduced using the automated HARPS and FEROS pipelines, and then corrected to
remove the offset in the barycentric velocity for the FEROS specra (M\"{u}ller et al. 2013)
and to remove the radial velocity instrumental offset betwen the FEROS and the HARPS spectra of -0.291 km/s (K14). 
Both effects are very small and thus have little effect on
the broad emission lines, but they could have some influence on the radial velocity measurements
of the narrow emission line components, of the order of few km/s.

\subsection{Stellar and disk properties revisited}

The spectra of EX Lupi in quiescence are consistent with a
M0 star (stellar mass $\sim$0.6 M$_\odot$; Gras-Vel\'{a}zquez \& Ray 2005) 
with numerous emission lines. The luminosity is highly variable due to variations in the accretion rate.
The estimated stellar radius is $\sim$1.6 R$_\odot$ (Sipos et al. 2009).
The rotational velocity ($v$sin$i$) is consistent with the
value of 4.4$\pm$2.0 km/s (Sipos et al. 2009). 

The lack of near-IR excess is interpreted as a small, dust-depleted hole in the disk, with size
$\sim$0.3AU (Sipos et al. 2009). CO observations (Goto et al. 2011; Banzatti et al. 2015) are also in agreement with
a sub-AU hole, mostly devoid of dust. The disk had been suggested to be relatively massive (0.025 M$_\odot$ for a star with mass
$\sim$0.6M$_\odot$) based on disk models containing only small dust grains (Sipos et al. 2009).
If instead we assume a more typical maximum grain size of 100 $\mu$m as it is usually found in other T Tauri stars
(D'Alessio et al. 2001, 2006; Andrews \& Williams 2007), the disk mass does not need to be so extreme
and could be below 1\% of the stellar mass (Appendix \ref{diskmass}).

The typical accretion rate during the quiescence periods has been estimated to be 
1-3$\times$10$^{-10}$M$_\odot$/yr (SA12, based on H$\alpha$ 10\% widths; Natta et al. 2004).
Estimates of the accretion rate based on Pa $\beta$ (Sipos et al. 2009) and 
on multiple emission lines (using the relations of Alcal\'{a} et al. 2014; Appendix \ref{acc-appendix})
result in a value of 4$\pm$2$\times$10$^{-10}$M$_\odot$/yr.
The observed H$\alpha$ width at 10\% of the peak (Natta et al. 2004) during our quiescence data are in agreement with 
these values, although for some of the spectra with stronger lines, the accretion rate 
increases up to $\sim$10$^{-9}$M$_\odot$/yr. The data discussed here thus includes variations in the
accretion rate within one order of magnitude, well below the outburst case where the
accretion rate increases by more than two orders of magnitude (SA12).

\section{Quantifying the line variability for a dynamical analysis\label{analysis}}

\begin{figure*}
\centering
\begin{tabular}{c}
\includegraphics[width=0.8\linewidth]{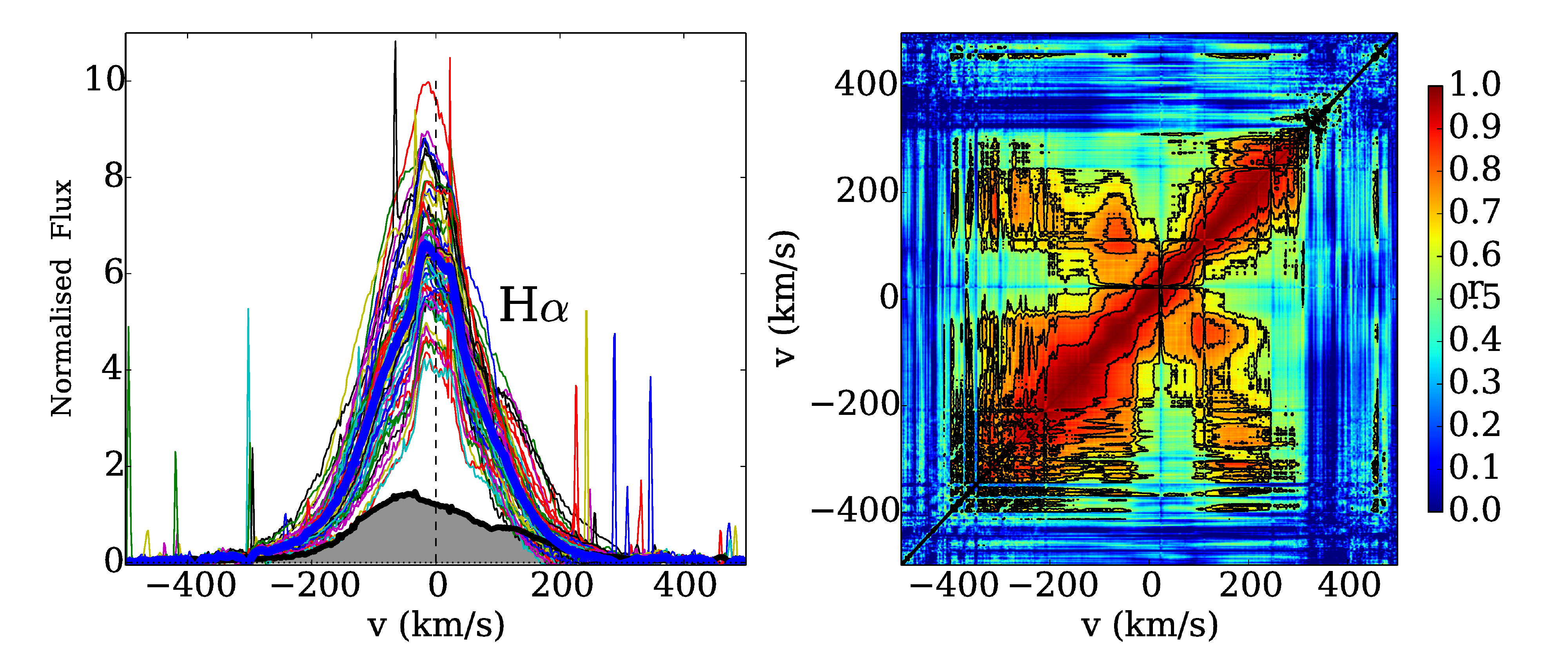}\\
\includegraphics[width=0.8\linewidth]{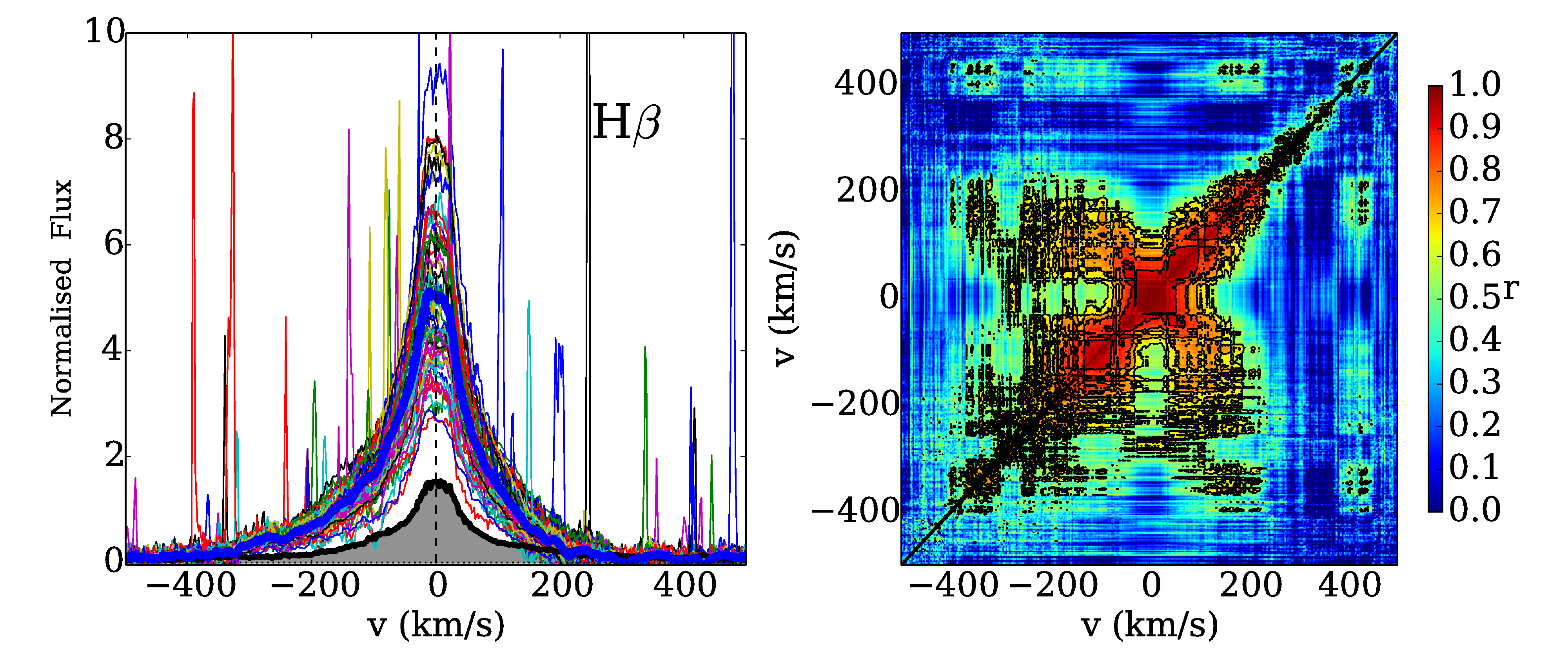}\\
\end{tabular}
\caption{H$\alpha$ and H$\beta$ line profile variability (left) and
autocorrelation matrix (right). The pixel-by-pixel median of the line 
is shown as a blue thick line. The pixel-by-pixel rms is shown as the thick black
line and shaded area. The zero velocities/wavelengths of the lines are marked as
dashed vertical lines. The correlation coefficient (estimated as a Spearman rank
correlation with value r and false alarm probability p) is shown on the right in the color
scheme, with the black contours marking the high significance (p$>$10$^{-5}$)
parts with r=0.6,0.7,0.8,0.9.  \label{hacor}}
\end{figure*}

The emission lines are very complex. To find a proxy to measure line variability
and dynamics, we take two different approaches:
\begin{itemize}
\item First, we explore the pixel-by-pixel variation of the broad components (BC) without assuming any particular
model for the line. This method was used by Lago \& Gameiro (1998) and Alencar et al. (2001)
to study variability patterns in T Tauri stars. It allows us to identify the
dynamical effects that dominate the changes in the line profiles,
which parts of the line have the highest variability, and whether variability
affects all parts or velocity components of the line similarly. It also allows us to
check for correlations between the different velocities and lines.

\item Second, we use a multi-Gaussian fit to reproduce the line emission and explore
the variations in intensity, radial velocity, and width, of the various Gaussian components.
This is particularly useful to study the behaviour of the narrow components (NC), whose velocity, amplitude, and
width do not depend on the particular choice of Gaussian fit. Since most of the metallic lines have weak (often undetectable) BC
with poorly resolved profiles, 
the metallic line analysis refers almost exclusively to the NC of the lines. 
\end{itemize}
 
\subsection{A pixel-by-pixel dynamical view of the broad component \label{corre}}

\begin{figure*}
\centering
\begin{tabular}{c}
\includegraphics[width=0.8\linewidth]{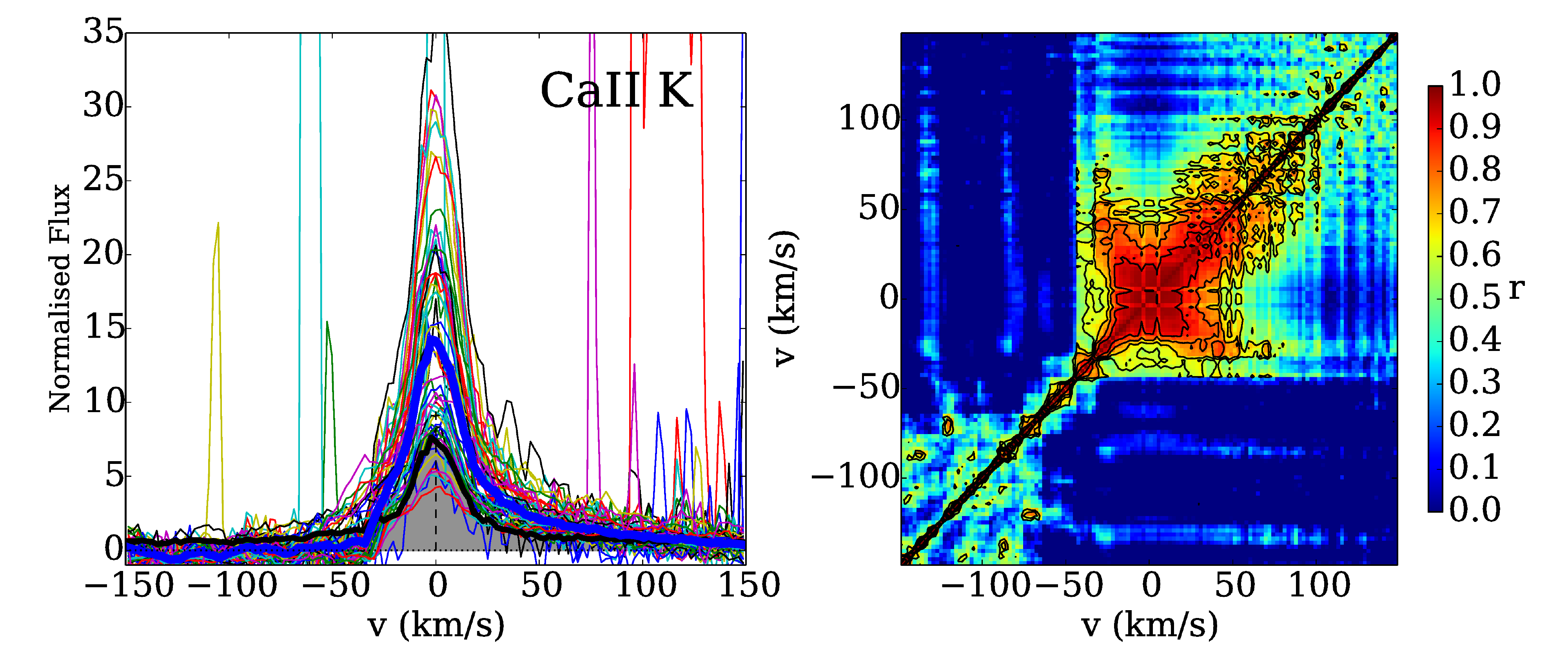}\\
\includegraphics[width=0.8\linewidth]{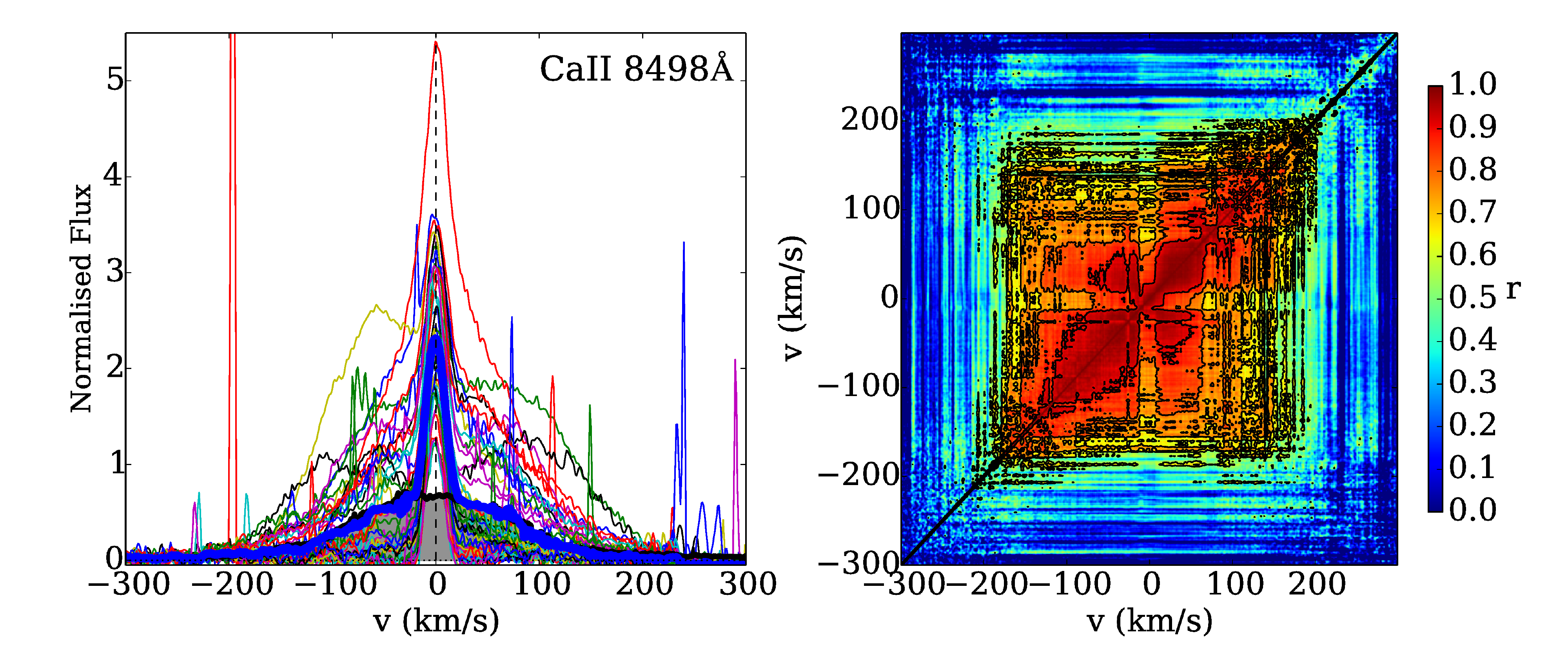}\\
\end{tabular}
\caption{Same as Figure \ref{hacor} for the Ca II K (top) and
8498\AA\ (bottom) lines. The pixel-by-pixel median of the line 
is shown as a blue thick line. The pixel-by-pixel rms is shown as the thick black
line and shaded area. The zero velocities/wavelengths of the lines are marked as
dashed vertical lines. The correlation coefficient (estimated as a Spearman rank
correlation with value r and false alarm probability p) is shown on the right in the color
scheme, with the black  contours marking the high significance (p$>$10$^{-5}$)
parts with r=0.6,0.7,0.8,0.9. \label{caiicor}}
\end{figure*}

\begin{figure*}
\centering
\begin{tabular}{cc}
\includegraphics[width=0.45\linewidth]{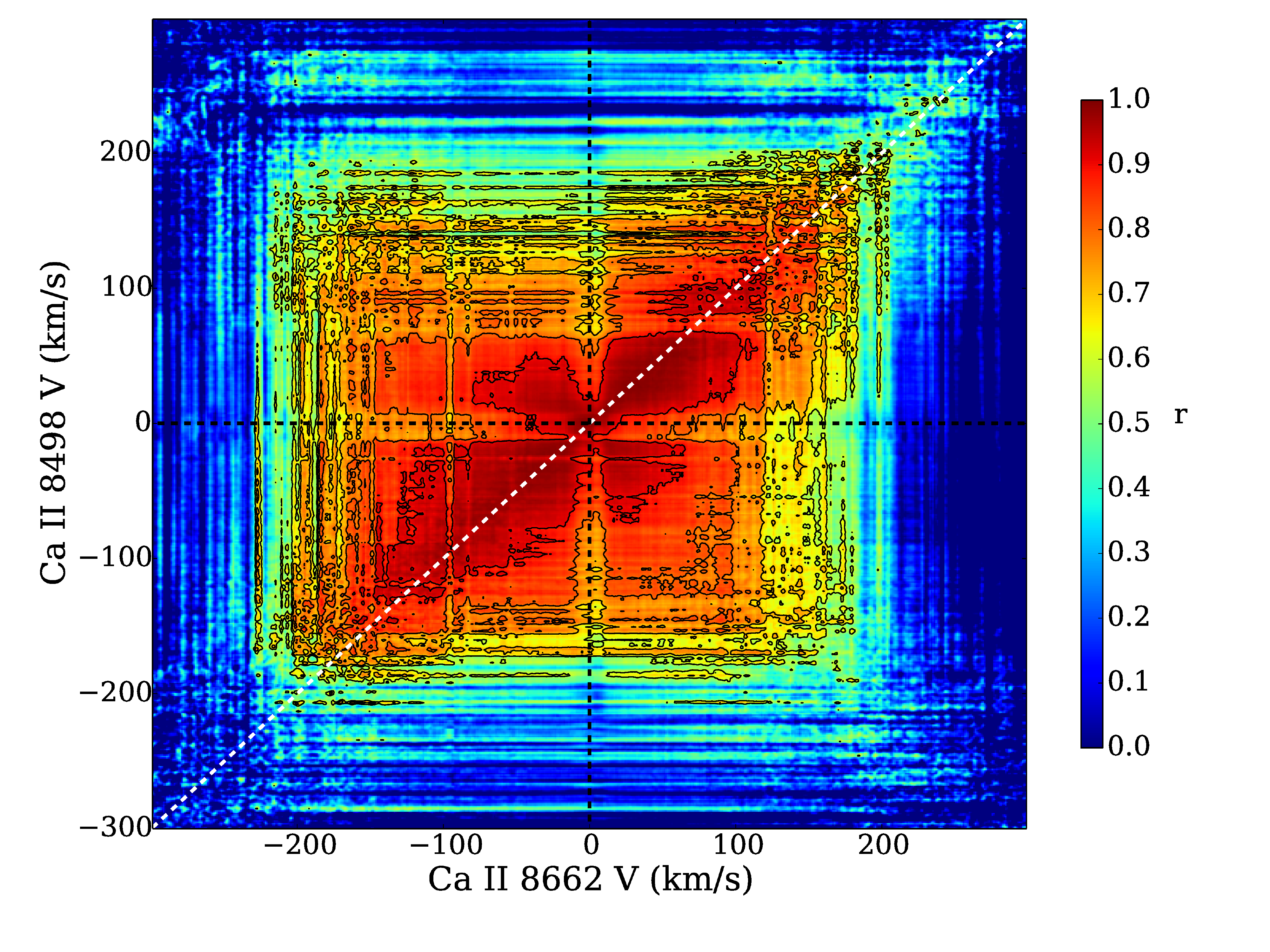} &
\includegraphics[width=0.45\linewidth]{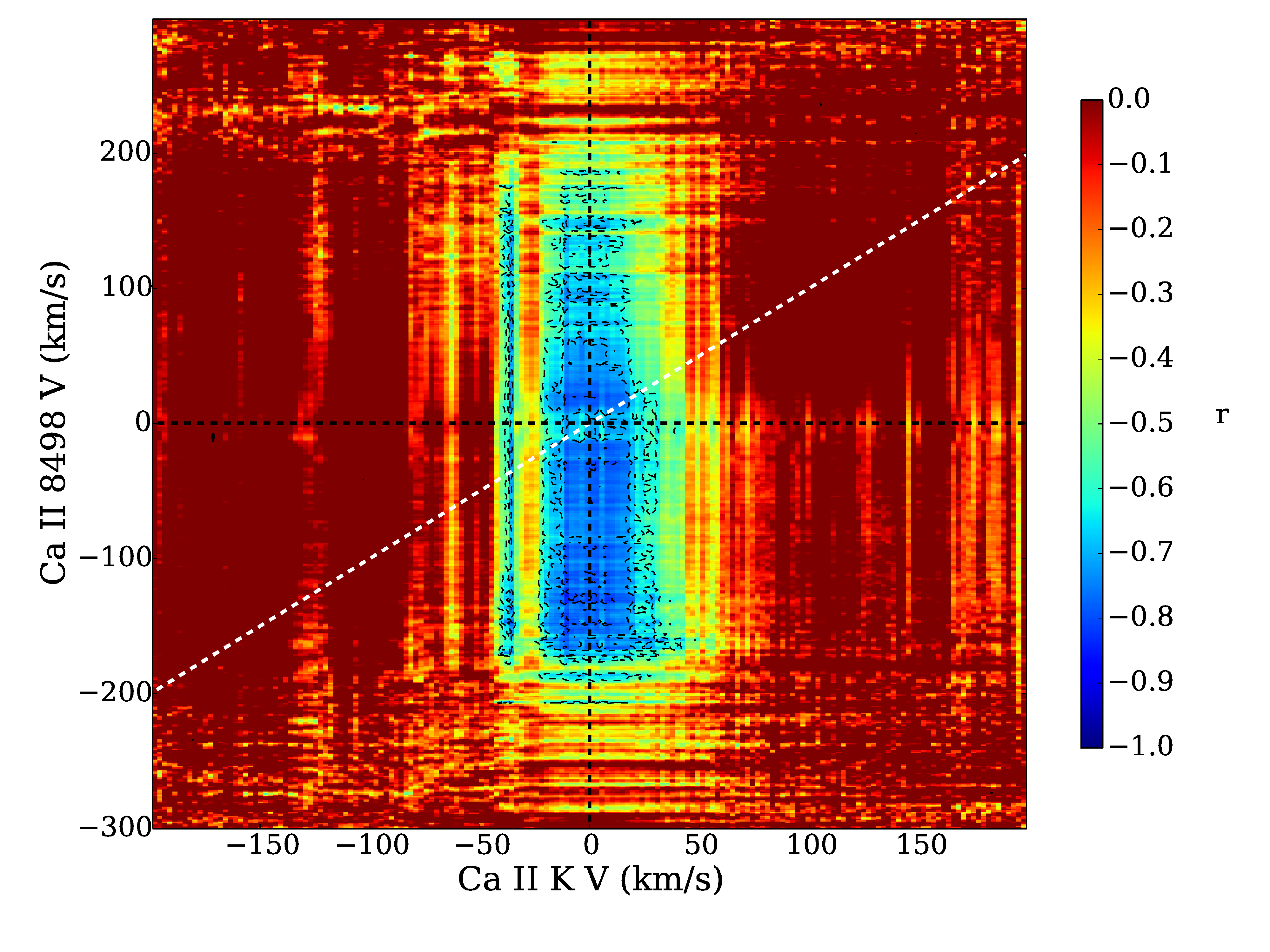} \\
\includegraphics[width=0.45\linewidth]{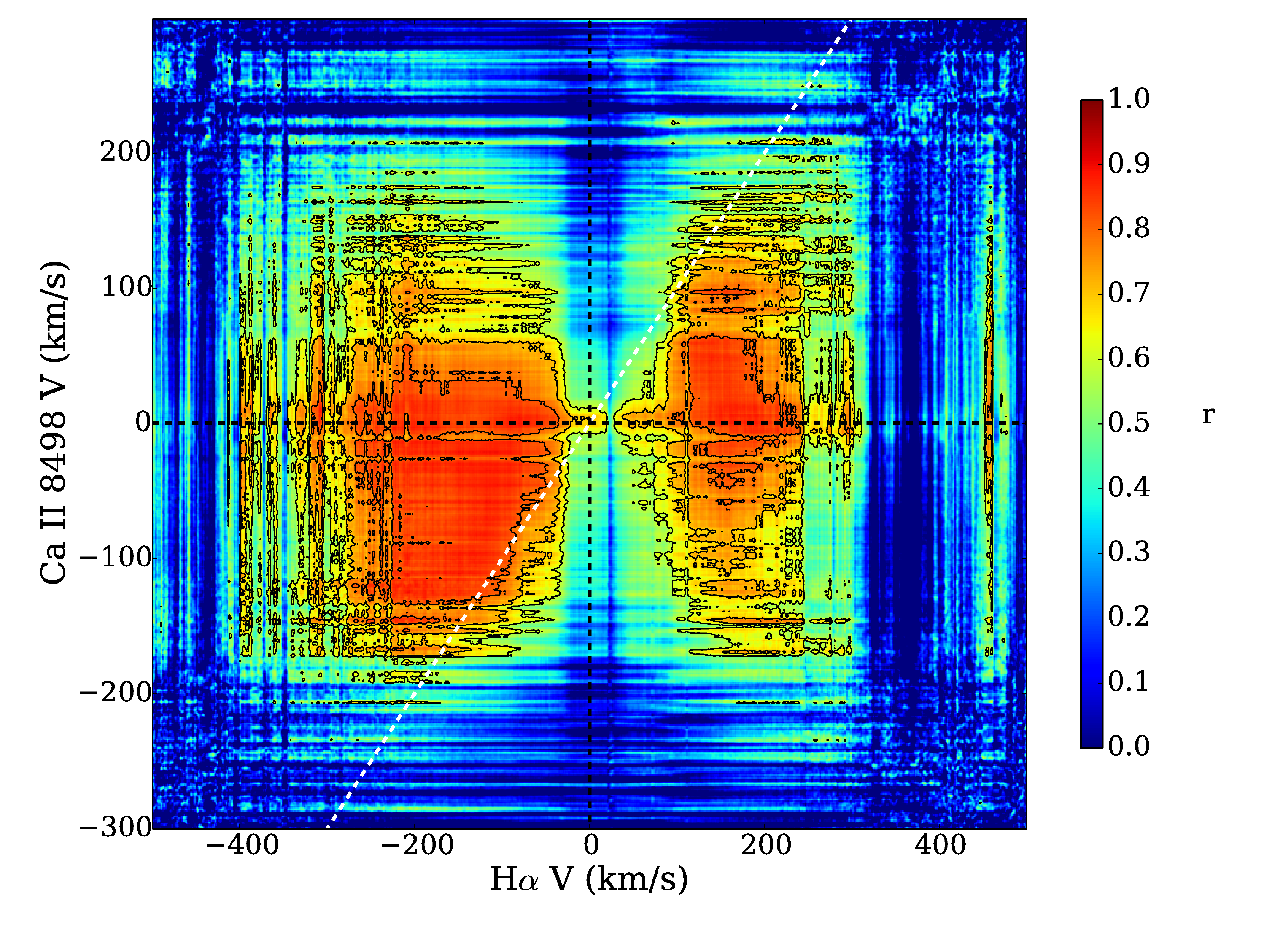} &
\includegraphics[width=0.45\linewidth]{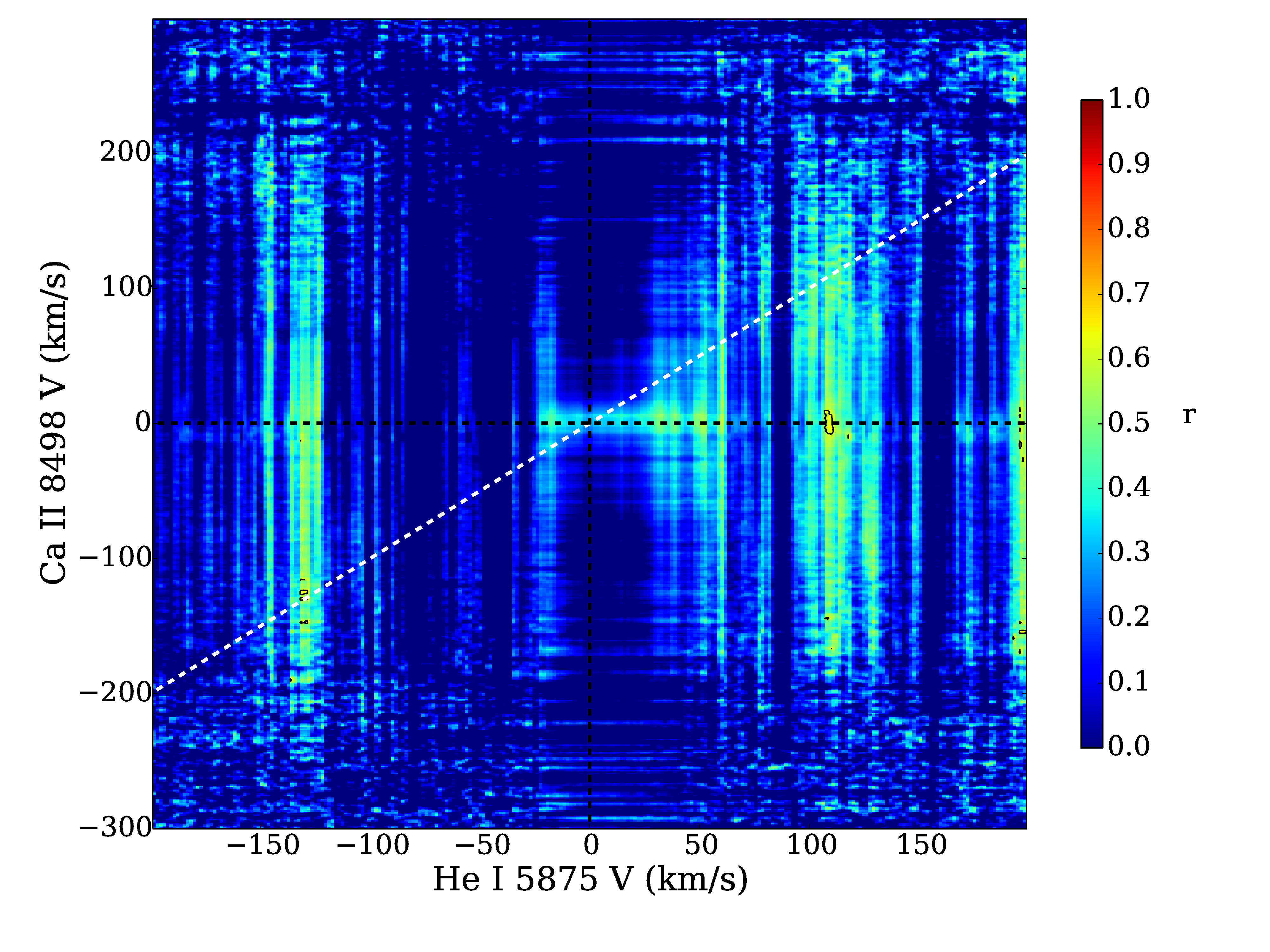} \\
\includegraphics[width=0.45\linewidth]{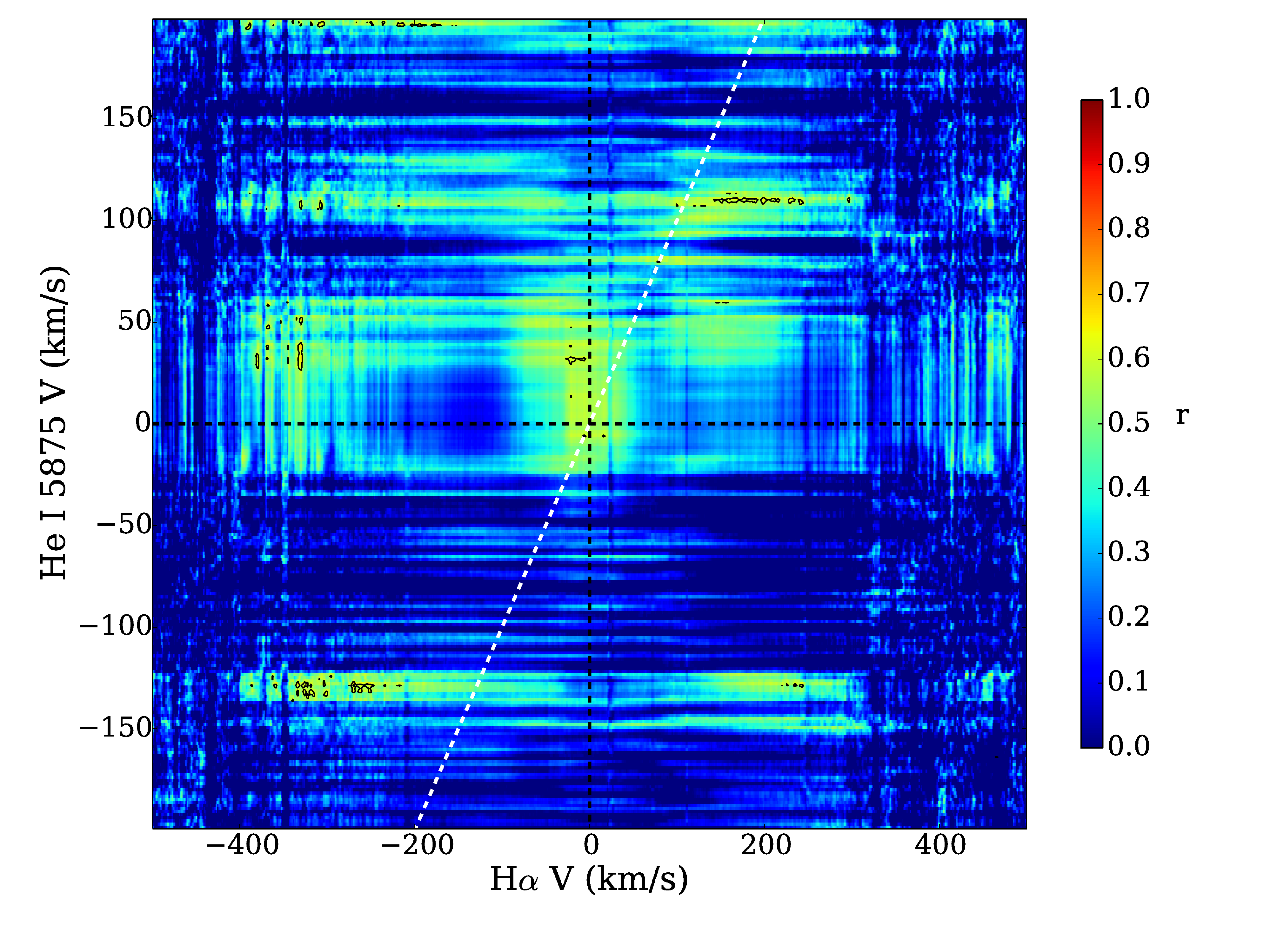} &
\includegraphics[width=0.45\linewidth]{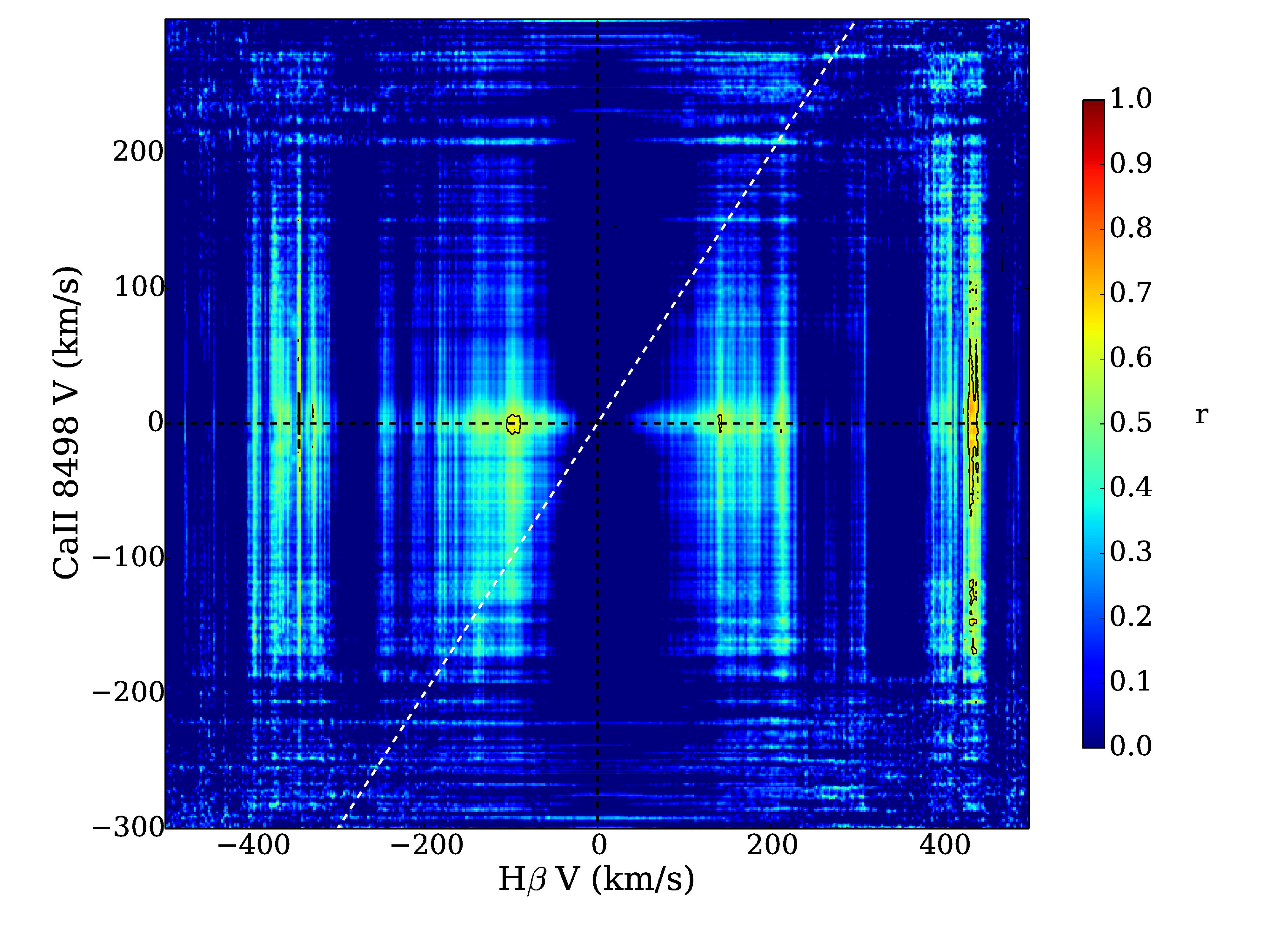} \\
\end{tabular}
\caption{Pixel-by-pixel velocity cross-correlations between different lines. From top to bottom, left to right:
Ca II 8498 vs 8662\AA, Ca II K vs Ca II 8498\AA, Ca II 8498\AA\ vs H$\alpha$, 
Ca II 8222\AA\ vs He I 5875\AA, H$\alpha$ vs He I 5875\AA\, and Ca II 8498\AA\ vs H$\beta$. 
The correlation coefficient (estimated as a Spearman rank
correlation with value r and false alarm probability p) is shown on the right in the color
scheme, with the  contours marking the high significance (p$>$10$^{-5}$)
parts with r=0.6,0.7,0.8,0.9.
\label{xcor}}
\end{figure*}

To explore the pixel-by-pixel
variation of the emission lines, we first resampled all the spectra to the (lower)
FEROS resolution.  The richness of features in the
EX Lupi spectra is such, that finding a proper normalisation at all wavelengths is very hard.
Therefore, for each line, the spectra are normalised to the local continuum levels,
measured on both sides of the line on parts of the spectrum that are not affected by 
emission or absorption lines. 
We then estimated the median and standard deviation of the emission line at
each pixel, after
clipping to remove bad pixels and cosmic rays. The procedure
was repeated for all the broad lines along the extent of the BC line wings.

We also explore to which extent the emission at a given velocity 
is correlated with the emission at any other velocity within the same line, and
how different lines are correlated with each other.
If we attribute each velocity bin to a particular gas parcel (or parcels) 
moving at the same rate, the autocorrelation matrix and the two-line cross-correlations
reveal which parts of the system may be connected. 
Nevertheless, there could be material moving
at the same projected velocity, but coming from different physical locations, 
which can  alter the pixel-by-pixel velocity correlations.
In addition, Dupree et al. (2012) demonstrated that there are significant temporal
variations of the order of hours between the various accretion indicators (including 
the Balmer H emission, the veiling continuum, 
and the He I lines) in a clumpy-accretion episode. Since the reaction time differences between
various lines are longer than our exposure times, this could
wash out the correlations.

Figures \ref{hacor} and \ref{caiicor} show both the line variance and median value
and the autocorrelation matrix at different velocities for hydrogen and Ca II lines. 
The Hydrogen Balmer lines are, as expected (Edwards et al. 1987; Appenzeller et al. 1989; Muzerolle et al. 2001), 
very complex and asymmetric.  They often
have self-absorptions that may correspond to colder wind/infall components, 
which can dilute the correlations or otherwise produce very complex patterns.
H$\alpha$
is particularly remarkable, with a profile with both
blueshifted and redshifted absorptions. The blue part of the spectrum 
shows a higher variability, probably due to the presence of a variable 
wind. Part of the redshifted emission may also be suppressed by self-absorption
in an optically thick accretion flow. The maximum velocities observed in the 
red and in the blue increase when the strength of the line increases, as
observed during outburst and expected in the case of fluctuations of the
accretion rate (Natta et al. 2004). If attributed to free-fall (infall-dominated) alone, the maximum
velocities ($\pm$300 km/s) would suggest material flowing onto the star from distances of
$\sim$2 stellar radii, considering M$_*$=0.6M$_\odot$ and R$_*$=1.6R$_\odot$. 
There is also a tendency of the line to 
become more symmetric when its intensity (and thus accretion rate) decreases,
also a sign of the complex profile being strongly affected by self-absorption.
The higher Balmer lines become increasingly symmetric, with similar
variability on their blue- and redshifted sides. 
For the H$\alpha$ and H$\beta$ lines, we find a strong correlation between 
the blue- and red-shifted parts, while the correlation of the line wings with
the central part of the line is significantly lower. This is a further sign that
the zero-velocity components may be saturated and/or produced in several
spatially different or very extended locations.

The Ca II lines, especially the IR triplet (only covered by FEROS) also show very broad
components. The Ca II IR triplet BCs have a remarkable profile
reminiscent of the outburst profiles of the metallic lines (SA12):
the BC of the line changes from blue- to redshifted within timescales of days, 
spanning a large range of velocities. The BC also decreases to be nearly undetectable when the accretion rate decreases
(see Appendix \ref{acc-appendix}). The median line profile is, in strong contrast to the
individual lines, very symmetric, as well as the pixel-by-pixel standard deviation,
showing that the variability in the blue and in the red is very similar, but
independent.
The Ca II 8542\AA\ line falls partially in the FEROS spectrograph gap, but it appears
similar the other two Ca II IR triplet lines.
The Ca II IR triplet is thus consistent with bulk motions of parcels of gas that are sometimes
blueshifted and sometimes redshifted, rather than to related processes (such as infall/accretion
and accretion-powered wind)
that produce correlated blueshifted and redshifted features. 
At first order, the lack of anticorrelations is also a problem for the companion scenario,
where we would expect the RV shift to increase one side of the line while decreasing the other one,
although the shift could be significantly smaller than the BC velocity widths and thus hard to detect.
Although the Ca II line forms closer to the star, this asymmetric motion
is consistent with the CO observations of Goto et al. (2011) and Banzatti et al. (2015) and 
with the day-to-day variations in the line profiles during the 2008 outburst (SA12),
suggesting that the asymmetric, rotating/infalling structure that fills in the inner disk 
probably continues onto the inner magnetospheric accretion region.
The Ca II H and K lines have asymmetric profiles with a
blueshifted absorption that is in part correlated with the wings on the redshifted
side of the line, although the velocity range is small compared to the H Balmer lines
or the Ca II IR triplet\footnote{Note that the Ca II H line is contaminated by
H$\epsilon$.}. 

We can take a further step exploring the connections between different lines and velocities
by calculating the pixel-by-pixel cross-correlation between different lines (Figure \ref{xcor}).
The cross-correlation matrix for the Ca II 8498 and 8662\AA\ shows the same
pattern as the autocorrelation matrix of each line, a sign that they are nearly
identical, as expected. 
There is an anticorrelation between Ca II K and Ca II IR, likely due to S/N effects
and variations of the accretion rate. A higher accretion rate produces strong Ca II IR, but also a higher 
blue continuum, and thus a lower peak over the continuum of all the UV and blue lines. 
At low accretion rates, the excess blue continuum decreases, and
since EX Lupi is a M0 star with little blue/UV emission, the continuum at short wavelengths is
noise-dominated and the strength of blue/UV lines over the continuum is artificially increased. 
A similar anticorrelation appears between H$\alpha$ and CaII K: the H$\alpha$ blue and red wings 
are anticorrelated with the Ca II K line. Since the red and
blue H$\alpha$ wings (and the near-UV continuum) grow when the accretion rate increases, 
the Ca II K peak over the continuum decreases if the
Ca II K line flux does not increase at the same rate. This is a 
sign that we need to be careful when comparing lines at different wavelengths.

Cross-correlation reveals that the wings of H$\alpha$ and the BC
of the Ca II IR triplet lines are correlated  (Figure \ref{xcor}), a sign that the strength of the Ca II IR triplet 
BC depends on the accretion rate, although the velocities
and line profiles change in an independent way.
There is no significant correlation between the H$\alpha$ peak
and the Ca II IR lines, probably because H$\alpha$ is strongly saturated.
The cross-correlation between the Ca II IR triplet and the He I lines does not
give any significant result. This suggests that both lines respond to
different physical processes that are likely to happen in different locations and at slightly different
times (Dupree et al. 2012).

\subsection{Multi-Gaussian fit and radial velocities \label{gaussfit}}

The second step involves studying and quantifying the strength, width, and radial velocity of the different 
emission lines. We fitted small sections of the
normalised and continuum-subtracted spectrum (typically, $\pm$100 km/s 
around the line for narrow features, and 
up to $\pm$300 km/s around features with BC) with a multi-Gaussian model. 
For the strongest, complex lines (H$\alpha$, H$\beta$, etc), any
attempt of simple multi-Gaussian fitting or interpretation is impossible. 
The He I and He II lines are less complex,
as well as the metallic lines (Ca II, Fe I, Fe II, Ti II, Si II,
Mg I, among others). These weaker emission lines are the main subject of this section.

\begin{figure*}
\centering
\begin{tabular}{cccc}
\includegraphics[width=0.24\linewidth]{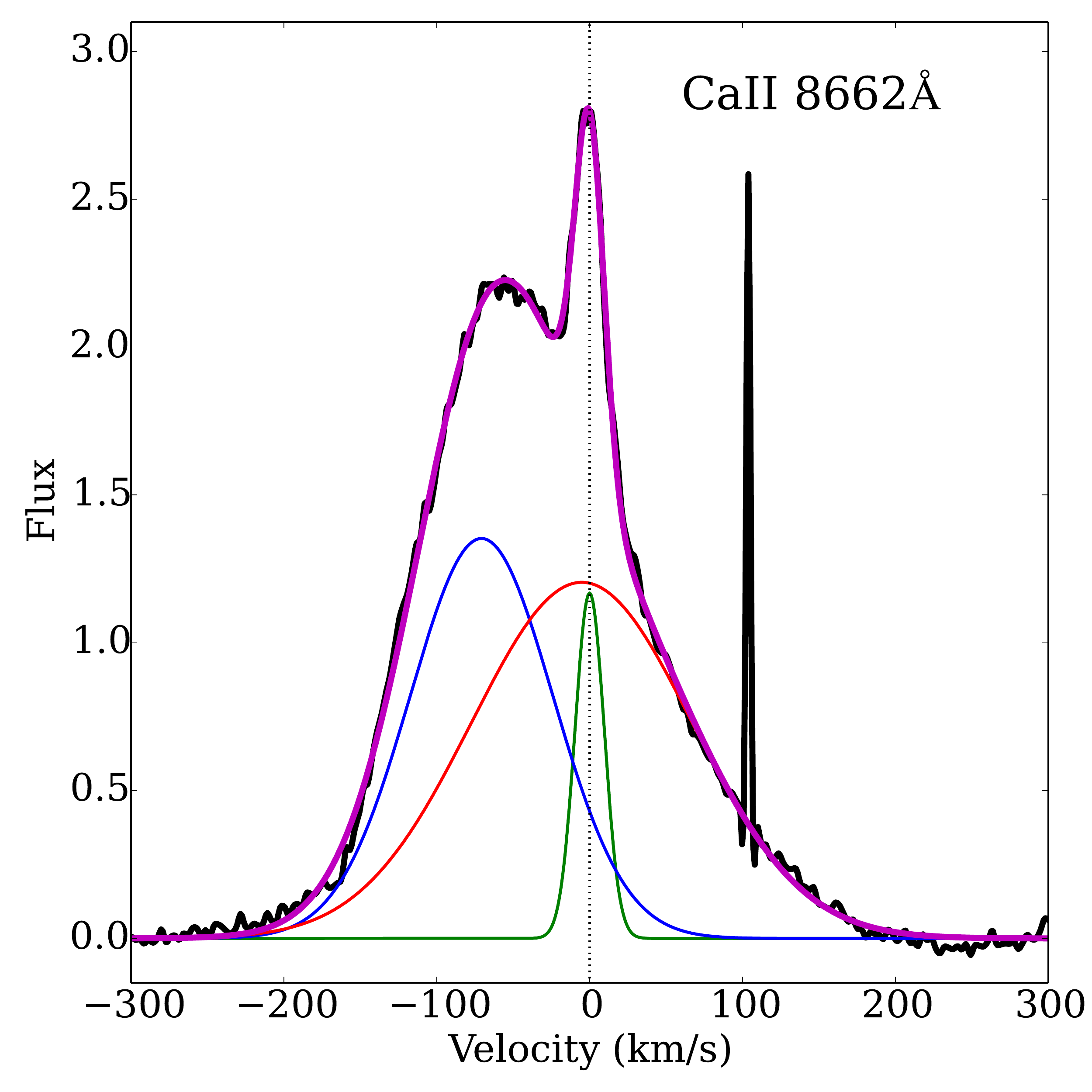} &
\includegraphics[width=0.24\linewidth]{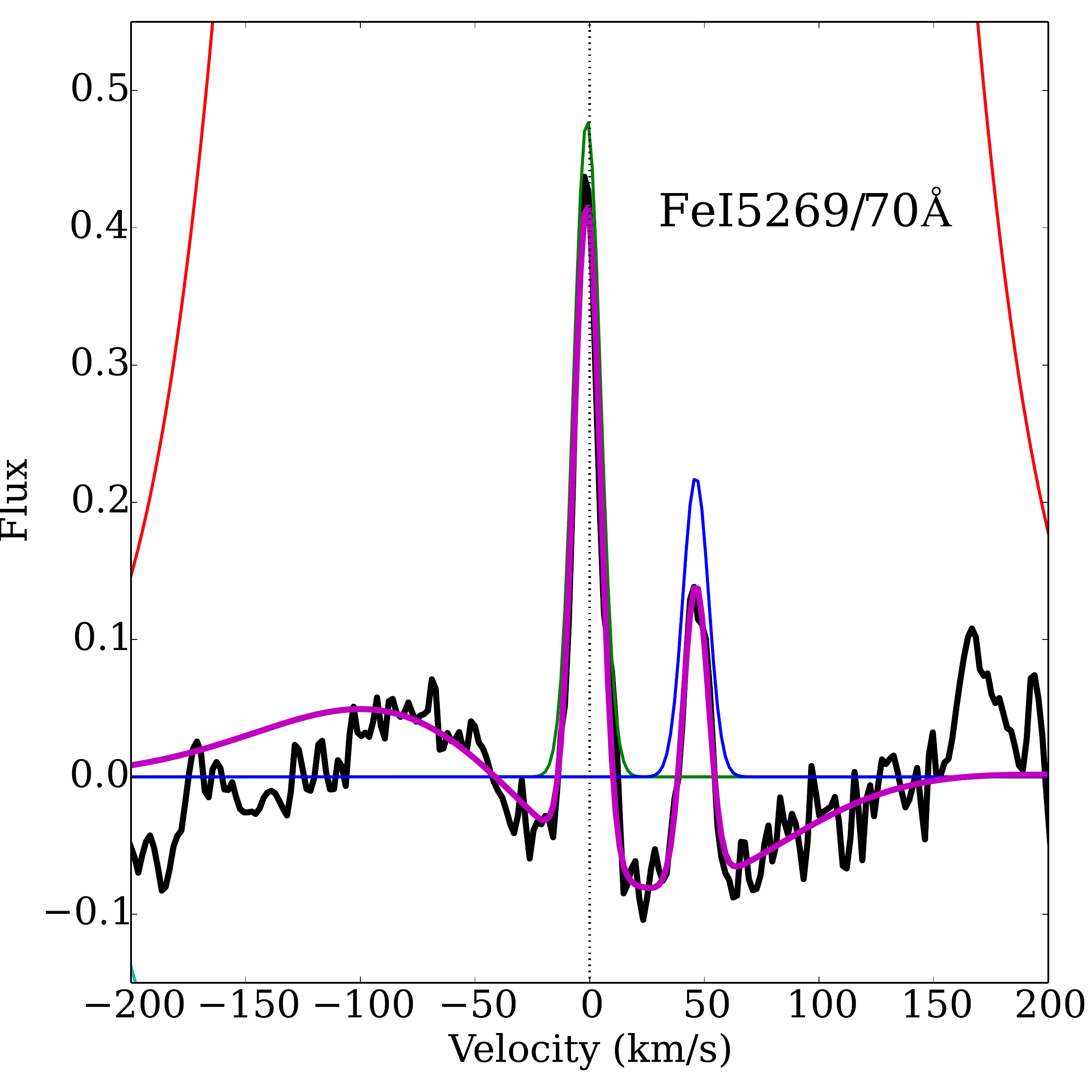} &
\includegraphics[width=0.24\linewidth]{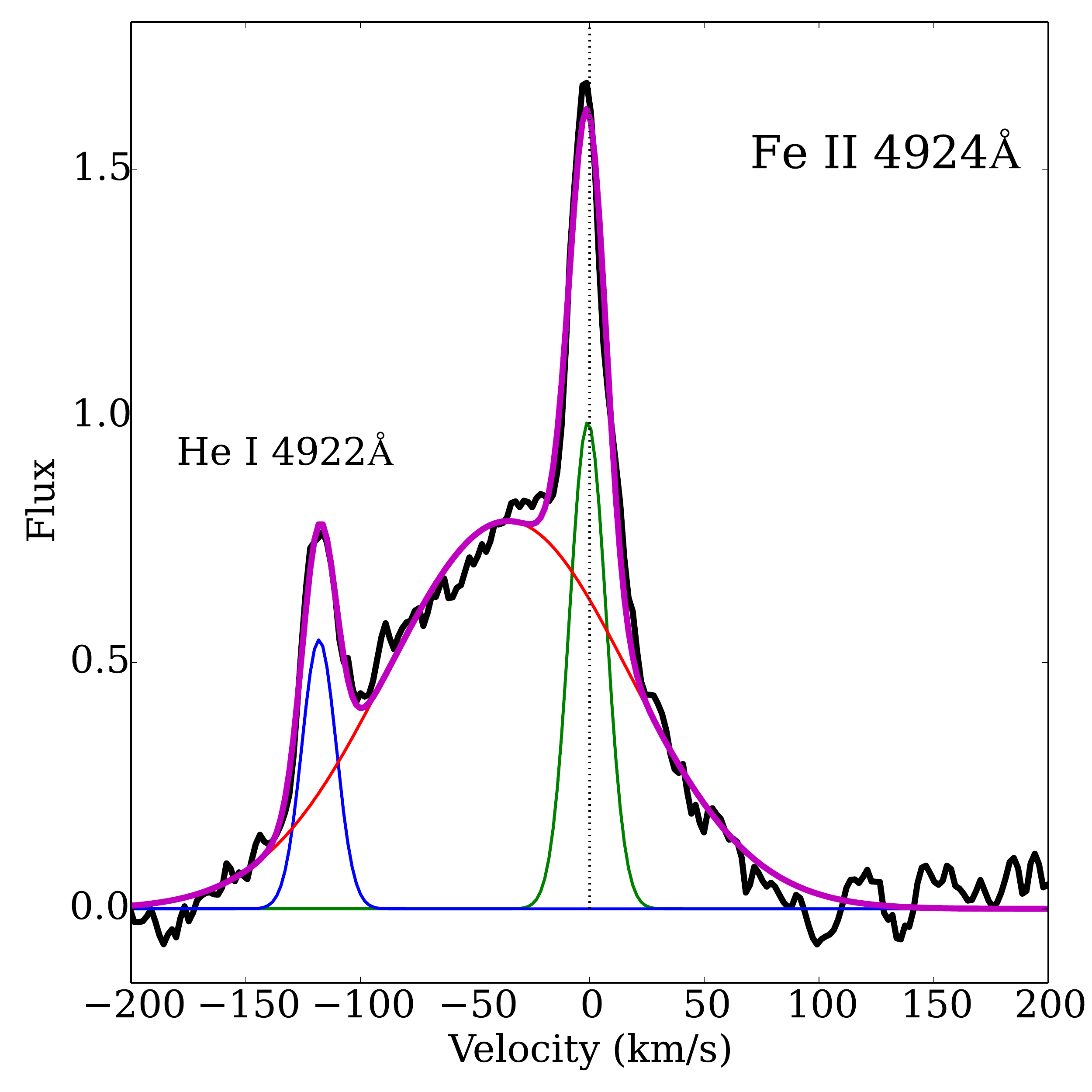} &
\includegraphics[width=0.24\linewidth]{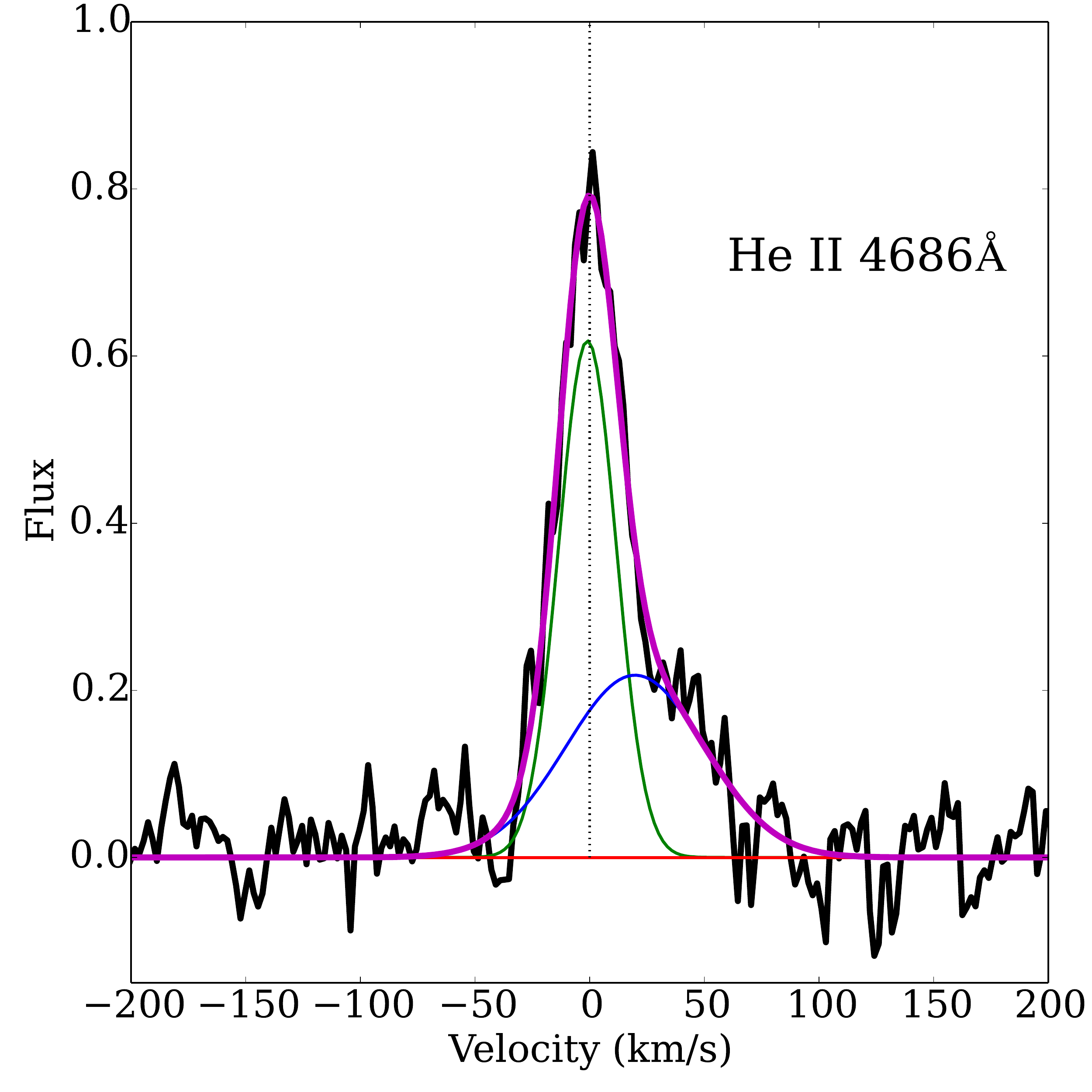} \\
\end{tabular}
\caption{Some of the multi-Gaussian fits for one of the dates (JD 2455674.82443). 
The normalised, continuum-subtracted observed spectra are displayed as thick black lines.
Individual Gaussian
components are plotted as thin lines (colours: green, red, blue, cyan). The combined fit is
plotted as a magenta thick line. The zero velocities of the central line are also marked
with a vertical dotted line. The plots are labelled with the line(s) elements and wavelength.
Note that the very complex shape of the continuum around the Fe I 5269/70\AA\ lines is better
fitted by two strong Gaussians, one in emission and the other in absorption (partly out of
the figure). These BC have no physical 
meaning/interpretation, but allow us to extract properly the strong NC of both lines.
\label{linefit}}
\end{figure*}

Our aim is not to find the best mathematical fit, but a simple and robust one that
can be used for all the data taken at different dates in order to compare them.
As a result, a direct physical interpretation may be difficult, but we will be able to detect,
quantify, and explore differences in velocities, widths, and amplitudes in the spectral lines.
A multi-Gaussian fit is strongly degenerated, so we imposed a few restrictions
based on the visual aspect of the lines and continuum
to make the models comparable. We used 3 Gaussians with independent amplitudes, widths, and wavelength
offsets. In some cases where more than one line is present within the 
considered section, we added a 4th Gaussian component (for instance, to fit the O I triplet at
7774\AA). 
The lines have a NC+BC structure, so we
request one of the Gaussians to be narrow, with a width between 8-30 km/s
(or two narrow Gaussians in case of two nearby narrow lines such as
Fe I 5269/70\AA).
The Gaussian amplitudes are 
forced to be positive, except for the He I lines, which are better fitted 
by including one Gaussian component in absorption, and  cases where the
emission line appears on top of a photospheric absorption feature (e.g. Fe I 5269/70\AA).
All fits were visually inspected, removing any parts of the line or continuum that were 
affected by spikes.
Figure \ref{linefit} shows some examples of the multi-Gaussian fits for various lines.

We find that the NC fit is independent of the assumptions made on the
fitted Gaussians.
Nevertheless, fitting the BC is strongly degenerated, as most spectra 
can be equally well fitted by various very different Gaussian components.
Therefore, the results of the BC fitting have to be handled with care,
while the amplitudes, velocities, and widths of the NC are very robust.
Details on individual lines are given in the Appendix \ref{appendix-lines}.

\subsubsection{Analysis of the narrow component \label{glsNC}}

\begin{figure*}
\centering
\begin{tabular}{cc}
\includegraphics[width=0.45\linewidth]{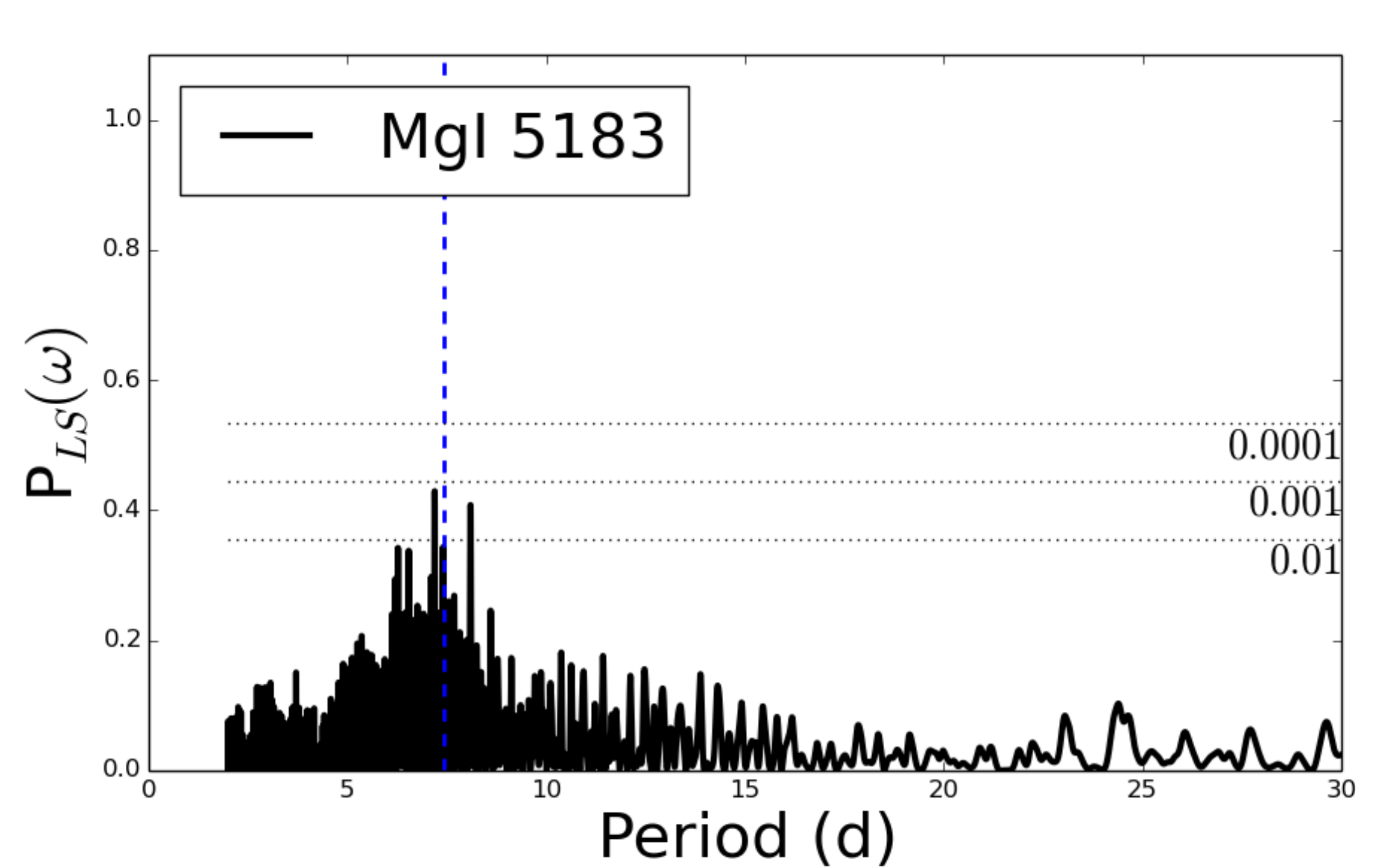} &
\includegraphics[width=0.45\linewidth]{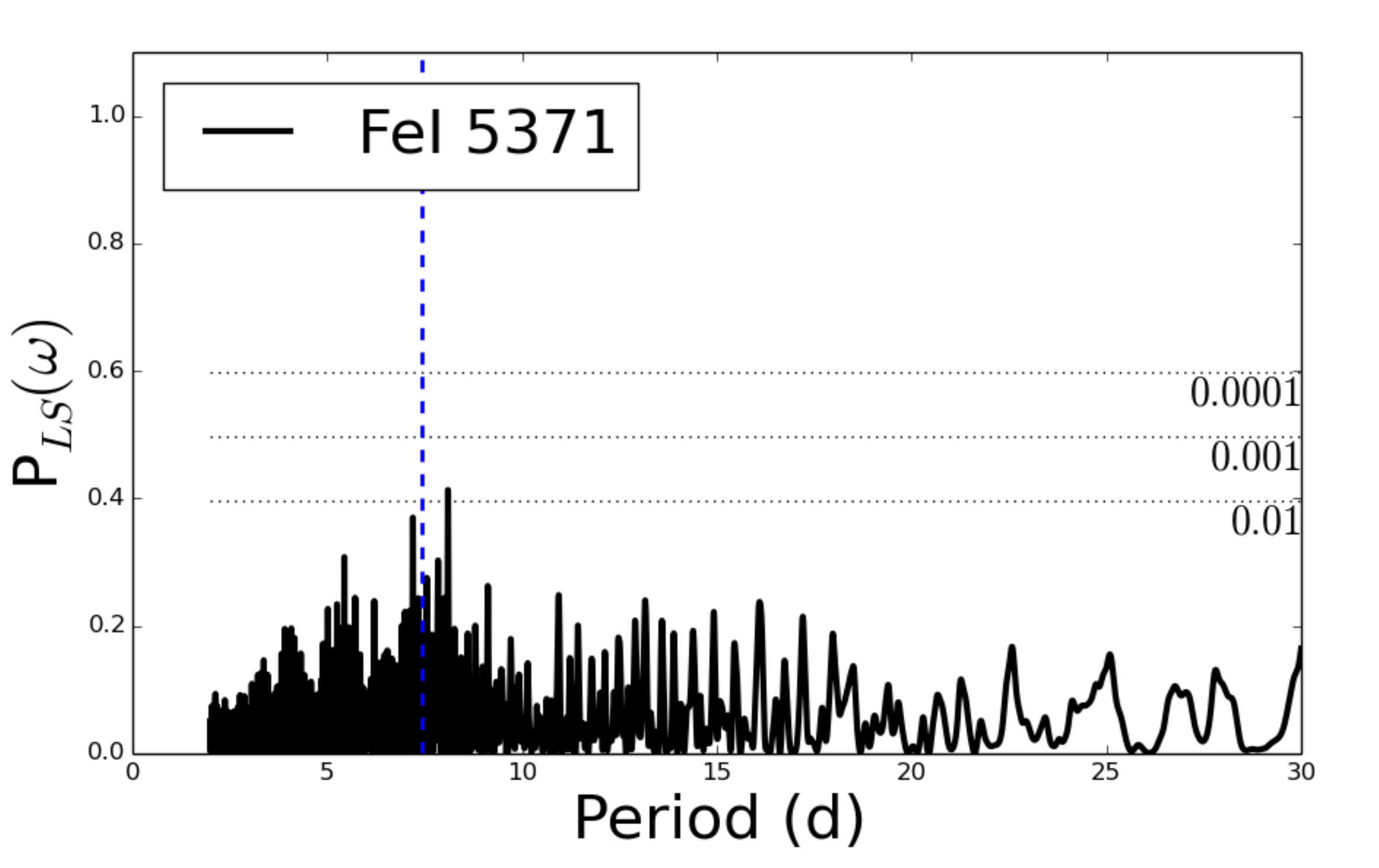} \\
\includegraphics[width=0.45\linewidth]{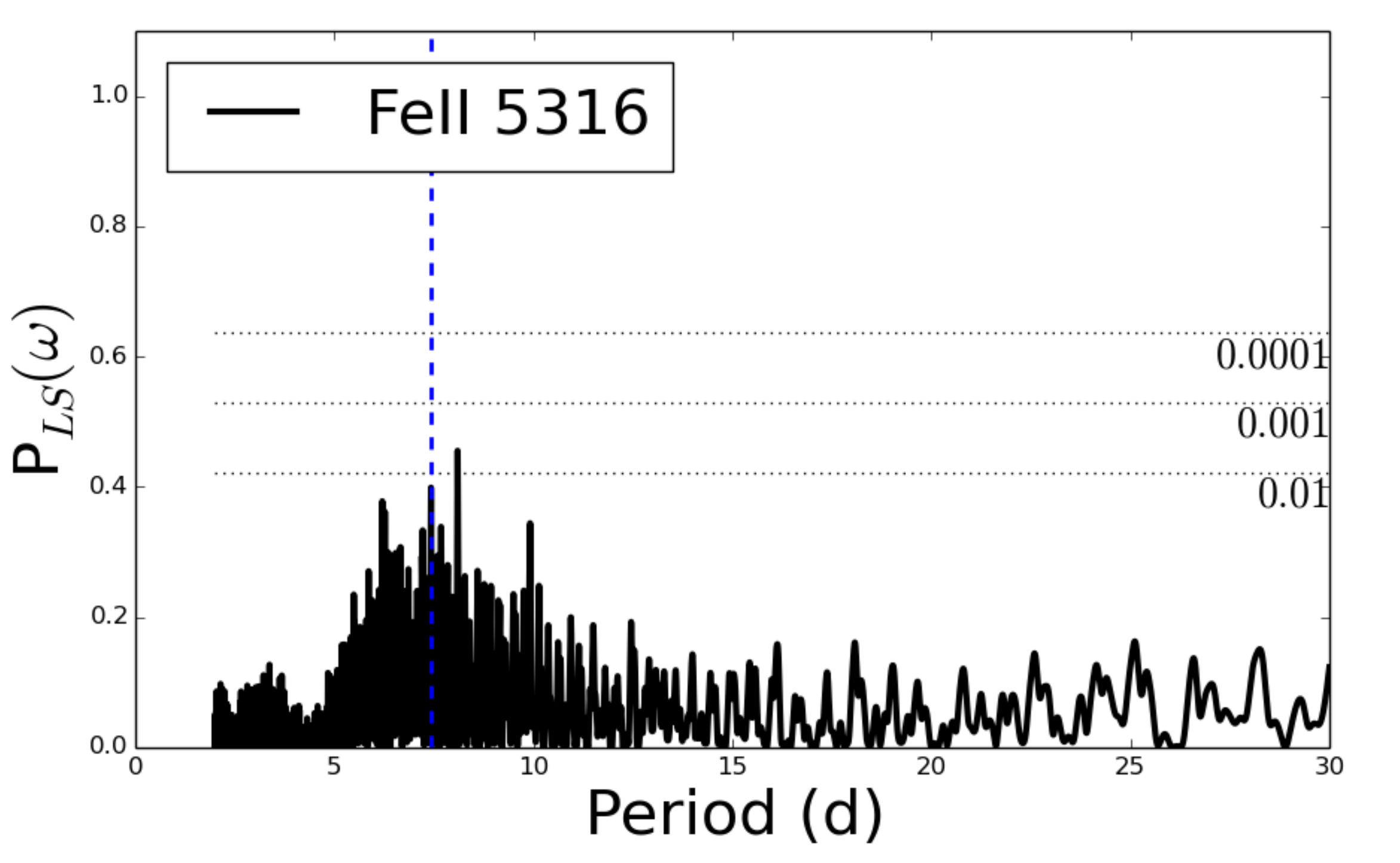} &
\includegraphics[width=0.45\linewidth]{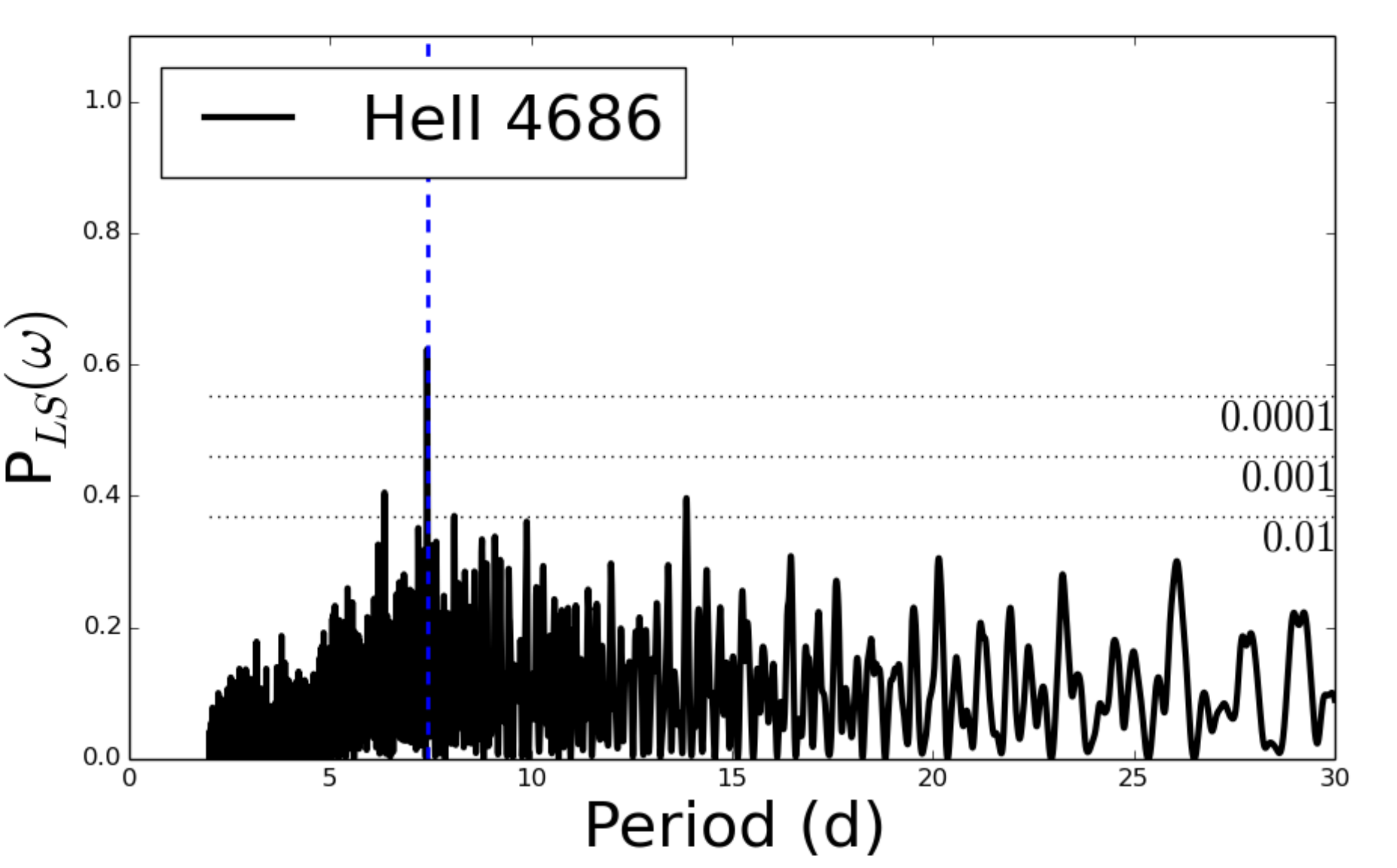} \\
\end{tabular}
\caption{Some examples of GLS periodograms for different lines. The 7.41d period 
(marked as a blue dashed line) becomes
increasingly significant as we move to lines with higher excitation potentials.
The FAP are marked as dotted horizontal lines and labelled accordingly. The situation observed for Mg I is 
typical of the rest of metallic lines labelled as having several peaks.
 \label{glsfig}}
\end{figure*}

Our detailed multi-Gaussian fit allows us to obtain
more accurate velocity measurements than the automated fiting and
averaging method in K14, allowing us to treat each line independently. 
Our discussion here is thus focused on the velocity modulation of the
strong NC emission lines. Weaker narrow emission lines are 
noisier, although they are consistent with the same modulation.
K14 noted that the NC of the emission lines
shows radial velocity patterns that are off-phase with respect to the RV
curves of the photospheric absorption lines. 
To quantify the velocity modulation of the NC emission, we
run a generalised Lomb-Scargle periodogram (GLSP; Scargle 1982; Horne \& Baliunas 1986; 
Zechmeister \& Kuerster 2009) to search for periodic signals. Each line has
a very stable characteristic line width, which can be as narrow as 9 km/s for some Ti II lines and
as broad as 20 km/s for He I lines. The data were clipped based on the NC line width, since
significantly broader lines are usually 
contaminated by the BC, while very narrow ones are often affected by spikes or
low S/N, leading in both cases to inaccurate NC velocities. 
The errors in the line fit can be very complex, considering that the sum of 
Gaussians is a degenerated fit. Therefore, we estimate the errors in the velocity as being proportional
to the line width and inversely proportional to the line strength, including a factor to
account for the finite spectral resolution, resulting in typical errors well below 0.5 km/s.
The clipping procedure leaves us between 30 to over 50 points per emission line, 
with the strongest lines having more valid points than the faint, low S/N ones.
The results of the GLSP (periods and false-alarm probabilities, FAP) 
are summarised in Table \ref{rvemission-table}.

\begin{figure}
\centering
\begin{tabular}{c}
\includegraphics[width=0.9\linewidth]{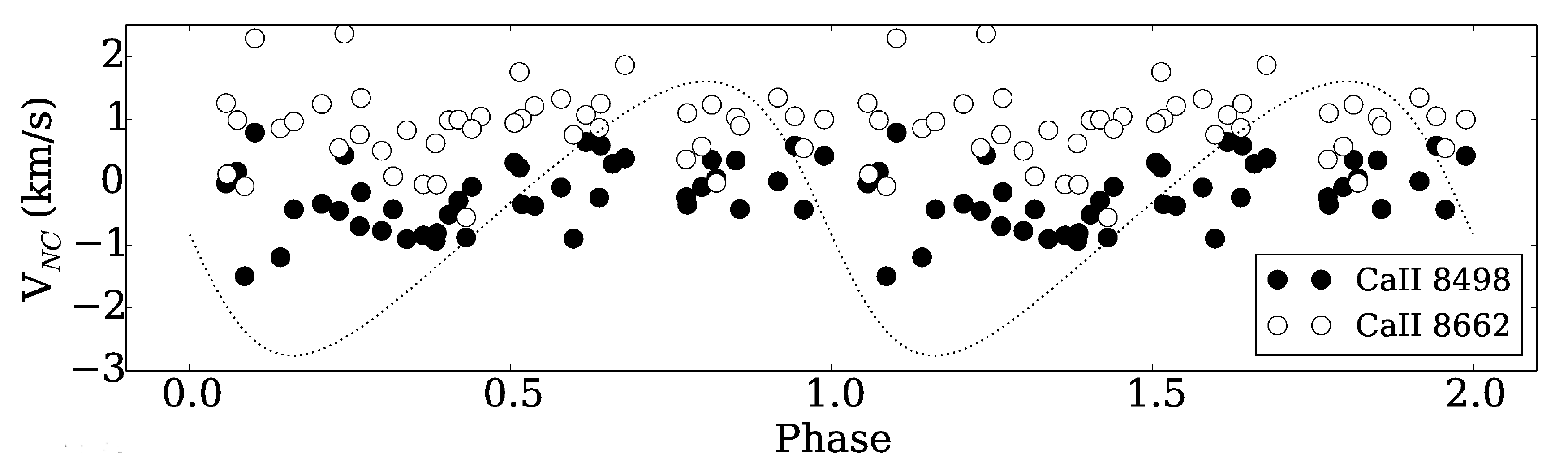} \\
\includegraphics[width=0.9\linewidth]{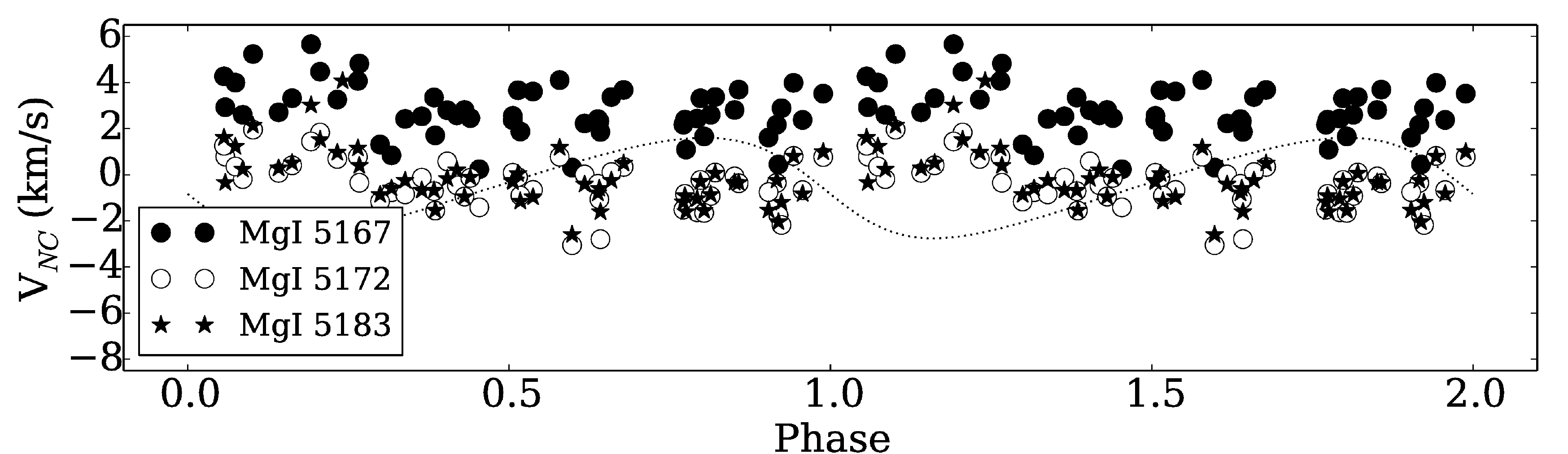} \\
\includegraphics[width=0.9\linewidth]{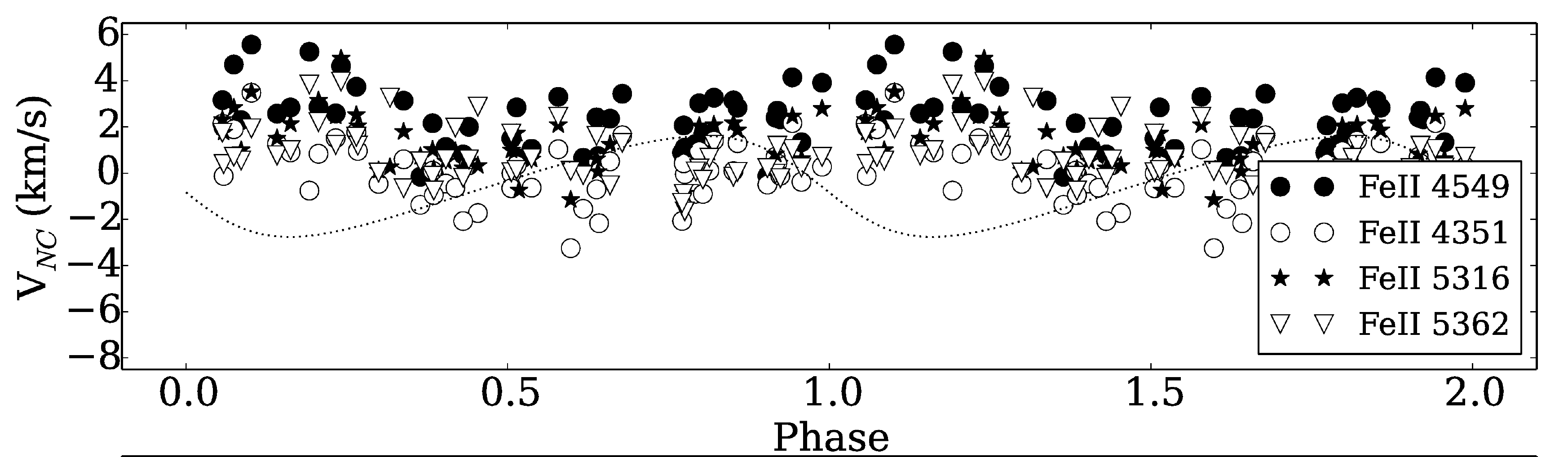} \\
\includegraphics[width=0.9\linewidth]{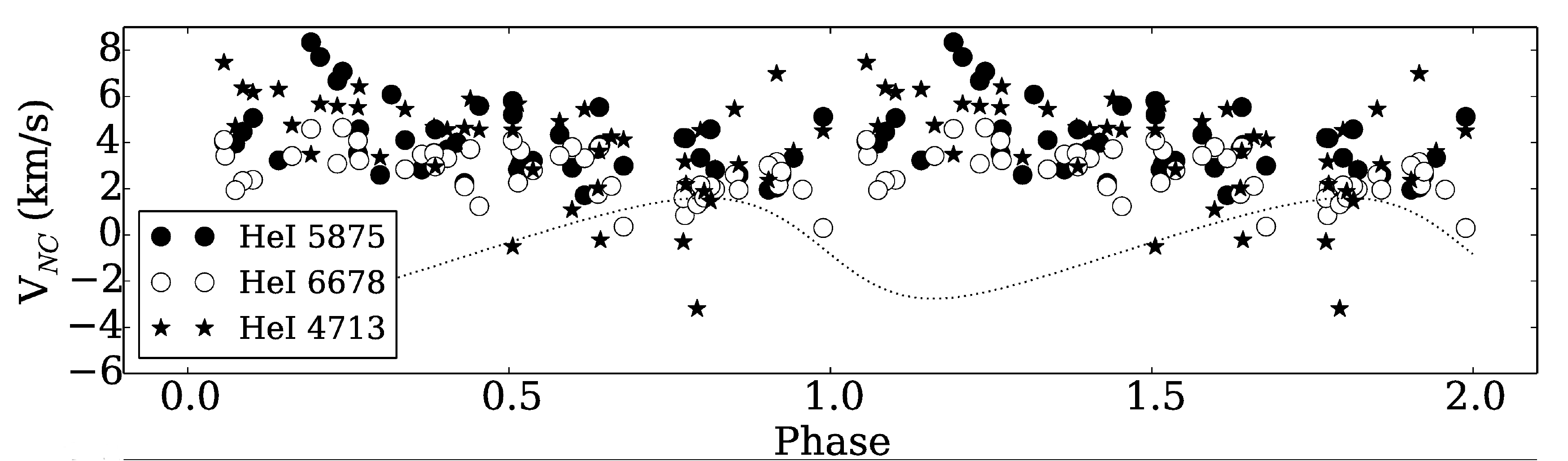} \\
\includegraphics[width=0.9\linewidth]{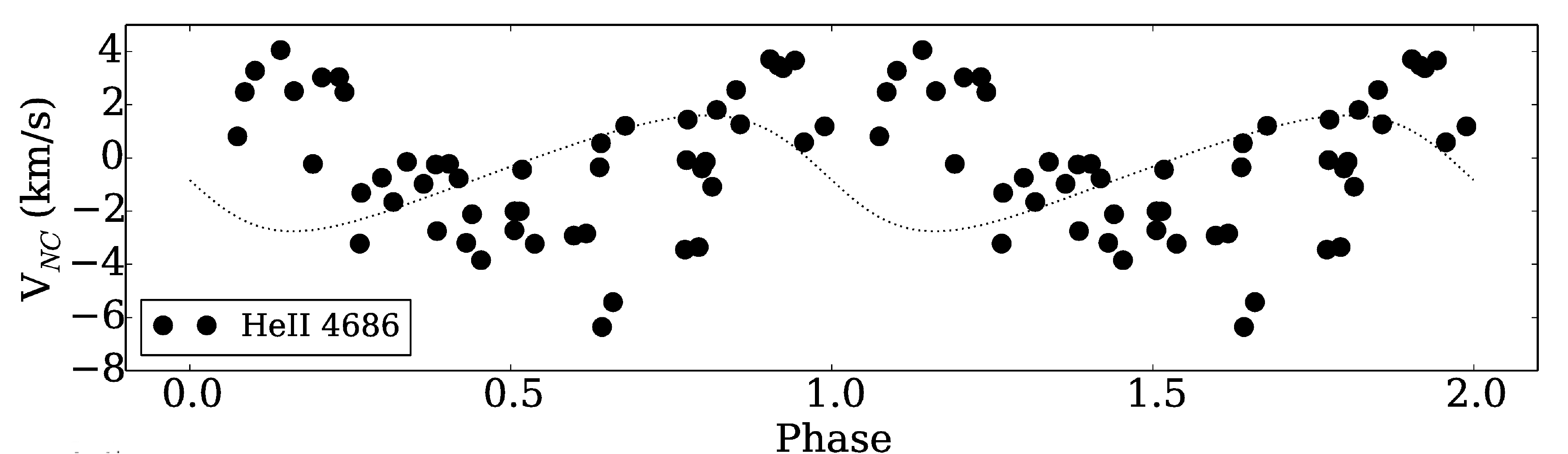} \\
\end{tabular}
\caption{Phase-folded curves (using the photospheric absorption line period and phase, which is nearly
identical to the He II 4686\AA\ one)) for several
of the observed NC of lines. Note that two 7.417d periods are plotted for clarity. 
The dotted line represents the RV curve measured for the photospheric absorption lines (K14).  
 \label{linemoduls}}
\end{figure}

We find that some of the lines show significant periods similar to the radial velocity period 
of the photospheric absorption lines, 7.41d (K14). In most
cases, rather than a single period, we observe a strong modulation and a 
signature of quasi-periodicity, with several similarly-significant
peaks around the 7-8d range. Periods of 8.09 and 7.19d are frequent, and could be aliases
of the 7.41d period. 
The strongest signature is observed for the He II 4686\AA\ line, for which we recover the period
obtained for the photospheric absorption lines by K14, although this line is
not particularly strong compared to others. Some of the strongest
lines, such as the Ca II 8498 and 8662\AA\ lines, show no significant periodicity (as expected
from the lack of modulation; Figure \ref{linemoduls}). Therefore, the presence (or lack of) 
significant periodic signatures is
not an issue of S/N\footnote{Considering the consistent opacity-related 
shift observed between the two Ca II IR lines,
we should be able to detect any modulation with amplitude at the $\sim$0.5 km/s level, if there was any.}, 
but rather due to the underlying structure of the
emitting material. Figure \ref{glsfig}
shows some examples of the resulting periodograms, and details about every individual line are summarized in
the Appendix \ref{support-appendix}, Table \ref{rvemission-table}. 

Plotting the
NC central velocities against the He II (or K14) period reveals that
most of the lines are consistent with similar sinusoidal periodic modulations
with the same period than the RV signatures of the absorption lines (Figure \ref{linemoduls}). 
The RV modulation depends on the line considered, ranging from lines that show hardly any
modulation (such as the NC of the Ca II IR triplet), to others where the amplitude of the RV
modulation is much larger than the RV amplitude of the photospheric absorption lines 
(e.g. He I, He II show modulations up to 8 km/s peak-to-peak amplitude, 
compared to 4.36 km/s for the photospheric absorption lines).
Among the  
parameters in the  Gaussian fit (RV of the center of the Gaussian, amplitude, 
and width), only the  RV shows a significant
modulation. The line width is remarkably stable within the same species and transition, and
the amplitude shows low-significance modulations that could be related to orientation
of the emitting structure (Section \ref{results}). Interestingly, although the BC appear more intense during
higher accretion phases, there is no evidence that the variations of the accretion
rate for the quiescence data produce any significant change in the properties and velocities of the NC 
(see Appendix \ref{acc-appendix}). 

Most of the lines are redshifted by 
different amounts (from less than $\sim$1 km/s for the Ca II
IR lines, up to $\sim$5 km/s for He lines) with respect to the zero-point stellar velocity
(-0.52 km/s; K14). This suggests that the lines are not produced
at the stellar photosphere, but in a slightly infalling structure. Since the redshifted velocities of
the lines also vary from element to element, the various species are not
produced in the same place of the structure. Significantly redshifted He I lines (compared to the stellar
velocity and to Fe I/II lines) have been also observed in other stars by
Beristain et al. (1998, 2001) and Gahm et al. (2013).

Figure \ref{linemoduls} also reveals a systematic offset between 
lines from the same multiplet. This is particularly remarkable for the 
8498 and 8662\AA\ lines of the Ca II IR triplet, and can be explained by
differences in the optical thickness of the lines. Although the third component of the triplet, at
8542\AA\, is very close to the FEROS spectrograph gap, the NC is often within the FEROS coverage, 
displaying a behaviour comparable to the 8498\AA\ line in peak intensity and zero-point velocity,
while the 8662\AA\ line is redshifted by $\sim$1km/s with respect to them and has a lower peak
intensity. The 8498\AA\ NC line has a tendency to being narrower than the other two.
Hamann \& Persson (1992) noticed
the difference in intensity and thickness of the 8498\AA\ vs 8542-8662\AA\ lines in CTTS (assuming
a chromospheric origin for the lines) and offer two
explanations:
The 8662\AA\ reduced intensity could be caused by its upper level 
being partly depopulated by interactions between the Ca II H line
(which shares the same level) and the H$\epsilon$ line. Regarding the line thickness (and, in
our case, also probably the offset velocity variation), an explanation may lie in the 9-times
lower opacity of the 8498\AA\ line, compared to 8662\AA. This can cause a change in the line
profile (the optically thicker line is broader) and also means that the 8498\AA\ line would
be produced in a deeper place, compared to the 8662\AA\ line. Hamann \& Persson (1992) suggest
a temperature inversion causing a source function inversion in the zone where the Ca II IR triplet originates, which 
would mean that the 8498\AA\ line would be produced in the deepest place, the 
8662\AA\ line would be produced at the minimum, and the 8542\AA\ line would originate where the temperature
raises again. Such a situation could also  arise in a post-shock cooling region (Section \ref{dynamics}). 
A similar explanation could also fit the differences between the Mg I 5167/5172/5183\AA\ triplet, which
also shows differences in strength and radial velocity between the 5167\AA\ line and the rest, although
the 5167\AA\ line coud be in part affected by a strong, nearby Fe I line.

\subsubsection{Analysis of the broad component}

We also searched for
periodic signatures in the strong, rapidly variable BC of the Ca II 8498 and 8662\AA\ lines.
The span of peak velocities is over $\pm$100 km/s, and similarly good fits may show variations
by several tens of km/s.  We first compared the independent fits for the 8498 and 8662\AA\ lines
to check how reliable and stable the BC velocity determinations are.
We estimated the Spearman rank correlation for all the three Gaussian fits components:
amplitude, central wavelength, and width. 
Cross-correlating the results for the two lines (see Appendix \ref{appendix-xcorCaII}
for details)
shows that there is a very strong correlation between the individual parameters of both lines.
This means that we are obtaining similar fits for these two very similar lines, despite
variations in the noise and continuum between both.
For most of our data points, a single broad Gaussian dominates the
BC fit, and the second broad Gaussian component is often marginal and/or poorly
constrained, which leads to degeneracy.

For a given line, we checked the cross-correlation between the fit parameters, to investigate
how changes in one of the components (NC or the two broad Gaussian components) are reflected in the other ones.
Appendix \ref{appendix-xcorCaII} details the procedure and results. 
We find that the NC is relatively independent in amplitude and central velocity
from the BC. Only the width of the NC seems to increase slightly when the intensity of the BC increases.
This could be a sign of an increase of the local turbulence at near-zero velocities for higher
accretion rates (expected by comparison with the outburst observations; SA12), but it
could also reflect contamination of the NC by a more intense BC. 
The centre and the amplitude of the two BC Gaussians are correlated, so their velocity
increases when the amplitude of the BC increases. This is similar to what we observed during outburst: at higher
accretion rates, the material seems to be dragged from more distant places or, conversely,
more distant parts reach the required temperature and density to produce emission lines (SA12).
This is consistent with accretion models for the H$\alpha$ line (Lima et al. 2010).
We do not find any significant correlation between the blueshifted and redshifted BC Gaussian components, 
which is a sign that they are independent and not caused by the same 
physical process (for instance, an increase in the accretion rate triggering an accretion-powered wind).
They seem only linked by temporal variability, as expected for
bulk motion of material rotating as it falls onto the star.

Finally, we searched for periodic signals in the BC. 
The centering errors for a broad Gaussian are much larger than 
for the NC, so we also checked the location of the peak of the BC 
emission (after subtracting the NC), which is less affected by the choice of fit. 
We only consider those
data points where the amplitude of the BC is $\geq$0.3 (over the normalised and subtracted
continuum, well above the noise level). 
The results of the GLSP for the BC
are listed in Table \ref{rvemission-table}. For the BC Gaussian components, the periodicity analysis is
not conclusive, mostly due to the small number of data points. The BC peak 
shows multiple low-significance periods, in agreement
with the periodic or quasi-periodic 
signatures in the 7-8d range observed in the NC of the lines. 
Examining the shape of the BC line (see Appendix \ref{appendix-lines}), we also find a trend with the phase of the
NC/photospheric absorption lines, especially for the cases with strongly bluesifted BC. 
This suggest
that there is a common mechanism causing the velocity modulations of all emission lines and 
that the NC and BC are physically related.

\section{Dynamical and physical models for the spot, the star, and the accretion column \label{models}}

\begin{figure*}
\centering
\includegraphics[width=0.8\linewidth]{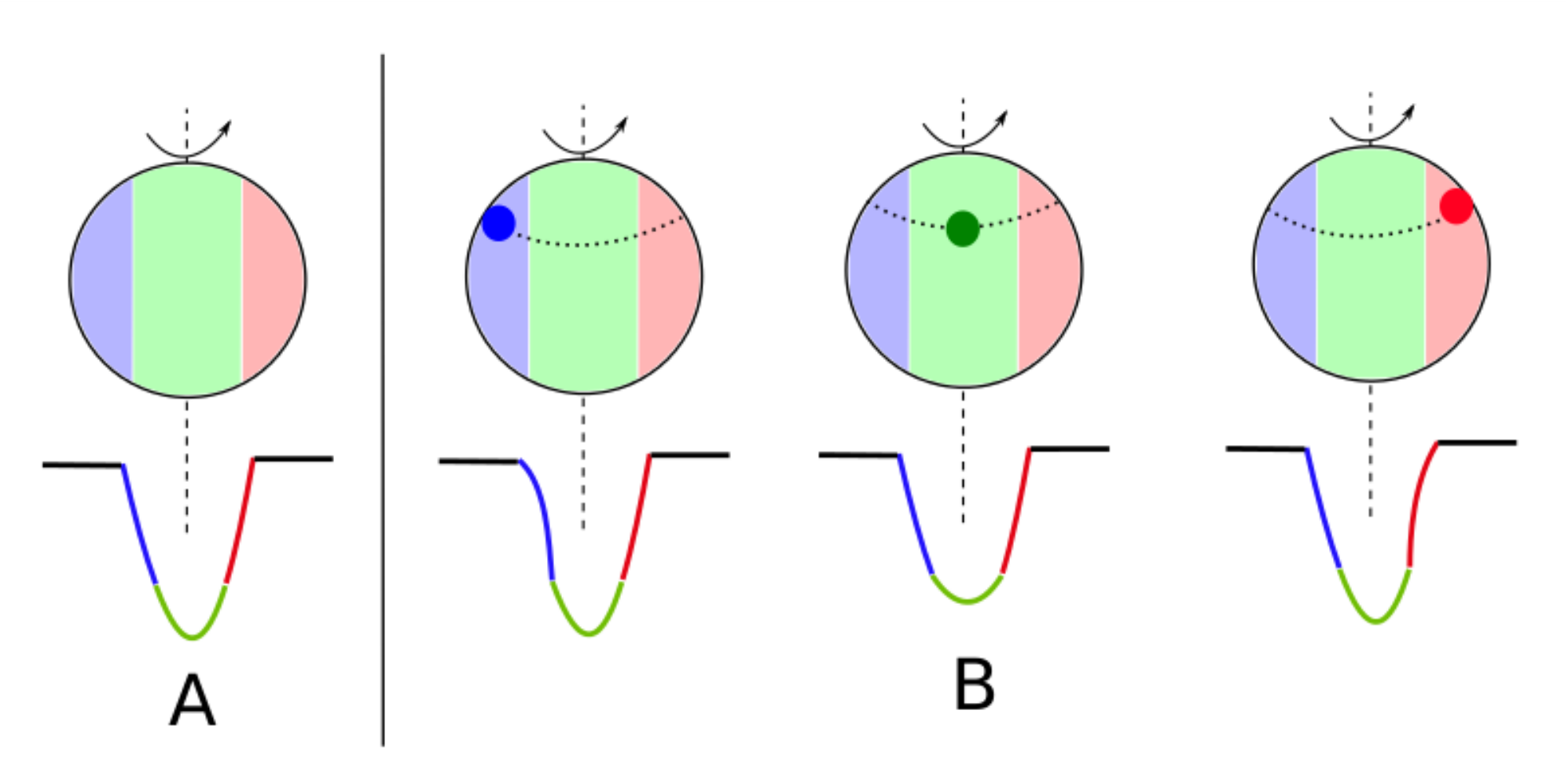} 
\caption{A scheme of how the photospheric absorption lines are affected when a hot
spot moves in and out of our line of sight when the star rotates. In A we show the different parts
of the stellar photosphere that produce the photospheric absorption lines. B shows how the
line is affected when a hot spot is included. When the absorption deficit
associated to the spot appears blueshifted, the centre of the photospheric lines seems to move
towards the red. When the spot appears redshifted, the centre of the absorption line appears
shifted towards the blue. Small emission lines are visible within some of the photospheric 
absorption lines, tracking the local rotational velocity of the part of the stellar surface
where the line originates. Non-photospheric lines or lines
that are particularly strong on the spot could appear simply as emission lines over the continuum,
without any underlying absorption.
 \label{thespotcartoon}}
\end{figure*}

In this section, we construct simple models 
to explain the RV signatures of the emission lines in EX Lupi,
and explore the possibilities of a stratified 3-D hot structure to reproduce the dynamics and 
the strengths of emission lines. We
assume that the  footprint of the accretion structure behaves like a hot spot.
A hot spot on the stellar surface would cause a clear modulation of the photospheric absorption lines
as shown in Figure \ref{thespotcartoon}, and could mimick the RV signature
caused by a companion. The main challenge for the spot scenario
is that different lines have different RV modulations,
which is inconsistent with plain flat hot spot models. Nevertheless, accretion can alter the vertical
temperature structure of the photosphere, suppressing normal absorption line formation in the accreting
region. The extent to which the absorption is suppressed will depend on the strength and depth of the
formation of the lines concerned (one example being the Li I 6708 \AA\ line, the only 
photospheric line that still presents a small absorption component during outburst; SA12).
The diversity of the
amplitudes and the zero-point offsets between lines  is a sign of a
more complex 3-dimensional (3-D) accretion structure,
extended in the vertical (radial) direction and having a temperature and density gradient within the
flow and/or accross the accretion column. This is consistent with the observations during outburst (SA12), and with evidence
found in other T Tauri stars (Dupree et al. 2012; Petrov et al. 2014).

\subsection{The dynamics of narrow and broad components and their RV signatures \label{dynamics}}

\begin{table}
\caption{Amplitudes and offsets of a sinusoidal fit to the emission line signatures.} 
\label{sinusoidal-table}
\begin{tabular}{l c c c }
\hline\hline
Line & A  & V$_z$   &   $\phi_0$ \\
 & (km/s) & (km/s)  &   (deg)  \\
\hline
He II 4686 & 3.06$\pm$0.28 & 0.66$\pm$0.19 & 11.8$\pm$0.8 \\
He I 4713 & 1.74$\pm$0.26 & 4.55$\pm$0.19 & 3.1$\pm$1.4  \\
He I 5875 & 1.03$\pm$0.19 & 4.46$\pm$0.14  & -0.3$\pm$1.8 \\
He I 6678 & 0.81$\pm$0.12 & 3.28$\pm$0.09 & -2.7$\pm$1.5 \\
Fe II 4351 & 1.06$\pm$0.16  & 0.70$\pm$0.11 & 9.8$\pm$1.3 \\
Fe II 4549 & 1.10$\pm$0.18  & 2.85$\pm$0.12 & 8.5$\pm$1.4  \\
Fe II 5316 & 0.99$\pm$0.16 & 1.86$\pm$0.11  & 8.0$\pm$1.4  \\
Fe II 5362 & 0.75$\pm$0.15 & 1.26$\pm$0.11 & -0.19$\pm$1.9 \\
Fe I 4215 & 1.05$\pm$0.16 & 3.14$\pm$0.10  & 4.8$\pm$1.2 \\
Fe I 4533 & 1.58$\pm$0.33 & 3.00$\pm$0.24 & 0.7$\pm$2.0 \\
Fe I 5270 & 0.86$\pm$0.14 & 0.12$\pm$0.09 & 6.2$\pm$1.4 \\
Fe I 5370 & 1.50$\pm$0.24 & 0.72$\pm$0.19 & 2.9$\pm$1.7  \\
Fe I 5371 & 1.15$\pm$0.18 & -0.06$\pm$0.12 & 4.4$\pm$1.3 \\
Mg I 5167$^1$ & 0.83$\pm$0.17 & 3.44$\pm$0.12 & 4.8$\pm$1.8 \\
Mg I 5172 & 0.71$\pm$0.13 & 0.21$\pm$0.09 & 4.8$\pm$1.3 \\
Mg I 5183 & 1.06$\pm$0.14 & 0.41$\pm$0.10 & 3.9$\pm$1.2 \\
Ca II 8498$^2$ & 0.31$\pm$0.07 & 0.33$\pm$0.05  & -2.6$\pm$2.3  \\
Ca II 8662$^2$ & 0.14$\pm$0.10 & 1.46$\pm$0.07  & 1.7$\pm$6.1 \\
\hline        
\end{tabular}
\tablefoot{Fitting the dynamical signatures (Figure \ref{linemoduls}) with a sinusoidal modulation
(RV=V$_0$+V$_z$+A$\sin$[2$\pi$($\phi$+$\phi_0$)]. "A" (amplitude) refers to the strength of 
the sinusoidal
modulation, "V$_z$" (zero-point velocity) is the general offset of the line, 
after taking into account the zero-point velocity of the star (V$_0$=-0.52 km/s; K14),
The phase is "$\phi$" (for a 7.417d period), and "$\phi_0$", the phase offset.
The values here correspond to the "best fit", but some of the parameters, such as the
amplitude, have a relatively large spread (0.5-1.0 km/s) comparing data taken at different dates and
similar phases. $^1$ The Mg I 5167\AA\ line may be affected by a nearby Fe I line. 
$^2$ The modulation of the Ca II lines is not significant, as
explained in the text. The fits for the Ca II lines are listed to show the evident differences
(especially, regarding the amplitude) with the rest of lines that have periodic modulations
consistent with those of He II and the photospheric absorption lines.}
\end{table}

\begin{figure*}
\centering
\includegraphics[width=0.8\linewidth]{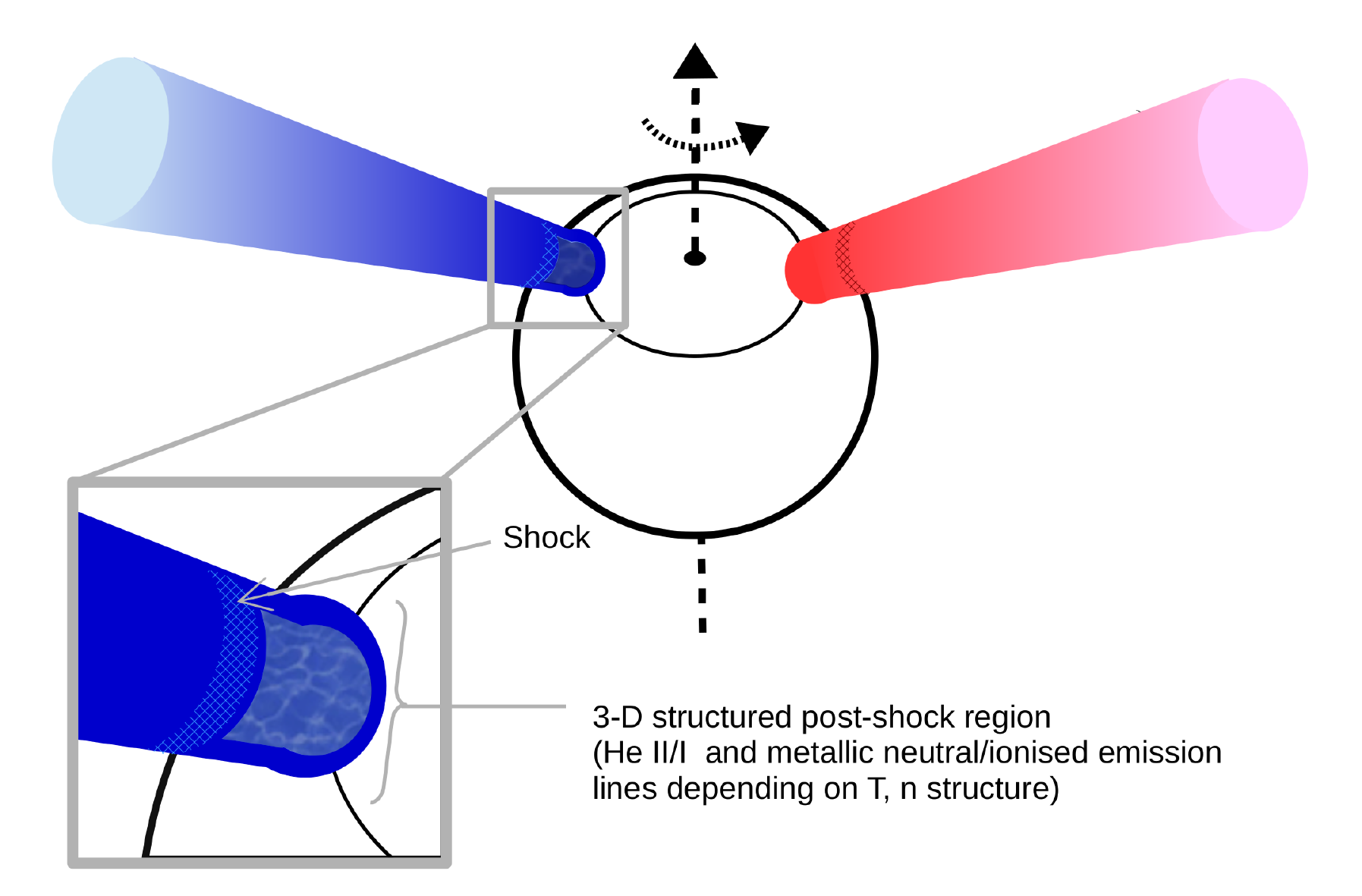} 
\caption{A cartoon of a star with 
an accretion column that is visible at all times due to the stellar inclination and whose emission
lines suffer Doppler shift as the star rotates (the observer is located perpendicular to the plane of the
paper). The blue (red) column shows the position at which the lines originated in it would appear blueshifted (redshifted) to an
observer looking from out of the page. The hashed areas mark the shock region. We 
suggest that the observed narrow emission lines are produced in the post-shock area,
which is stratified in density and temperature.
Lines produced in the shock region (He II, requiring very hot and dense conditions) originate 
in a nearly flat structure, so their periodic signatures are much stronger and less messy than other lines
produced in the extended, infalling post-shock region. The BC is produced in the accretion column before the
pre-shock region, showing higher velocity offsets as they rotate with/around the star. 
Not to scale: note that the shock region is at a very small distance over the photosphere, compared to
the stellar radius. \label{cartoon}}
\end{figure*}

The location of the line-emitting region can be inferred by modeling the observed 
dynamical signatures and radial velocities
(Figure \ref{linemoduls}). If the NC lines are produced in a localised accretion spot structure,
stellar rotation would affect their radial velocities imprinting a sinusoidal modulation. A similar
scenario  has been proposed for UV lines, for which some stars are also consistent with 
accretion-related spots getting in and out of view as the star rotates, and showing different lines-of-sight
depending on whether they form in the accretion column or at the associated spot (Romanova et al. 2004; Ardila et al. 2013).
Simulations predict that the coverage of the spot decreases as the density of the accretion column increases 
and at small misaligment angles between the stellar rotation and the magnetic moment (Romanova et al. 2004).
Given that the accretion structures in EX Lupi are expected to be very dense to produce the observed emission lines
(see Section \ref{physics}), accretion models would favour very localised spot(s). 
The NC emission lines present a sinusoidal modulation
with the same periodicity as observed for the absorption lines (Figure \ref{linemoduls}), 
but variable amplitude and zero-point shift 
(Table \ref{sinusoidal-table}). For lines with lower excitation potentials, 
the significance of the periods is lower and the 
signature is quasi-periodic rather than purely periodic. This can happen if
they are produced in more
extended and potentially messy structures (Figure \ref{cartoon}).

The amplitude is rather constant within a given 
species (e.g. He I, Fe I, Fe II), although there are significant shifts in the zero-point
velocity, which is always redshifted.
The maximum shift is observed for the He I lines, which is similar to the results of
Beristain et al. (2001) on a large sample of CTTS, although the fact that this line could
be in part affected by a redshifted absorption (due for instance, to wind; Section \ref{gaussfit}) could
also led to a higher redshift. Differences in zero-point velocity can also result from differences in 
the optical thickness (and thus depth) of the lines (Section \ref{glsNC}).
The differences in profile/velocity between high- and low-temperature lines would be
associated to different parts of the accretion structure. Detailed models (Romanova et al. 2004, 2011; 
Ardila et al. 2013) have suggested various possibilities,
ranging from vertical stratification (e.g., in an accretion structure with an aspect ratio where photons can easily
escape through the walls), to horizontal stratification accross the accretion structure 
(where the low-density, cooler material at the edges of
the accretion column would also show a slower-than-free-fall motion). A further possibility, the presence of
several accretion columns with various densities around the stellar surface, appears unlikely for
EX Lupi, given that all lines show a very similar modulation (period and
phase). An exception could be the Ca II IR NC lines, which do not show modulation and could be
produced in a more uniform way over the stellar surface. 

We can use the observed sinusoidal modulation to investigate the typical location of the emitting
structure (Romanova et al. 2004). The post-shock region at the bottom of the accretion column would 
be similar to a plage as seen in active stars (Ardila et al. 2013; Dumusque et al. 2014),
although in this case, our "plage" would be 3-D and stratified in density, velocity, and temperature.
If a plage (or accretion structure) 
is at a low latitude, it would only be visible during about half of the rotational
period, producing an incomplete sinusoidal modulation (from blueshifted to redshifted; the part from redshifted to
blueshifted would not be visible). We 
observe a full modulation, which requires the plage-like structure to be visible during the whole
rotational period.   If we observe the star under an inclination angle $i$, this happens for
latitudes higher than $i$ (or, in spherical coordinates, $\theta$<90-$i$). 
The observed modulation would be caused by rotation and the 7.41d period would
correspond to the rotational period of the star.

For an emission line produced in a small spot located on the surface of the star 
at spherical coordinate $\theta_s$, the rotation 
of the star will change the central velocity of the line following:
\begin{equation}
	RV_{line}=V'_0+\frac{2\pi R_*}{P}\,\sin i \, \sin \theta_s \sin \phi = V'_0 + v\sin i \,\sin\theta_s \sin\phi \label{RVline}
\end{equation}
\noindent Here, V$'_0$ is any zero-point velocity that the system or the emitting zone may have (including the stellar
velocity, V$_0$, and any systemic velocity in the accreting structure, such as a systematic redshift in an infalling component,
V$_z$ in Table \ref{sinusoidal-table}),
P is the rotational period of the star, $\phi$ corresponds to the phase angle (from 0 to 2$\pi$),
and $v$sin$i$ is the projected rotational velocity of the star. 
If the spot does not change its latitude
during the time we observe it, we obtain a sinusoidal modulation with a radial velocity amplitude that
depends on the viewing angle $i$ and the spot location $\theta_s$. A purely polar spot ($\theta_s$=0;
such as the one proposed by Dupree et al. 2012 for TW Hya) suffers no
modulation. For an equatorial spot, the radial velocity amplitude is maximal,
although it will not be visible at all times. 
The maximum radial velocity that can be observed in the emission lines produced \textit{at the surface}
of the star is thus $v$sin$i$. Lines originating at some distance (d) over the
surface of the star could have radial velocity amplitudes up to 2$\pi$(R$_*$+d)$\sin i$/P,
assuming that the line-emitting structure rotates as a solid with the star\footnote{This may not hold as we move 
away from the object. Note that from the V$_z$ of the lines, if infall is assumed, they would be
produced very close to the stellar surface, so d$<<$R$_*$.}.
Similarly, if the extended structure where the lines are produced is displaced off the radial direction
(for instance, trailing the star, or extended along the stellar surface), 
we would expect to observe some phase shift between lines
originated at various altitudes. Although significant differences in zero-point velocity and
phase are found comparing various lines (Table \ref{sinusoidal-table}), optical depth differences between lines 
and unresolved absorption (e.g. wind components) could also produce changes in both 
measured quantities that cannot be
properly estimated without detailed radiative transfer models.

Assuming that the period is 7.417d and $v$sin$i$=4.4 km/s, the system inclination angle $i$
would be 23.8-25.5 degrees (for radius 1.5-1.6R$_\odot$).
For the He II line, we observe a mean amplitude of about 3 km/s, corresponding
to $\theta_s\sim$43 degrees (latitude $\sim$47 degrees). For an inclination of 23.8-25.5 degrees, this is compatible with
a spot that is visible at all times, in agreement with the observed modulation. 

The information in Table \ref{sinusoidal-table} can be used to infer the relative
location of the various species along the emitting region, although for species with a large range 
in excitation conditions and critical densities (e.g. Fe I and Fe II), 
line strength and opacity differences can make this stratification not so evident.
We explore two options for stratification: along the extended spot vs across the
spot (Ardila et al. 2013). In the first case, material closer to the shock region would be expected
to be hotter and denser compared to material closer to the stellar photosphere, and the relative velocities
would depend on a mixture of infall and magnetic/gas pressure. In the second, the
edges of the post-shock structure would be cooler and less dense, and simulations predict it to show slower
infall velocities compared to the hot and dense column centre (Romanova et al. 2004). This situation is
analog to the observations of UV lines in CTTS (Ardila et al. 2013).

The He II line has by far the strongest requirements regarding high temperature and density, which would require
formation relatively close to the shock and/or deep within the post-shock structure. In both cases, this
spatial restriction would favour a stronger rotational modulation, compared to other lines, which is
also observed. Depending on
the velocity structure in the shock/post-shock region, its null zero-point velocity shift could suggest
excitation  very close to the shock and with a strong magnetic/gas pressure support
accounting for the lowest infall velocity. In case of stratification accross the column,
He II would be expected to form in the innermost part, although in this case its expected infall velocity would be
higher than for lines formed in the periphery (Romanova et al. 2004). The difference in phase could be caused by
location higher on the post-shock structure (compared to the stellar photosphere) or by a bow-shaped spot where
the density varies along the stellar longitude (Romanova et al. 2004).

The He I and metallic lines would be produced in cooler, less dense parts of the post-shock region. 
Their small redshifted velocities suggest that the material, despite being basically
at rest (due to magnetic/gas pressure), is still accreting or infalling.
Taking into account the phase differences, after He II, most of the Fe II emission would be produced closer to the
shock area or to the denser part on a bow-shaped spot (having the next large phase 
difference and lower infall velocity). He I lines also require hot and dense material. The
nearly zero phase difference observed for He I line and the relatively large infall velocities (if blueshifted absorption
due to wind is negligible) would in fact suggest an origin very close to the
stellar surface, although the higher He I abundance (compared to
metals) can also result in emission over larger areas/volumes, including more extended parts over the stellar surface and
eventually, winds. Fe I and Mg I lines would be mainly emitted in an area with lower temperature
conditions, comparable to photospheric temperature/conditions of G/F stars.
For an infall velocity of 4.5 km/s (typical for the  He I lines), the material should be falling 
from 1.4$\times 10^{-4}$R$_*$ ($\sim$160 km) over the stellar surface. Although gas/magnetic pressure and turbulence
counteract the infall, models suggest that fully ionised post-shock columns have sizes of the
order of 10$^2$-10$^5$ km (Dupree et al. 2012; Ardila et al. 2013), compatible with our findings of a relatively compact
region compared to the stellar radius.
Figure \ref{cartoon} shows a scheme of the proposed configuration.

We can also use Equation \ref{RVline} to investigate the effects of such emission lines on the radial velocity
of the absorption lines. We can model a photospheric absorption line using a Voigt profile that reproduces
a typical, unblended photospheric absorption line.
We then add a small emission component with a central velocity that has a sinusoidal modulation
such as shown in Equation \ref{RVline}. A continuum veiling contribution can also be added.
We estimate the apparent radial velocity variation of the resulting absorption line 
by cross-correlation. The modification of the stellar photosphere by the proximity of an accretion hot spot can be very
complex, depending on plasma parameters, and can also contribute to powering a stellar wind
(Orlando et al. 2010, 2013).
Therefore, this analysis does not pretend to derive strong constraints on the properties of the
line-dependent veiling, since it involves too many free (or poorly
constrained) parameters, but to demonstrate that a simple accretion-plus-rotation model can
explain simultaneously the observed RV signatures in the emission and the absorption lines.

We first considered the effect of the narrow emission lines, finding that although the presence of a small 
NC within the line can produce radial velocity offsets as large as observed, they also produce a very strong distortion
of the line profile, which is not observed (some lines have narrow emission-line
cores, but they were excluded from the radial velocity analysis in K14). 
We then consider the effect of a broad component in emission on the absorption line. Although broad components with
emission over the continuum are only 
seen in certain lines (e.g. Ca II IR triplet, strong Fe II lines such as
the 5018 and 4923 \AA\ lines), we also see clear evidence of line-dependent veiling in EX Lupi, 
as proposed by Dodin \& Lamzin (2012) for other CTTS.
A single veiling factor (or a slowly changing one, such as resulting from a black-body
approximation to the veiling) cannot reproduce all the lines, even if we restrict ourselves to a very small portion
of the spectrum (see more details in Section \ref{physics}). This is a sign that many of the absorption lines,
if not all, have a line-dependent veiling, or extra emission that fills part of the absorption line
and that is at least as broad as the absorption line itself (to avoid distorting the line profile). 
As the accretion rate increases, the line-dependent veiling would increase,
filling in the whole absorption line and eventually showing as a broad emission profile for very high
accretion rates, as observed in
outburst (SA12).

\begin{figure*}
\centering
\includegraphics[width=1.0\linewidth]{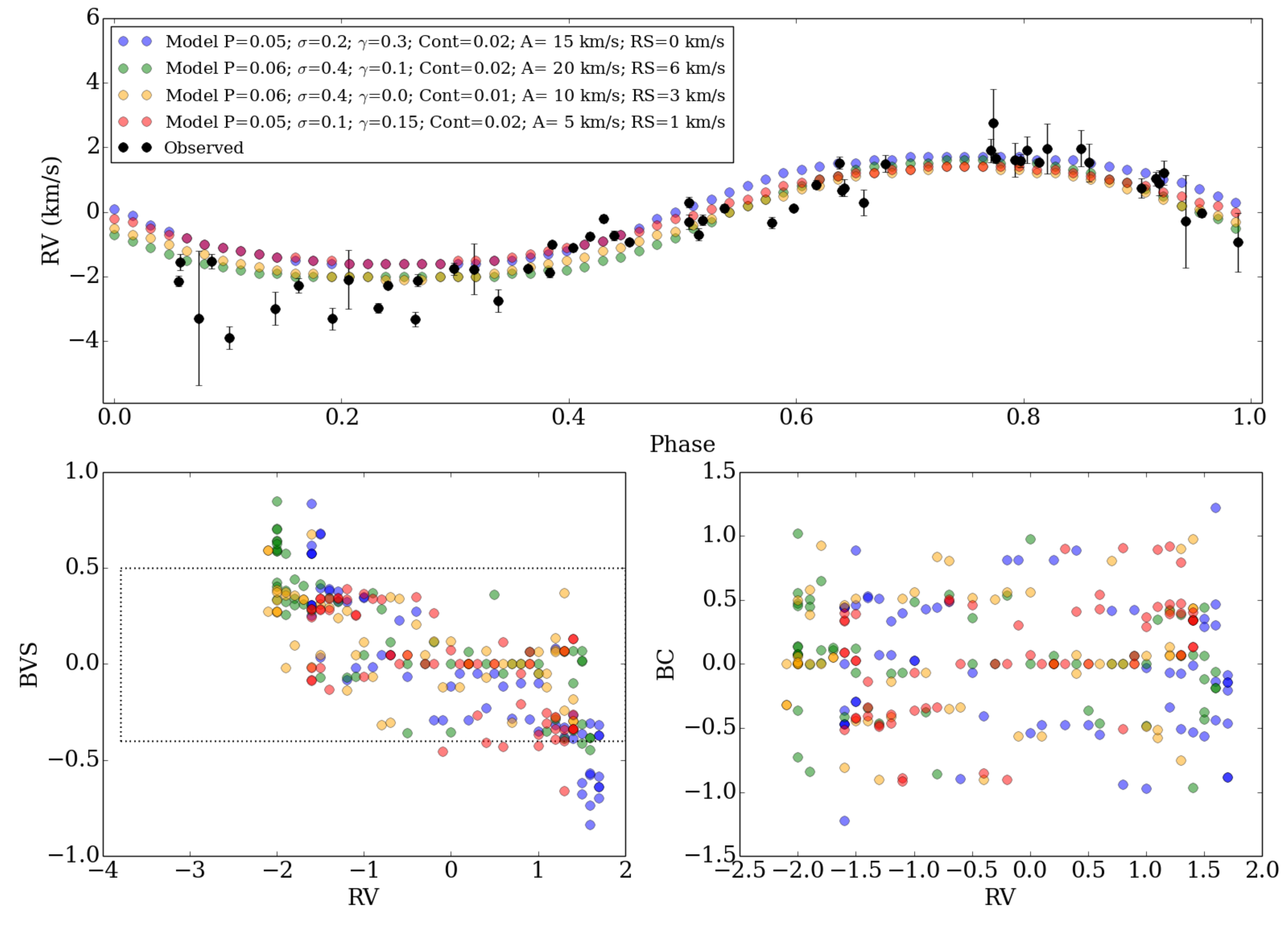} 
\caption{Results of our toy model for the radial velocity signatures (upper panel), and bisector velocity shift (BVS) and
curvature (BC; lower panels). It shows the kind of 
radial velocity modulations that can be induced on a photospheric absorption line
by an emission component (broader than the absorption line) that shows a rotational modulation. 'P' denotes the peak value of  an emission
component with a Voigt profile with Gaussian width $\sigma$ and Lorentzian contribution $\gamma$.
'Cont' marks the classical veiling continuum added to the model. 'A' denotes the radial velocity amplitude of the emission line 
(maximum RV offset for a sinusoidal modulation), and 'RS' is the zero-point redshift of the line. The dashed line box
shows the values of the bisector velocity as measured by K14, which are similar to those derived from the toy model.
Although there is a slight BVS slope, the typical errrors at this level can make it very hard to distinguish.
 \label{dynamicalmodel-fig}}
\end{figure*}

Figure \ref{dynamicalmodel-fig} shows that, for
 reasonable values for the BC width and amplitude, it is possible to 
reproduce the observed radial velocities while avoiding a significant distortion of the line profiles and symmetry.
Our simple model 
assumes  that the rotationally-modulated BC in emission has a Voigt profile (with various Gaussians and Lorentzian
contributions, $\gamma$). A Gaussian core about 3-4 times broader than a 
typical photospheric absorption line avoids introducing large
distortions in the line profile. We also consider 
a BC strength that results in veiling factors for the line between 0.25-0.65, and that
it has a peak-to-peak velocity amplitude for in the range of 10-30 km/s 
and a few km/s net redshift (as observed; Table \ref{sinusoidal-table}). A velocity amplitude of $\sim$20 km/s would correspond
to a distance of about $\sim$5.7 stellar radii (assuming solid-body rotation and $\theta_s\sim$43 degrees),
compatible with material near the edge of the stellar magnetosphere. We also include a "classical"
continuum veiling as a constant emission factor that dilutes both the photospheric absorption line
and the wings of the emission line. Blending of nearby lines (which often occurs in the blue part of the
spectrum) could also result in a relatively flat extra emission over the real continuum. 
The broad line causes a rotationally modulated "tilting continuum" 
that changes the apparent radial velocity of the line. With the strongest change being in the broad line wings that always
fall out of the absorption line, the line bisector distortion is minimal and comparable 
to the distortion observed in K14. Checking the two bisector parameters explored by K14 (the bisector velocity
BVS and bisector curvature BC, see K14 for details), the variations observed in our model are comparable to the observations 
of the photospheric absorption lines.

This simple model can reproduce a
radial velocity modulation very similar to the one observed. A small net redshift and/or a non-Gaussian
profile (adding Lorentzian wings) can produce the
observed red-blue asymmetry in the RV curve (interpreted as
eccentricity in the companion scenario). Further asymmetries in the observed curve could be caused by
the emitting region not being entirely symmetric\footnote{Note that for instance, the Ca II IR BC are not entirely symmetric but deviate
from pure Gaussian profiles.} (due to occultation effects, or to variations
in the observed projected velocity for an extended spot as the star rotates), or by having
a small phase offset (which would happen if the accretion structure is trailing the star or
if the spot is extended along the stellar surface). As long as the period of the line-dependent veiling is maintained,
small variations in the strength of the lines would not destroy the global modulation.

We thus conclude that the most complete explanation of the observed RV modulation
of the photospheric absorption and the emission lines
is offered by assuming a modulation induced by line-dependent veiling
throughout the absorption-line spectrum. It also shows that the phenomenon of line-dependent veiling by
broad lines may
be one of the worst-case scenarios to detect the presence of companions in accreting stars, given how robust the signal is
to small variations in the accretion rate, and how weak the induced line asymmetries can be if the BC is
significantly broader than the photospheric absorption lines, being diluted with emission from nearby lines and/or
extra "classical" continuum veiling. If most photospheric lines are affected by line 
veiling, distinguishing this effect from any other
periodic modulation (such as the one induced by a companion) may be very hard.

This model requires that the accretion-related structure is stable over 
several years. The signature does not seem affected by the strong 2008 outburst
(although there are only 3 points taken before the outburst), 
suggesting that the accretion structures infalling onto this star are extraordinarily well-fixed to the
stellar structure/magnetic field. Even though the accretion rate varies in time and is probably very
clumpy,
it would be a case of remarkably stable accretion columns with very little instability (Kurosawa \& Romanova 2013).
Checking the long-term stability of the signal will be also a strong test for the
proposed accretion column scenario.

\subsection{Physical conditions in the accretion column \label{physics}}

We also explore whether the observed mean spectrum (obtained by combining all
the normalised observed spectra) can be reproduced as the sum of a stellar template, a classical veiling continuum (constant
over small wavelength ranges) and emission lines originated in a region with a given temperature
(T) and electron density (n$_e$). Dynamics and BC emission are not considered in this simple model. 
We do not attempt to model the H Balmer series, Na I D, Ca II IR triplet or
other broad lines (or BC) since they are very complex and self-absorbed. 
We also do not have high enough S/N to constrain
line ratios and physical conditions for the BC (which was nevertheless possible during outburst; SA12).
Lines with different excitation potentials and critical densities
allow us to explore the structure of the line-emitting accretion column in quiescence (S12; Petrov et al. 2014). 

\begin{figure*}
\centering
\includegraphics[width=0.8\linewidth]{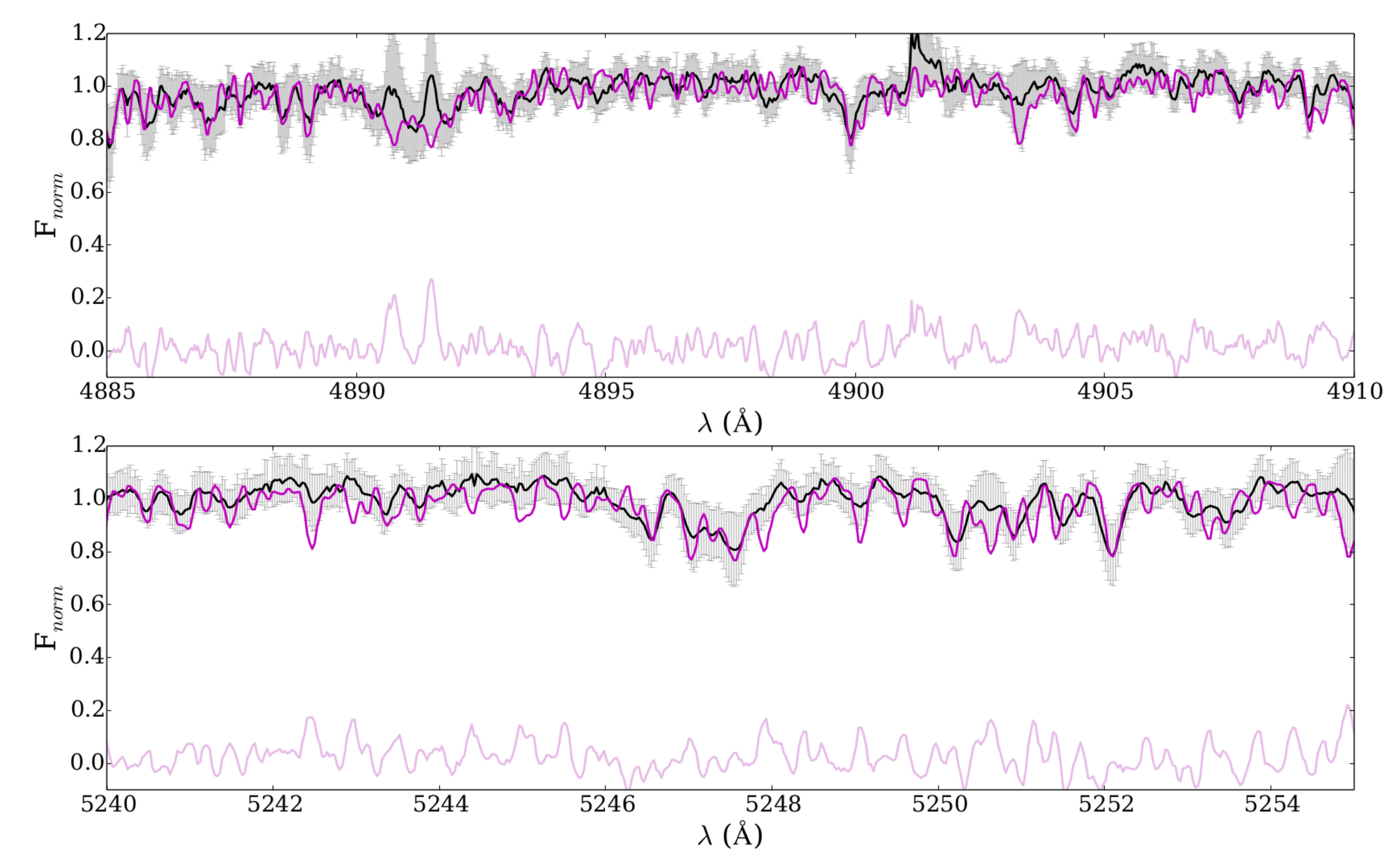} 
\caption{An example of the attempt to reproduce the median EX Lupi spectrum (black line, with
standard deviation per pixel marked in grey) by using a rotationally-broadened standard veiled by
a continuum emission (magenta). The difference between the observed and veiled spectra is shown in
pink. A constant veiling cannot reproduce the depths of all lines, even if we only
consider a very small wavelength range. The differences are too large to be explained by the standard deviation
or time variation observed between the different spectra.
 \label{veilinginlines-fig}}
\end{figure*}

For the photospheric model we consider an effective temperature T$_{eff}$=3750 K, logg=4.0,
and $v \sin i$=4.4 km/s
(Sipos et al. 2009), using the standard template from Coelho et al. (1995).
The veiling is considered as a continuum with a certain value over a small wavelength range. 
The atmospheric model with the veiling continuum should reproduce the observed absorption spectrum, 
but we find that a simple continuum veiling
with a slow dependency on the wavelength cannot reproduce simultaneously all the photospheric lines,
even if we only consider a small wavelenght range
(Figure \ref{veilinginlines-fig}). This suggests 
a line-dependent emission component filling in part many (if not all) the emission lines
with different strengths.

For the emission line spectrum, we consider only the metallic and narrow lines,
including He I and He II, which are dominated by narrow components. 
We first estimate whether the line emission is
optically thin or thick, and obtain approximate values (or ranges) for the
temperature and electron density in the emitting region.
In broad lines, the high velocity gradients ensure that the lines are optically thin
(Beristain et al. 1998; Petrov et al. 2014). For EX Lupi, the NC the line ratios
are nevertheless consistent with partially optically thick emission. We use the 
atomic data from the National Institute of Standards and Technology database (NIST;
Kelleher et al. 1999, 
Ralchenko et al. 2010\footnote{http:\/\/physics.nist.gov/PhysRefData/ASD/lines\_form.html})
to estimate line ratios in the optically thin approximation, which are
proportional to the atomic $gf$ factors for permitted lines (Martin \& Wiese 1999). For
a transition with upper level k and lower level i, the line strength g$_i$f$_{ik}$ can be estimated as:
\begin{equation}
	g_i f_{ik}= \frac{m_e c \epsilon_0}{2 \pi e} \lambda^2 A_{ki} g_k,
\end{equation}
\noindent where g$_i$ and g$_k$ are the level multiplicities (2J+1), A$_{ki}$ is the transition probability,
c is the speed of light, m$_e$ and e are the mass and charge of the electron, and $\epsilon_0$ is the
dielectric constant. 
Comparing line pairs with similar upper-level energy (Table \ref{lineratio-table}, see Table \ref{narrow-table} for a complete
list of unblended lines) shows that although some of the lines
(e.g. He I) are optically thick and saturated, the metallic neutral and ionized lines
(Mg I, Fe I, Cr I, Mn I, Fe II, Cr II) span a broad range where
the stronger lines tend to be optically thick and the weaker ones are sometimes optically thin. 
We also observe significant differences compared with the case of DR Tau in Beristain et al. (1998),
such as the suppression of the lines with lower transition probabilities (A$_{ij}$) and lower
excitation temperatures, which point to higher temperatures and densities and/or stronger high-energy
irradiation for EX Lupi than for DR Tau, despite its lower accretion rate.

The density can be estimated from the accretion rate (SA12):
\begin{equation}
	n = \frac{\dot{M}}{A\,f\,v\,\mu} \label{density-eq}
\end{equation}
\noindent Here, n is the number density, \.{M} is the accretion rate, A is the area of the stellar surface of which
a fraction f is covered by the accretion-related hot spot, v is the velocity of the accreted material as it
arrives to the shock region, and $\mu$ is the mean particle mass. If we consider the observed values for
the accretion rate (\.{M}$\sim$10$^{-10}$-10$^{-9}$ M$_\odot$/yr) and the velocity at which the material
arrives to the shock region (v=100-200 km/s), we arrive to n=6-120/f $\times$10$^9$ cm$^{-3}$.
We assume that the spot covers a small part ($\sim$1\%) of the stellar surface 
(suggested by Grosso et al. 2010 based on X-ray observations, and
 expected in case of higher density; Romanova et al. 2004). At relevant temperatures 
all the metals would be ionised and the H and He would be mostly neutral, producing
electron densities of n$_e \sim$10$^{8}$-10$^{10}$ cm$^{-3}$. If the material arrives to the
shock region at high velocities (v=100-200 km/s) but is deccelerated onto the star at few km/s
(Table \ref{sinusoidal-table})
in the post-shock region where NC are originated, we can expect the density to increase by
up to two orders of magnitude, arriving to n$_e \sim$10$^{10}$-10$^{12}$.
These higher values are more in agreement with what we would expect from the nearly equal ratios
observed for the Ca II IR triplet, among others. 
Extra ionization due to energetic radiation in the  accretion shock
or in the stellar chromosphere could also increase the electron density in the post-shock area. 
The presence of extra sources of ionization were evident during the outburst (SA12),
and some extra contribution of ionising UV emission (accretion-related) and by the 
shock itself (Orlando et al. 2010) is also likely to happen in quiescence.

We also follow Hamann \& Persson (1992) to evaluate the electron density based on
the requirements to achieve optically thick emission lines in the spectrum (similar for 
chromospheric emission  and an accretion column or
post-shock region scenario). The condition for equal
flux in a multiplet of emission lines (such as the Ca II IR triplet, or the Mg I triplet) is 
satisfied when collisional decay dominates over the transition rate, n$_e$C$_{ki} \gg$A$_{ki}$/$\tau$.
This mean $\tau$\,n$_e \gtrsim$ 10$^{13}$ cm$^{-3}$ for Ca II 8542/8662\AA. 
Grinin \& Mitskevich (1988) suggest that values of $\tau >$1 for the 8498\AA\ line, $\tau >$10 
for the 8542\AA\ line are enough for this. Considering the lower $\tau$ values, Shine \& Linsky (1974) 
estimate the minimum threshold density for detection of Ca II emission to be N$_{CaII} \sim$ 2.5$\times$10$^{15}$(V/50 km/s) cm$^{-2}$.
Assuming a solar composition and that most of the Ca is ionised, 
this results in N$_{(HI+HII)} \sim$ 2.5$\times$10$^{21}$(V/50 km/s) cm$^{-2}$.

In EX Lupi, the BC of the Ca II IR triplet becomes
undetectable when the accretion rate is at its minimum (\.{M}$\sim \times$10$^{-10}$ M$_\odot$/yr).
This suggests that the Ca II BC is very close to its optically thin
limit, so that n$_e \sim$10$^{12}$-10$^{13}$ cm$^{-3}$ in the region where the BC originates. 
If we assume that the accretion rate is constant for the amount of matter that flows through the BC-region
into the NC-region,  \.{M}=n$_{BC}$A f$_{BC}$ v$_{BC}$=n$_{NC}$A f$_{NC}$ v$_{NC}$, the relation
between the density depends on the velocity ratios of the material, 
and the areas covered by the BC and NC zones. If there are no different sources of extra ionization between the two
regions, the relation can be transformed into a relation for the electron density:
\begin{equation}
	n_{e,NC} = n_{e,BC} \bigg( \frac{v_{BC}}{v_{NC}} \bigg)  \bigg( \frac{f_{BC}}{f_{NC}} \bigg).
\end{equation}
\noindent The velocity widths are about 12-14 km/s for the NC and 50-100 km/s for the BC, which means by itself an
increase in density by a factor of 3.5-8.5. If in addition the accreting structure gets thinner/compressed
as it moves into the shock/post-shock region, a further increase in the density is to
be expected. This means that in the NC emitting region, n$_e \gtrsim$ 5$\times$10$^{12}$-10$^{13}$ cm$^{-3}$.
We will use this estimate as a starting point in the subsequent calculations. Such density estimates
are in agreement (slightly higher) with the values derived by Petrov et al. (2014).

Following SA12, we apply Saha's law for various temperature and electron density
values, in order to constrain the temperature and density structure in the accretion columns (Mihalas 1978).
Saha's law works under the assumption that the system is in local thermodynamic equilibrium (LTE).
The LTE assumption is questionable, although acceptable in an approximate toy model. We use the NIST line emission calculator
to estimate line intensities and ratios for a given T, n$_e$.
As in outburst, a single temperature/density pair is unable to
reproduce all lines at the same time. The lack of significant emission from Ti I lines puts a very strong constrain to the typical temperatures
in the accretion column. For the relevant densities, a minimum temperature of $\sim$7000 K is needed
to make the Ti I lines undetectable compared to Ti II. Regarding the He I/II emission, a
temperature between 8000-10000 K can explain the strength of the 5875 and 6678 \AA\ lines, but the
formation of more energetic lines (such as the He I 4920\AA\ and especially the He II 4686 \AA\ lines)
require temperatures of the order of 20000-23000 K, which would render other elements such as Fe, Cr, and Mg fully ionised.

\begin{figure*}
\centering
\begin{tabular}{c}
\includegraphics[width=0.7\linewidth]{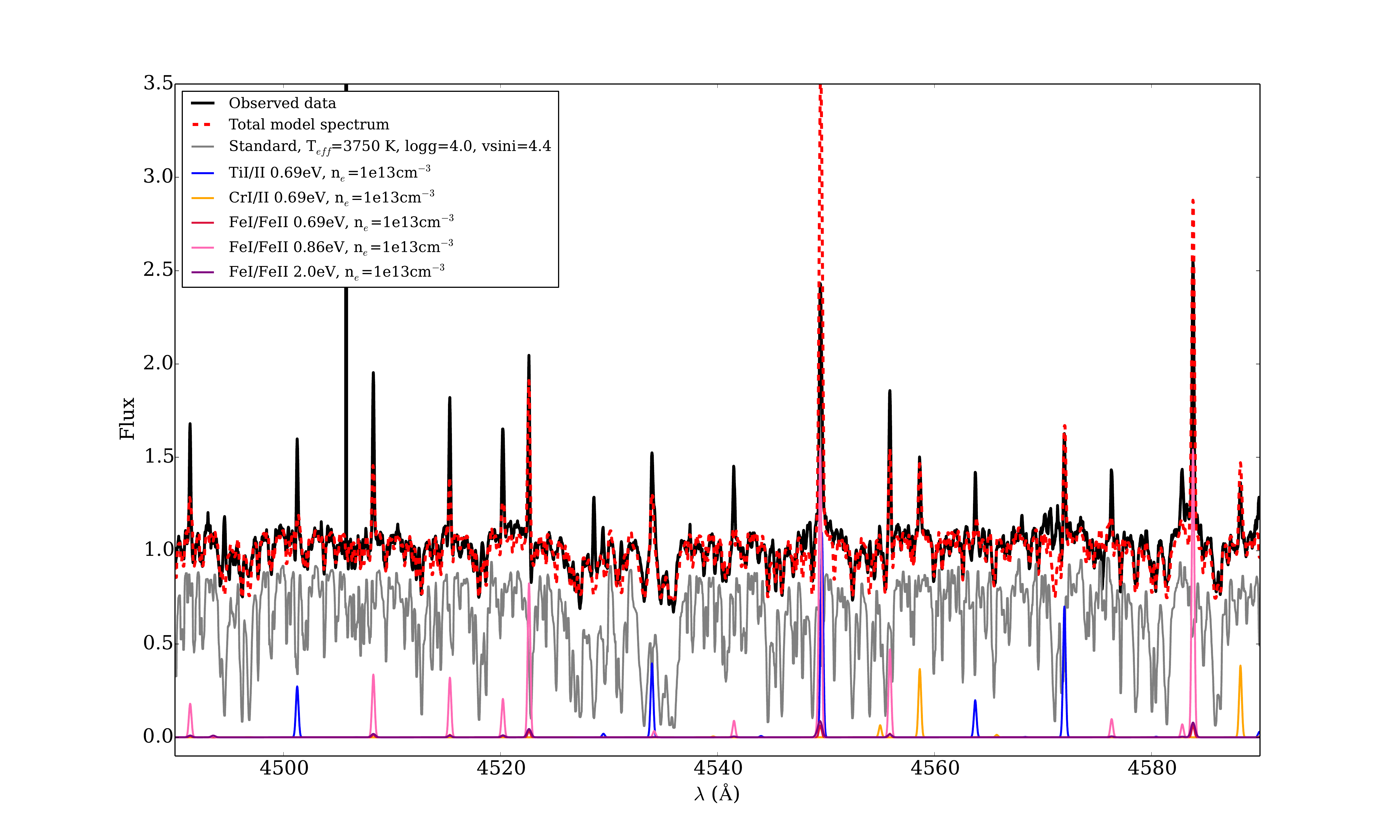} \\
\includegraphics[width=0.7\linewidth]{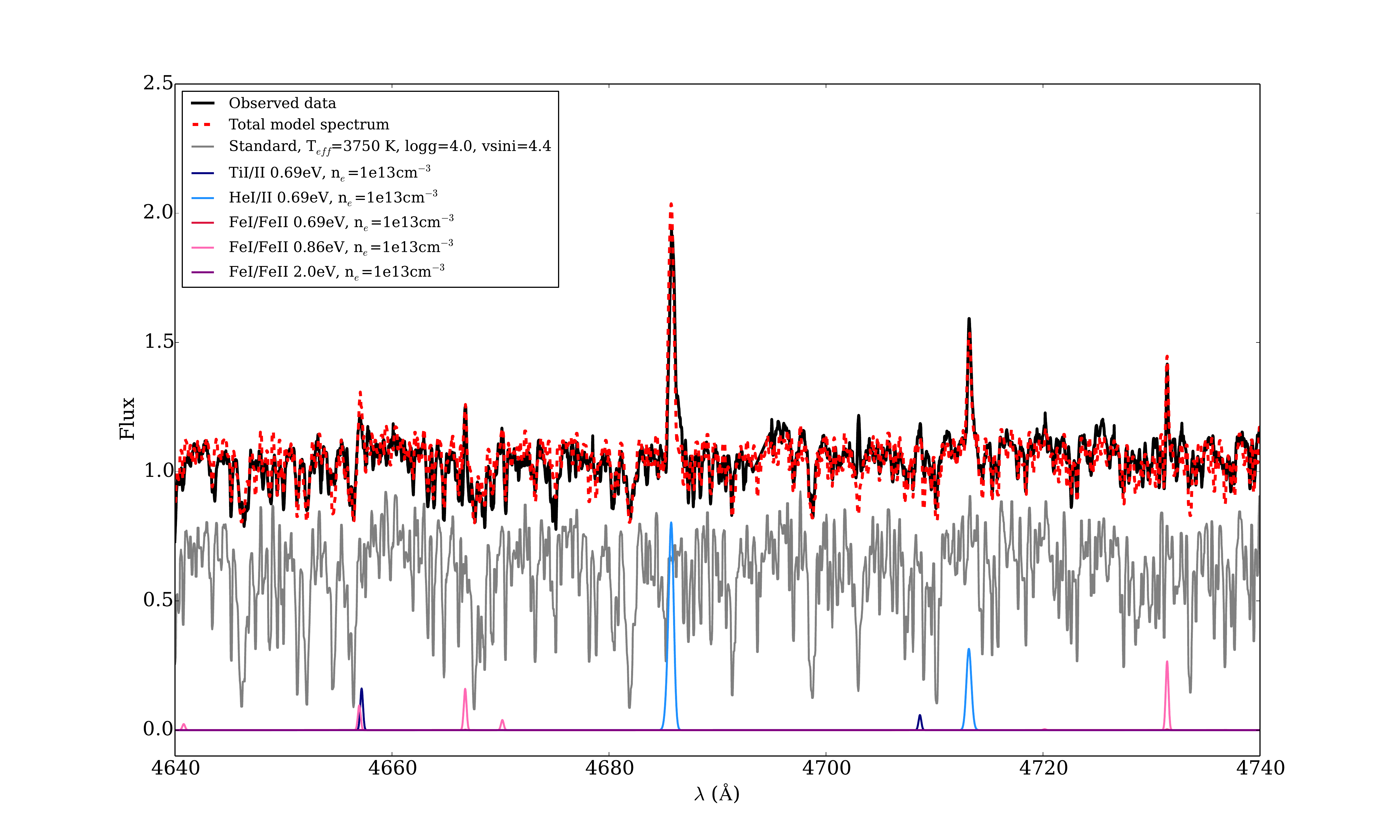} \\
\includegraphics[width=0.7\linewidth]{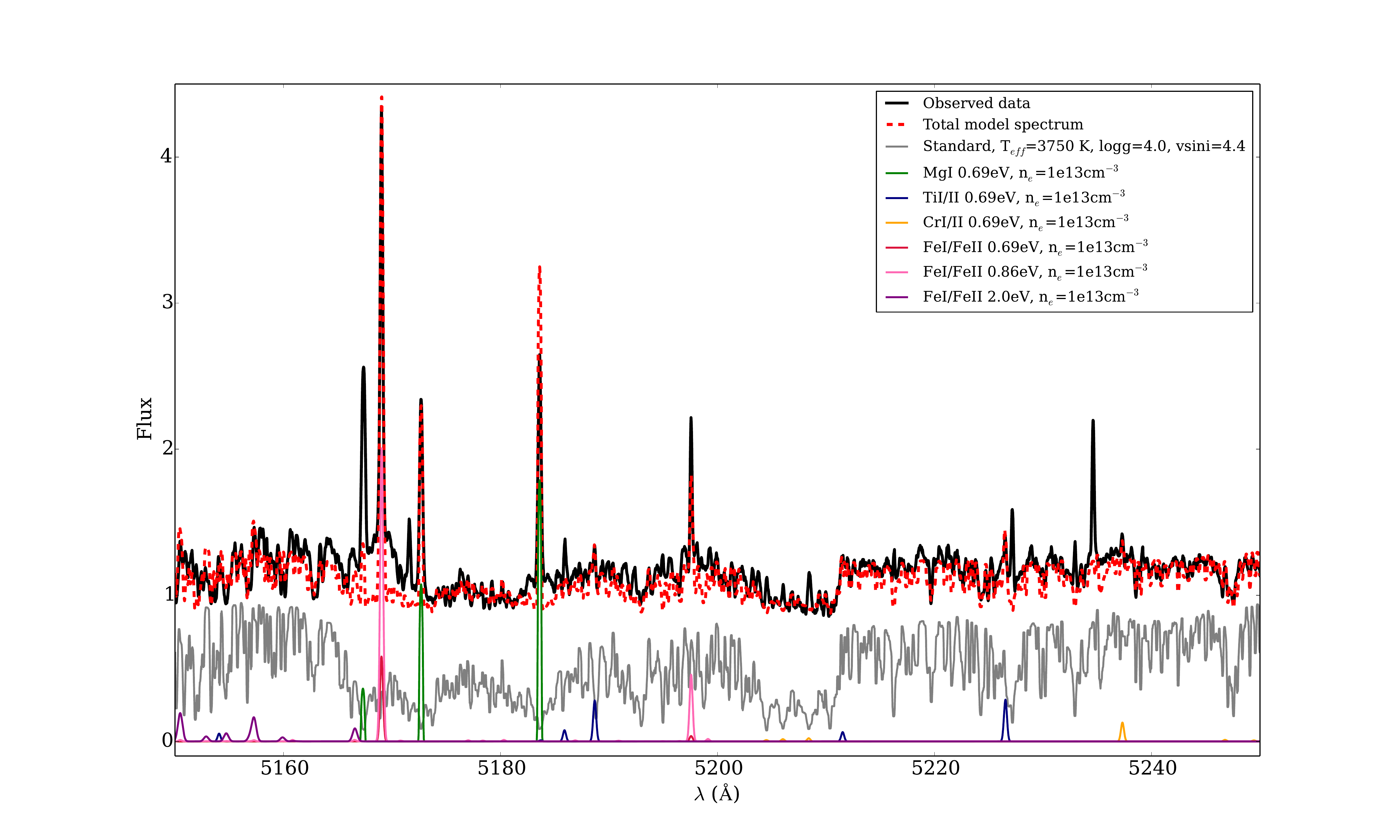} \\
\end{tabular}
\caption{Toy model for the narrow emission lines, obtained by considering an electron
density 10$^{13}$ cm$^{-3}$ and three different temperatures along the accretion
column (8000, 10000, 23000 K). The rotationally-broadened standard star spectrum is shown in 
grey, and the individual lines are shown in various colors. The final model, including a constant
veiling, is obtained by summing all three components. \label{linemodels}}
\end{figure*}

For our final toy emission-line model we thus consider a constant electron density n$_e$=10$^{13}$ cm$^{-3}$
and three different temperatures: 8000, 10000, and 23000 K (0.69, 0.86, and 2.0 eV, respectively). 
Note that in the real situation, both
the electron density and the temperature are expected to vary within the accretion column.
Since most of the lines are optically thick, 
the line intensity is simply scaled to match the observations.
The final spectrum is obtained by summing the photospheric component, a continuum veiling, and the lines.
Figure \ref{linemodels} shows the results of our simple model in several wavelength ranges. 
Despite its simplicity, our toy model confirms that a temperature stratification is
needed to fit the observed emission. 
A range of temperatures and electron densities is also consistent with our naive picture in
Figure \ref{cartoon}, with the most energetic He I and He II lines being dominated by the
high temperature component, closer to the shock region, and Fe II and lower-excitation He I lines  being dominated by
higher temperature emission than Mg I/Fe I lines. Further exploration would require a proper line radiative transfer 
modeling with detailed physical conditions along the accretion structure, which is beyond the
scope of this paper.

\begin{table}
\begin{footnotesize}
\caption{Line ratios in the optically thin limit.} 
\label{lineratio-table}
\begin{tabular}{l c c l }
\hline\hline
Lines & Observed   &    Opt.Thin  & Comments  \\
 &  Ratio   &   Limit & \\
\hline
HeI 6678/5875 & 0.41  &  9e-5 & Very opt thick \\
MnI 4034/4033 & 0.76 & 0.64 & Nearly opt. thin  \\
MnI 4034/4030 & 0.72 & 0.47 &   Rather opt. thick \\
MgI 5173/5184 & 0.92 &  0.59 & Rather opt. thick \\
TiII 4313/4300 & 0.53 &  0.40 & Nearly opt. thin  \\
TiII 4307/4302 & 1.50 & 0.76 & Optically thick\\
TiII 4054/4028 & 0.63 &  0.63  & Opt. thin \\
TiII 4450/4443 & 1.0 & 0.18 &  Opt. thick  \\
TiII 4590/4572 & 0.78 & 0.15 &   Rather opt. thick \\
SiII 6371/6347 & 0.63 &  0.59 & Nearly opt. thin \\
AlI 3944/3961 & 1.2 & 0.50   & Opt. thick  \\
CrI 4290/4275 & 0.80 & 0.74 &  Nearly opt. thin  \\
CrI 4290/4254 & 0.73 & 0.57 &  Nearly opt. thin  \\
CrII 4617/4588 & 0.43 & 0.22 &   Rather opt. thick  \\
CrII 4617/4559 & 0.38 & 0.23 &   Rather opt. thick  \\
CrII 4876/4848 & 1.0 & 0.47  & Very weak lines   \\
CrII 4876/4824 & 0.75 & 0.58  &  Very weak lines   \\
OI 7775/7774 & 0.88 & 0.60  &  Rather opt. thick   \\
OI 7775/7772 & 0.65 & 0.43  &   Rather opt. thick  \\
FeI 3896/3900 & 0.68 & 0.73  & Opt. thin    \\
FeI 3920/3928 & 0.66 & 0.60 &  Opt. thin  \\
FeI 3923/3930 & 1.07 & 0.70  & Opt. thick   \\
FeI 4144/4045 & 0.96 & 0.16 &  Opt. thick   \\
FeI 4216/4376 & 0.82 & 0.47  &  Rather opt. thick   \\
FeI 5329/5172 & 0.80  & 0.88 & Contaminated by Mg I  \\
FeI 5456/5406 & 0.82 & 0.57 &   Rather opt. thick  \\
FeII 4576/4351 & 0.30  &  0.15 & Rather opt. thick \\
FeII 4520/4549 & 0.39  & 0.13 &  Rather opt. thick   \\
FeII 4555/4549 & 0.68 & 0.30  &  Rather opt. thick  \\
FeII 4583/4576 & 0.76 & 0.72 &  Opt. thin  \\
FeII 4923/5018 & 1.16  & 0.73 &  Opt. thick  \\
FeII 5235/5198 & 1.16 & 0.70 & Opt. thick   \\
FeII 6238/6248 & 0.50 & 0.47 & Nearly opt. thin   \\
\hline        
\end{tabular}
\tablefoot{Observed and optically-thin limit line ratios from similar-energy
(calculated using the atomic data from the NIST database). }
\end{footnotesize}
\end{table}

\section{Discussion: The star, the spot, and the accretion column \label{results}}

\subsection{A critical revision of the origin of the line and RV modulations}

The velocity modulations observed in EX Lupi emission lines were first described in SA12, 
concentrating on the most conspicuous modulations (tens of km/s) observed during
the 2008 outburst. For the RV variations in the photospheric absorption lines, K14
proposed two possible origins: RV signatures induced by a
BD companion (with m$\sin i$=14.7M$_{jup}$ and semimajor axis 0.063\,AU), vs rotational modulation induced by
possibly two hot spots on the stellar surface. Both the model of the BD companion
and the model of "classical" hot spot(s) had problems to produce a complete
explanation of the observations.
Here we propose a 3-D accretion
column as the origin of the RV modulation of both emission and absorption lines. 
An accretion hot spot or accretion column is necessary 
to produce the temperature inversion that gives rise to the 
emission lines. Therefore, if accretion alone
is able to explain all observations, it would offer the most simple explanation,  
without the need to add further elements.

The main caveat of the companion model is that a companion-only 
scenario cannot explain why the emission and absorption line
velocities are off-phase.  A tentative solution could be sought if the accretion would mostly 
land on the BD companion or in a structure connecting both objects (e.g. Artymowicz \& Lubow 1996). 
In that case, we would detect the photoshere of the
bright primary (a M0 star, the BD photosphere would not be detected due
to contrast) and the hot spot on the secondary or on the connecting structure. This would produce 
an anticorrelation between the RV of absorption
and emission lines, similar to what has been observed in X-ray studies from
accretion shocks in interacting binaries  (Donatti et al. 2011, Argiroffi et al. 2012).
Since the masses of EX Lupi and the proposed companion are very different, the
amplitudes of the RV signatures for both would be very different. Nevertheless, this
scenario is inconsistent with the observations. 
Considering the line with the strongest RV signature (He II 4686\AA\ line), there is only a very
mild anticorrelation: a Spearman rank test gives a correlation coefficient r=-0.20 and 
false-alarm probability p=0.20, which is non-significant.
Figure \ref{linemoduls} reveals that the emission line modulation is not simply the (scaled) stellar
RV with a changed sign (which would be strongly asymmetric), but a redshifted symmetric sinusoidal 
modulation (Section \ref{dynamics}).
The timing of the RV cycle as observed in the absorption and emission lines is also in
contradiction with the expected signatures in case of accretion in a binary system (Artymowicz \& Lubow 1996),
since it would request a (variable) difference in phase of approximately 1/4 of cycle, while accretion is most likely to
occur along the line connecting both objects. The phase variation between the companion and the accretion would also
require the accreted material to move at unphysical, variable speeds, which change the relative position between
the spot and the companion during the phase. A scenario with spots on both
objects would similarly fail, as it would either dilute the signature of the RV of the emission
lines or modulate the line thickness (by summing a redshifted and a blueshifted components that change
along the orbital phases). It also fails to explain why the RV amplitudes are often larger for
the emission lines than for the photospheric absorption lines, and why the shapes of both curves
are so different (asymmetric for absorption lines, symmetric for emission lines).

The main caveat of the hot spot scenario in K14 is that it would require the 7.41d period
to be the rotational period of the star, which would be hard to reconcile with the $v\sin i$ and
the constraints on inclination from disk models and CO observations. We nevertheless find that 
$v \sin i$=4.4$\pm$2.0 km/s (Sipos et al. 2009) reproduces better the observations than the upper limit
value of 2-3 km/s given in K14. With this higher rotational value, a rotational period of 7.41d would
require an inclination angle of 23-25 degrees (consistent with the disk models
of Sipos et al. 2009).
A low viewing angle was also predicted by Grosso et al. (2010) based on X-ray absorption.
The observed velocity amplitude of the NC is also in
agreement with an origin in a high-latitude structure, in a star that is seen at a 
relatively low angle (Section \ref{dynamics}). Moreover, all the NC velocities are redshifted, also in
agreement with a relatively low viewing angle where the line-emitting region is always seen
infalling (never blueshifted) as the star rotates.

The problem for the hot spot hypothesis in K14 
is that models for a classical, continuum-emitting, flat hot spot
would require a coverage as large (or more) than the stellar
surface to produce the radial velocity signatures observed. Such a spot would also produce a luminosity modulation
that is not observed in the photometry records. K14 assumed that the spot would 
suffer occultations as the star rotates, which would increase the luminosity contrast, but a high-latitude spot
visible at all times would not induce dramatic variations in the luminosity.
In addition, if most or a large fraction of
the energy released in the shock structure is emitted in lines (including many
metallic UV lines) and not in the optical
continuum, photometric rotational modulations may not be
evident, while RV modulations would be significantly enhanced with respect to the
continuum-only scenario. This situation has been described in other T Tauri stars
(usually, with strong accretion) and named "veiling-by-lines" 
(Dodin \& Lamzin 2012, 2013).

The detailed line-by-line dynamical analysis breaks one of the main counter-arguments 
in K14 against the spot/column explanation,
that the RV shifts observed in the emission lines had a smaller amplitude than the RV of
the photospheric absorption lines and that the signal periodicity was not significant. This happened
because the relative offsets between lines and the various degrees of modulation (for instance,
comparing He II with Ca II NC) dilute the global signature when many lines are combined. Once the lines
are treated individually (including detailed individual fits), the modulation is strong
and significant. This is an extension of what we observed during outburst (SA12),
with different lines having various velocity offsets depending on their excitation and
physical conditions in the extended non-axisymmetric accretion structures.
Petrov et al. (2014) also demonstrated that the stratification in velocities and physical conditions in the
accretion columns result in different velocities for lines with various excitation potentials, 
although in this case the lines were seen in absorption and were not time-resolved. 

If the emission lines are formed in a hot spot near/on the stellar
surface whose velocity is modulated by the rotation of the star, 
the largest extent of the spot is visible at zero velocity, so the line would be stronger
when the spot is facing us. Despite the substantial variations from day-to-day 
and epoch-to-epoch in the line strength, there is a low-significance modulation 
of the amplitude/strength in several Fe II, Fe I, Ti II and Mg I lines. The
strength of the lines is smaller at zero velocity, and
maximum when both the redshifted and blueshifted velocity offsets are maximal.
Although this is the opposite situation than above mentioned, it 
is similar to the case of rotationally modulated X-ray emission (e.g. Flaccomio et al. 2005; 
Argiroffi et al. 2011, 2012). X-ray emission is produced at the bottom
of a 3-D accretion shock, which results in the accretion column obscuring most of it 
at zero velocity, while most of the emission should escape through the sides of the
accretion column at maximum redshift/blueshift. This effect has been observed for EX Lupi
(Grosso et al. 2010). The same explanation could apply to the NC lines, 
if formed within a
vertically extended structure where the emission can escape better along the "sides"
of the accretion structure, instead of from the top. The Ca II IR lines lack these occultation effects and tend to
be stronger when the structure is facing the observer, in agreement with their origin in a
less embedded, more delocalised structure with lower density and temperature.

The rotational modulation gets increasingly smaller as we go to lines that are produced at lower-density,
lower-temperature conditions, such as the Ca II IR line (Figure \ref{cartoon}). Since lower-excitation
lines require less restrictive conditions for their formation that can be met over a larger area (more similar to the conditions at the 
photosphere of the star), the RV signature of these NC lines could be diluted and/or appear rather as a quasi-periodic modulation.
Differences in the rotation in various parts of the structure are also expected from models, and could
lead to quasi-periodic modulations as observed here for the low-excitation lines (Romanova et al. 2004).
Deviations from pure solid-body rotation in an extended structure can also dilute the 
modulation\footnote{The region where the NC are produced is expected to be very small
compared to the stellar radius, which means that the nearly-solid-body rotation would not
need to be extended over a large vertical structure. }. The lower temperature, lower density
areas will be larger, diluting the signature with a broader range of velocities.
Although the periodicity of the BC of the Ca II lines (the only BC detected in most of the
spectra) is not significant, it is also consistent with modulations in the 7-8d range. 
If the material is fed
to the star in a non-axisymmetric way, as proposed in outburst (SA12) and confirmed by CO observations
(Goto et al. 2011; Banzatti et al. 2015), it could also help to keep accretion confined to a
relatively small area on/near the stellar surface, compared to cases where the material approaches
the star from all directions. An accretion column/structure that trails the star (or 
material that is both accreting and rotating) could induce phase shifts in the rotational
modulation, which is also observed (Table \ref{sinusoidal-table}).

\subsection{Origin and implications of stable accretion columns in EX Lupi}

Accretion in EX Lupi appears to follow a combination of the standard accretion column scenario 
as an origin for the emission lines in CTTS (Muzerolle
et al. 1998, 2001), and the post-shock scenario
(Dupree et al. 2012). We find evidence of both broad components
consistent with an origin before the shock region in an extended accretion column, and narrow components produced in a post-shock,
slightly infalling structure extending down to the stellar photosphere.  The line widths observed in the NC would
suggest a less turbulent environment than proposed by Dupree et al. (2012) for TW Hya, being 
close to the expected thermal broadening for the relevant temperatures. 
The velocity differences observed between BC and NC are also consistent with the 
differences between a pre-shock and post-shock region (Ardila et al. 2013). The 
BC lines require structures with a large velocity span,
which could be due to spatially extended rotating/infalling packages of gas,
although line broadening contributions from turbulence cannot be excluded. The lack of correlation 
between H$\alpha$/Ca II IR and He I lines is in agreement with the
time delay observed by Dupree et al. (2012), although we do not have enough time-resolved coverage
to follow the accreting packages on a short (hours) timescale.

Cooling in the post-shock region would be dominated by line emission (including the metallic and He lines
observed). Given the observed line velocities and stratification, 
the possibility that the narrow emission lines are produced in the post-shock region 
is more likely than an origin in active regions of the star, which would tend to be
more distributed over the stellar surface (Ardila et al. 2013; Brickhouse et al. 2010), even though many of the lines
observed in emission are also common in active stars. A strong contribution from distributed 
chromospheric NC emission would also tend to dilute the rotational signatures observed. Although this does not
happen for most of the metallic lines, it could affect some of the typical activity indicators such as the Ca II NC.
At this point, we cannot thus determine whether EX Lupi is more or less active than other similar
young stars. 

It also remains unknown why most of the shock energy in EX Lupi is released in
spectral lines. 
Stars with very rich emission line spectra have typically high accretion 
rates (Hamann \& Persson 1992; Fang et al. 2009; Sicilia-Aguilar et al. 2010).
This is not the case for EX Lupi in quiescence, since most M0 young stars are
accreting at levels higher than $\sim$10$^{-10}$ M$_\odot$/yr (e.g. Fang et al. 2009; Sicilia-Aguilar et al. 2010).
A plausible
explanation would be a significant structural difference in the accretion channels,
since the emission of many of the observed lines (Fe II, He II) requires particularly hot and
dense material. A similar accretion rate being channeled through and landing on a more extended part of the stellar
surface may thus not be enough to produce the rich forest of lines we observe in EX Lupi.
The comparison of EX Lupi and DR Tau (from Beristain et al. 1998) shows that despite its
lower accretion rate, the accretion structures in EX Lupi probably reach higher densities and temperatures.
A lower temperature contrast between the stellar photosphere and the accretion structure
would also tend to prevent the formation of the high energy lines. 
Interestingly, some of the line-rich
objects identified by Fang et al. (2009) also show, like EX Lupi, low 2MASS near-IR excesses and
large Spitzer mid-IR fluxes. This could suggest a correlation between the strong line emission and the
presence of inner holes or optically thin inner disks, although the sample of objects is too 
small to draw any statistically significant conclusion. 
In addition, only ASASSN13db (Holoien et al. 2014), also an outbursting
star, has a comparable spectrum. Whether the rich emission line spectra reflect
a particular type of accretion structures in these
EXors, or is a consequence of the accretion variability of the stars, is to be explored in the future.

If line veiling (instead of continuum emission) is
dominant for a significant number of stars, accretion rates are likely to be overestimated (Dodin \& Lamzin 2012).
Veiling is the ultimate calibration of accretion rate tracers, including U-band excess (Gullbring et al. 1998) and 
emission lines (Natta et al. 2004; Fang et al. 2009; Alcal\'{a} et al. 2014). 
This could have led to an uncertain estimate of the accretion rate of EX Lupi, especially in
quiescence. In outburst, the accretion luminosity overwhelms the stellar luminosity,
so the measured rate should be more accurate. The outburst accretion
rate (few times 10$^{-8}$ M$_\odot$/yr; SA12; Juh\'{a}sz et al. 2012) is high, but not
extreme, for a low-mass star, which is also in agreement with a very low rate during the
quiescence phases.

The observed velocities of the NC and BC can be compared with the different 
velocities associated to the star, which can be due to stellar rotation (assuming
a structure would rotate as a solid body together with the star),
infall (considered as free-fall\footnote{Note that in the presence of a strong magnetic field
near the stellar photosphere, the
infall is probably caused by downward motions along the magnetic field structure that corotates with the star.}), 
and Keplerian rotation. Table \ref{velocity-table}
shows a small summary of velocities and associated locations for these three cases (note
that not all velocities are expected to be relevant at all locations).
Taking into account the period of the star, the corotation radius is located at
$\sim$8.45 R$_*$, or $\sim$0.063 AU, corresponding to a velocity of $\sim$92 km/s.
This velocity is not too different from the maximum span of the peak of
the Ca II BC ($\sim \pm$100 km/s). As we move away from the star, Keplerian
rotation (probably combined with infall) is probably dominant. 
Solid-body rotation with the star (even if the column is trailing the star) probably would
break at relatively short distances, depending on the strength and extent of the
magnetic field. Considering the observations of BC
during outburst (SA12) and the CO data from Goto et al. (2011) and Banzatti et al. (2015), the 
non-axisymmetric, rotating/infalling structure of gaseous material may extend well into the few tenths of AU within the inner
disk dusty hole. Feeding the star in a non-axisymmetric way may help to keep the
rotational modulation visible over time (for instance, bringing material only onto certain
areas over the stellar magnetosphere).

An open question is whether the magnetic field structure of a star like EX Lupi can be stable over such a long
period of time ($\sim$5 yr). Although 
rapid changes in the magnetic field configuration have been proposed for young stars
(Brown et al. 2011), some observational evidence suggests that magnetic field reversals 
are not so common in young (10-50 Myr), rapidly rotating, solar-type stars, at least
in timescales of 3-7 years (Marsden et al. 2009). Rotational modulation is observed in 
some cases (Johns-Krull et al. 2013), but there are large differences in the behaviour of
different low-mass stars, with some showing strong variability, while others have persistent fields
over several years (Symington et al. 2005). 
As a young, low-mass star, EX Lupi 
should be completely convective, favoring a simple dipolar field (Gregory et al. 2014),
which could lead to a simple accreting structure within a well-defined part of the stellar surface,
as proposed here. Young stars with simple
magnetic fields tend to have large disk truncation radii and spots at high latitudes
(Johnstone et al. 2014), both of
which would be in agreement with the location of the accretion structure and corotation radius for EX Lupi. 
On the other hand, close-in companions can have very complex field structures, leading to accretion
distributed over large surfaces on the stars (Donati et al. 2011). Although very different in mass, a
hypothetical BD companion near EX Lupi could have a strong field (Morin et al., 2010)
and lead to a complex total field, which would wash out the velocity signatures in the accretion-related lines.
There are no constraints on the long-term magnetic
field variability of EX Lupi (nor of large samples of CTTS), but in principle a magnetic field stability
over several years would not be exceptional. Future 
studies of the magnetic field of EX Lupi and its variability would
be a very strong test of the validity of our stable accretion hypothesis.

\begin{table}
\caption{Different velocities associated to the system.} 
\label{velocity-table}
\begin{tabular}{l c c c}
\hline\hline
Velocity & R$_{rotation}$ &  R$_{freefall}$ & R$_{kepler}$  \\
(km/s)  &     (R$_*$)     &    (R$_*$)    &   AU          \\
\hline 
5    &      0.14 	  &  1.7e-4       &	21.3	\\
10   &      1.3           &  7.0e-4       &	5.3     \\
20   &      3.5 	  &  2.8e-3       &	1.3	\\
50   &      10.4 	  &  1.8e-2       &	0.2 (29R$_*$)\\
100  &      21.7 (0.17 AU) &  7.5e-2       &	0.053 (7.2R$_*$)\\
200  &	    44. (0.34 AU)  &  0.39	  &	0.013 (1.8R$_*$)\\
300  &	    67.	(0.51 AU)  &  1.7          &	5.9e-3 (0.79R$_*$)\\
\hline
\end{tabular}
\tablefoot{Estimated for a rotational period 7.417d, M$_*$=0.6M$_\odot$, R$_*$=1.6R$_\odot$.
"Rotation" refers to solid-body rotation (locked to the star). The 
distances are measured from the stellar photosphere at R$_*$. Note that for the rotation options,
the velocities are corrected by the sin$i$. Also note that
at any given location, not all three motions are expected to be relevant, (e.g.
solid-body rotation is unlikely very far from the star; Keplerian rotation is
unlikely very close to the star; infall near the star is probably dominated by the
magnetic field and thus not pure free-fall),  so this table is merely informative.}
\end{table}

\section{Summary and conclusions \label{conclu}}

We analyse the rich emission line spectra of EX Lupi to
investigate the origin of the emission lines and whether they could be responsible for the RV
signatures previously observed. 
Our study shows how we can turn a well-known problem, namely the
impossibility to obtain reliable radial velocities for young stars 
with emission lines to search for companions, into a useful
tool to trace the structure and properties of the accretion columns.
This tool is able to probe into regions over the stellar photosphere 
that are smaller than the stellar radii, which are inaccessible by other means,
including interferometry.

\begin{itemize}

\item As previously observed, the EX Lupi spectra in quiescence display a
very rich collection of metallic lines, most of which are narrow, although
some strong lines have a NC+BC structure. The BC becomes
more prominent during periods of
increased accretion. 

\item The spectra show line-dependent veiling. If a faint BC partly fills in
many (or most) of the photospheric 
absorption lines, it can account for a substantial fraction of the accretion-related
released energy. If most of the veiling in EX Lupi is emitted in lines
and not as a hot continuum, there could be some uncertainties in the accretion 
rate estimates. For line-veiling, the total
energy released would be smaller than in the continuum case. Depending on whether 
the stars used for the calibration are dominated by continuum or line veiling,
the accretion rate could be lower.
Although veiling-by-lines tends occur in strong
accretors (Dodin \& Lamzin 2012; Ingleby et al. 2013), EX Lupi demonstrates
that objects with low accretion rates can also have significant line-dependent veiling.

\item Line-dependent veiling can explain the origin of the remarkable spectrum observed
during the accretion outburst (SA12). The BC-dominated outburst spectrum,
which hides all photospheric features under a whealth of broad emission lines, 
would correspond to
the reaction of the line-dependent veiling to a higher accretion rate. 
During quiescence, the BC of most lines would remain undetected as it does not
completely fill in the photospheric lines. The same effect could be expected also
in the outbursting star ASASSN13db (Holoien et al. 2014), which has a similarly
rich outburst spectrum.

\item The NC emission is consistent with an origin in a hot
post-shock region,
deep in the accretion channels and close to the stellar photosphere.
The dynamical analysis suggests that the accretion "spot" is a 3-D structure similar
to a stellar chromosphere, where the temperature goes through a minimum at the top of the stellar photosphere and
then rises afterwards, although on a localised place near the stellar surface. 
The profiles of the stellar absorption spectral lines can have narrow emission
cores if they are opaque enough to be formed in the region where the temperature is increasing outwards.
The BC is consistent with being formed at larger distances from the star in rotating/infalling 
non-axisymmetric accretion structure(s), in agreement with the outburst observations.
In quiescence, the emitting structure appears closer to the stellar
surface than in outburst. 

\item  There is a clear difference in velocity modulation between the NC and the BC. The BC
shows velocities of several tens to hundreds km/s, while the material associated
to the NC only shows infall at a few km/s rate. 
There are also clear differences between the various emitting species, consistent with 
stratification in temperature and density within an
extended, 3-D accretion structure. The analysis of the dynamics of lines with different excitation 
potentials gives a self-consistent view of the accretion column structure and can be applied to further
objects in order to trace spatial scales which are beyond direct spatial resolution.

\item The RV signatures observed in emission and absorption lines
can be both explained with rotationally-modulated signatures of an accretion 
column that cools mostly by emission lines. We show that the line-dependent veiling 
emitted by the accretion column, velocity-shifted due to the stellar rotation, can produce a RV signature
in the photospheric absorption lines that mimics the signature of a companion.
The resulting "tilting continuum" fills in the photospheric absorption lines,
leading to an apparent RV shift.
The velocity structure and phased RV curve in the emission 
lines can be used to distinguish companion-related from accretion-related
phenomena in young stars. In case of very stable accretion structures and
line-dependent veiling, it can be nevertheless very hard to distinguish 
both scenarios.

\item The line velocity modulations observed for EX Lupi are suggestive of the stable
accretion scenario proposed by Kurosawa \& Romanova (2013). Our 
long-term monitoring is consistent with accretion occurring through remarkable stable 
channels, despite the frequent variations in the accretion rate.
The only difference between the quiescence and the outburst phases
is the amount of matter channelled onto the star. 
In order to maintain stable accretion channels over
years, a very strong mechanism (maybe magnetically controlled or controlled by the disk) has to rule the
flow of material onto the star. Given that the gaseous material within the inner disk 
hole also consist of infalling/rotating, non-axisymmetric flows of
matter (Goto et al. 2011; K\'{o}sp\'{a}l et al. 2011; SA12; Banzatti et al. 2015), 
part of the flow regulation could happen within the inner disk.

\item The cause of the outbursting behaviour in EX Lupi remains unknown.
Its disk is not substantially
different from the disks around other stars. We cannot 
conclude that the star is anomalously active. The only remarkable fact
is the stability of the accretion channels, which should be explored in the coming
years.
The potential relation of "overregulated" stable accretion onto the star
and the violent outbursts and variability is an intriguing possibility 
that needs to be explored in the future (either as a cause or consequence of
the outbursts, or as 
star-controlled irregular accretion). A mismatch between the rate at which the disk
feeds material onto the inner disk and the rate at which the star 
lets accretion proceed could be a possibility to explain the outbursts and the
observations of Banzatti et al. (2015) of material accumulated in the inner disk. 
Extending the detailed metallic line emission analysis to other (non-outbursting)
stars may reveal whether this is a typical behaviour of some CTTS, or
whether it may have a deeper relation to the accretion variations observed in
EX Lupi and other EXors.
\end{itemize}

Acknowledgments: We thank the La Silla observers, specially J. Setiawan and A. M\"{u}ller, 
for collecting the spectra. We also thank the anonymous referee for his/her valuable comments that helped clarifying this paper,
and A. Mortier for her comments on spot-related rotational modulations. 
This research made use of Astropy, a community-developed core 
Python package for Astronomy (Astropy Collaboration, 2013). A.S.A. and M.F. acknowledge support
from the Spanish "Proyectos de investigaci\'{o}n no orientada" project number AYA2012-35008.
A.K. is partly supported by the Momentum grant of the MTA CSTK Lend\"{u}let Disk Research
Group. P.\'{A}. acknowledges support from the Hungarian Research Grant OTKA 101393.

\onecolumn

\Online

\begin{appendix}

\section{Observations and summary tables \label{support-appendix}}

\begin{table}
\caption{List of observations during EX Lupi quiescence analysed in this paper.} 
\label{spectra-table}
\begin{footnotesize}
\begin{tabular}{l c c c}
\hline\hline
JD & Instrument  & Accretion & BC shape \\
\hline        
2454309.615 & FEROS  & M &   \\
2454310.657 & FEROS  & M &   \\
2454311.687 & FEROS  & L &   \\
2454874.838 & HARPS  & H &   \\
2454875.850 & HARPS  & H &   \\
2454876.808 & HARPS  & M &   \\
2454877.791 & HARPS  & M &   \\
2454877.906 & HARPS  & L &   \\
2454891.800 & HARPS  & H &   \\
2454891.882 & HARPS  & H &   \\
2454892.774 & HARPS  & M &   \\
2454894.760 & HARPS  & L &   \\
2455053.675 & FEROS  & L &   \\
2455058.493 & FEROS  & M & B  \\
2455257.820 & FEROS  & L &   \\
2455259.877 & FEROS  & L &   \\
2455262.807 & FEROS  & L &   \\
2455263.868 & FEROS  & L &   \\
2455309.894 & FEROS  & L &   \\
2455338.791 & FEROS  & L &   \\
2455353.827 & FEROS  & L &   \\
2455396.589 & FEROS  & H & C  \\
2455396.637 & FEROS  & H & C  \\
2455397.611 & FEROS  & M &   \\
2455399.658 & FEROS  & M & C  \\
2455400.536 & FEROS  & H & B  \\
2455401.506 & FEROS  & H & B  \\
2455402.584 & FEROS  & M &   \\
2455403.612 & FEROS  & M & C  \\
2455405.661 & FEROS  & H & C  \\
2455407.620 & FEROS  & H & C  \\
2455586.871 & FEROS  & H & D  \\
2455587.853 & FEROS  & H & C  \\
2455588.858 & FEROS  & H &   \\
2455630.816 & FEROS  & H & B  \\
2455637.895 & FEROS  & M & B  \\
2455667.701 & FEROS  & H &   \\
2455668.691 & FEROS  & H & D  \\
2455669.737 & FEROS  & H & C  \\
2455670.794 & FEROS  & H & D  \\
2455671.698 & FEROS  & H & C  \\
2455672.876 & FEROS  & M &   \\
2455673.849 & FEROS  & M & R  \\
2455674.824 & FEROS  & H & B  \\
2456111.513 & FEROS  & L &   \\
2456113.565 & FEROS  & L &   \\
2456114.477 & FEROS  & L & R  \\
2456115.477 & FEROS  & L &   \\
2456116.524 & FEROS  & L &   \\
2456117.562 & FEROS  & M &   \\
2456118.670 & FEROS  & L &   \\
2456119.494 & FEROS  & L &   \\
2456120.508 & FEROS  & L &   \\
2456121.579 & FEROS  & L &   \\
\hline
\end{tabular}
\tablefoot{The "Accretion" column correspond to the classification as high/medium/low accretion given in the text,
based on the strength of the BC of the emission lines (see Appendix \ref{acc-appendix}). 
The "BC shape" field indicates the shape of the Ca II IR broad component,
which can be classified as redshifted (R), blueshifted (B), centred (C), or double-peaked (D;
see Appendix \ref{appendix-lines}). Only spectra with strong BC
can be classified (see text).} 
\end{footnotesize}
\end{table}

\begin{landscape}
\begin{table}
\caption{Periodic signals in the emission line radial velocities.} 
\label{rvemission-table}
\begin{footnotesize}
\begin{tabular}{l c c c c c l}
\hline\hline
Line & E$_k$+E$_i$$^1$ (eV) & A$_{ki}$ (s$^{-1}$) & N & P(d) & FAP  &   Comments  \\
\hline        
FeII 4549 & 7.87+5.55 & 1.00e+06 & 49 &  8.094$\pm$0.020 & 0.012   & Further peaks at 7.19 with FAP=0.02  \\
FeII 5018 & 7.87+5.36 & 2.0e+06 & 53 &  7.457$\pm$0.017  & $<$0.01 & Several peaks, one at 5.20 nearly as strong    \\
FeII 4923 & 7.87+5.41 & 4.28e+06 & 54 &  7.194$\pm$0.013, 6.207$\pm$0.010   & $<$0.01 &     \\
FeII 4351 & 7.87+5.55 & 4.86e+05 & 52 &  7.672$\pm$0.016   & $<$0.01 &  Lower-significance peaks at 7.4, 9.4, and 8.6   \\
FeII 5316 & 7.87+5.48 & 3.89e+05 & 55 & 8.100$\pm$0.027, 7.429$\pm$0.035   & $\leq$0.01 & Multiple peaks   \\
FeII 5362 & 7.87+ -- & -- & 54  & 8.100$\pm$0.017   & 0.012 & Several similar low-significance peaks    \\
CaII 8662 & 6.11+3.12 & 1.06e+07 & 45 &  NSP   & -- &     \\
CaII 8498 & 6.11+3.15 & 1.11e+06 & 43 &  NSP   &  -- & A FAP$\sim$0.1 non-significant peak at 7.42 and 9.45  \\
CaII 8248 & 6.11+9.02 & 6.10e+07 & 43 &  NSP   & -- & Not enough data, low S/N \\
CaII 8927 & 6.11+ -- & -- & 42 & NSP & -- & Not enough data, low S/N \\
OI 7772 & 10.74 & 3.69e+07 & 43 &  6.208$\pm$0.017   & $<$0.01 &  Some other peaks at lower significance  \\
SiII 6347 & 8.15+10.07 & 5.84e+07 & 51 &  8.097$\pm$0.020   & 0.001 & Other peak at 7.19 ($<$0.01)     \\
SiII 6371 & 8.15+10.07 & 6.80e+07 & 43 &  NSP  & -- & A peak at 8.09 with FAP=0.05   \\
TiII 4300 & 6.82+4.06 & 7.70e+06 & 46 &  13.89$\pm$0.05   &  $<$0.01 & Other peaks around 7.41  \\
TiII 4302 & 6.82+4.04 & 6.2e+06 & 43 &  NSP   & -- &  Not enough data, low S/N  \\
TiII 4307 & 6.82+4.04 & 4.6e+06 & 48 &  4.584$\pm$0.006   & $<$0.01 & Several peaks with FAP$\sim$0.01 at 6.54 and 8.09    \\
HeI 5875 & 23.07 & 5.30e+07 & 49 &   6.546$\pm$0.012  & $<$0.01 & Other peak at 6.75    \\
HeI 6678 & 23.07 & 6.37e+07 & 48 &   7.410$\pm$0.022  & 0.02 & Other low-significance peaks at 5.14, 6.33, 7.02   \\
HeI 7065 & 22.72 & 9.28e+06 & 42 &   NSP & 0.01 & 0.02-FAP peaks at 8.09, 6.35, 5.43  \\
HeI 4026 & 24.04 & 8.70e+06 & 40 &   NSP  & --- &    \\
HeI 4713 & 23.59 & 5.28e+06 & 49 & 7.411$\pm$0.021 & 0.05 & \\
HeII 4686 & 24.59+51.02 & 2.21e+08 & 53 & 7.414$\pm$0.018 & $<$0.0001  &   \\
MgI 5167 & 5.11 & 1.13e+07 & 55 & NSP & --   &   \\
MgI 5172 & 5.11 & 3.37e+07 & 54 &  7.278$\pm$0.013, 7.193$\pm$0.011 & 0.01   & Several peaks, all low significance   \\
MgI 5183 & 5.11 & 5.61e+07 & 54 &  7.195$\pm$0.012 & 0.001   & More peaks at 7.40 and 8.09 with FAP$\leq$0.01    \\
FeI 3856 & 3.27 & 4.64e+06 & 32 &  NSP   & -- & Not enough data, low S/N  \\
FeI 3859 & 3.21 & 9.69e+06 & 45 &  NSP   & -- &   \\
FeI 4215 & 5.93 & -- & 51 & 7.550$\pm$0.016  & $<$0.01  &  Further lower-significance peak at 8.09  \\
FeI 4216 & 2.93 & 1.84e+04 & 45 &  NSP & -- & Not enough data, low S/N   \\
FeI 5270 & 3.96 & 3.67e+06 & 54 &  NSP  & -- & Non-significant peak at 17.6  \\
FeI/TiII 4533 & 5.68/6.82+3.97 & ---/9.20e+06 & 51 & NSP   & -- &     \\
FeI 5269 & 3.21 & 1.27e+06 & 55 & 7.195$\pm$0.014,8.095$\pm$0.016   & $<$0.01 & Two similarly strong peaks   \\
FeI 5371 & 3.27 & 1.05e+06 & 49 & 8.099$\pm$0.073   & $<$0.01 & Further peak at 7.19, nearly as strong  \\
CrI 4254 & 2.91 & 3.15e+07 & 41 & 6.546$\pm$0.024   & 0.02 & Further peak at 7.40 at nearly the same level    \\
CrI 4274 & 2.9 & --- & 38 &  NSP & -- & Many peaks with FAP$\sim$0.05    \\
\hline        
CaII 8498 Peak BC & 6.11+3.15 & 1.11e+06 &  38 & 8.859$\pm$0.072 & 0.015 & Other periods at 11.49, 15.10 with FAP$>$0.02 \\
CaII 8662 Peak BC & 6.11+3.12 & 1.06e+07 & 37 & 7.62$\pm$0.06/7.01$\pm$0.05/9.50$\pm$0.09 & $<$0.05 & Several FAP$\sim$0.05 peaks \\
CaII 8662 -BC  Blue  & 6.11+3.12 & 1.06e+07 & 26 & NSP & --  & Not enough data \\
CaII 8662 - BC Red & 6.11+3.12 & 1.06e+07 & 33 & NSP & -- & Not enough data \\
CaII 8498 - BC Blue  & 6.11+3.15 & 1.11e+06 & 24 & 8.79$\pm$0.02: & $<$0.02  & Some low-significance peaks (incl. 14.74 and 7.4) \\ 
CaII 8498 -BC  Red & 6.11+3.15 & 1.11e+06 & 35 & NSP & -- & Not enough data \\
\hline
\end{tabular}
\tablefoot{Except when otherwise stated, the signature correspond to the line NC. The field 'N' indicates the number
of lines used in the GLSP estimation, after removing low S/N and anomalous ones. NSP stands for "no significant
period".
Energy includes excitation potential (E$_k$) and (for ions) ionization potential (E$_i$), which are
shown separately as a sum for clarity. Transition probabilities and line excitation potentials
 are not available for some of the lines. 
All atomic data are from the NIST database. }
\end{footnotesize}
\end{table}
\end{landscape}

\begin{longtable}{l c c c c c c c}
\caption{\label{narrow-table} Non-blended, narrow emission lines observed in EX Lupi.
Based on line identifications by SA12, and using the atomic constants fron the NIST database. 
Only lines dominated by a NC and not blended with other emission lines
are listed here.} \\
\hline\hline
$\lambda_{obs}$ & EW & FWHM & Spec. & $\lambda_{theo}$ & A$_{ki}$ & E$_i$ - E$_k$ &  J$_i$ - J$_k$  \\
(\AA) 		& (\AA) & (\AA) &   &  (\AA)           & s$^{-1}$ & (eV)          & 			\\
\hline
\endfirsthead
\caption{Continued.}\\
\hline\hline
$\lambda_{obs}$ & EW & FWHM & Spec. & $\lambda_{theo}$ & A$_{ki}$ & E$_i$ - E$_k$ &  J$_i$ - J$_k$  \\
(\AA) 		& (\AA) & (\AA) &   &  (\AA)           & s$^{-1}$ & (eV)          & 			\\
\hline
\endhead
3964.71 & -0.20 & 0.37 & HeI & 3964.73 & 6.95e+06 & 20.61 - 23.74 & 0 - 1 \\
4026.28 & -0.53 & 0.51 & HeI & 4026.19 & 3.22e+05 & 20.96 - 24.04 & 2 - 1 \\
4471.57 & -1.06 & 0.49 & HeI & 4471.48 & 2.46e+07 & 20.96 - 23.73 & 2 -3 \\
4713.24 & -0.20 & 0.44 & HeI & 4713.15 & 5.29e+06 & 20.96 - 23.59 & 2 - 1 \\
4921.99 & -0.17 & 0.28 & HeI & 4921.93 & 1.99e+07 & 21.22 - 23.74 & 1 - 2 \\
5015.69 & -0.43 & 0.35 & HeI & 5015.68 & 1.34e+07 & 20.62 - 23.09 & 0 - 1 \\
5875.74 & -1.67 & 0.74 & HeI & 5875.62 & 7.07e+07 & 20.96 - 23.07 & 2 - 3 \\
6678.23 & -0.69 & 0.59 & HeI & 6678.15 & 6.68e+03 & 21.22 - 23.07 & 1 - 2 \\
7065.30 & -0.49 & 0.79 & HeI & 7065.19 & 1.55e+07 & 20.96 - 22.72 & 2 - 1 \\
4685.84 & -0.59 & 0.66 & HeII & 4685.3 & 1.47e+07 & 48.37 - 51.01 & 5/2 - 5/2 \\
\\                   
4287.89 & -0.11 & 0.29 & NaI & 4287.84 & --- & 2.10 - 4.99 & 1/2 - 1/2 \\
\\                   
4226.72 & -0.46 & 0.22 & CaI & 4226.73 & 2.18e+08 & 0.00 - 2.93 & 0 - 1 \\
6456.86 & -0.04 & 0.34 & CaII & 6456.87 & --- & --- - --- & --- - --- \\
8248.84 & -0.05 & 0.38 & CaII & 8248.80 & 6.10e+07 & 7.51 - 9.02 & 3/2 - 5/2 \\
8912.07 & -0.11 & 0.61 & CaII & 8912.07 & --- & --- - --- & --- - --- \\
8927.33 & -0.14 & 0.64 & CaII & 8927.36 & --- & --- - --- & --- - --- \\
\\                   
6587.65 & -0.01 & 0.25 & CI & 6587.61 & 5.09e+06 & 8.54 - 10.42 & 1 - 1 \\
\\                   
7772.00 & -0.23 & 0.69 & OI & 7771.94 & 3.69e+07 & 9.15 - 10.74 & 2 - 3 \\
7774.18 & -0.17 & 0.63 & OI & 7774.17 & 3.69e+07 & 9.15 - 10.74 & 2 - 2 \\
7775.30 & -0.15 & 0.63 & OI & 7775.39 & 3.69e+07 & 9.15 - 10.74 & 2 - 1 \\
8446.46 & -0.58 & 1.65 & OI & 8446.25 & 3.22e+07 & 9.52 - 10.99 & 1 - 0 \\
\\                   
3905.52 & -0.69 & 0.17 & SiI & 3905.52 & 1.33e+07 & 1.91 - 5.08 & 0 - 1 \\
4130.88 & -0.12 & 0.23 & SiII & 4130.89 & 1.74e+08 & 9.84 - 12.84 & 5/2 - 7/2 \\
6347.10 & -0.16 & 0.38 & SiII & 6347.10 & 5.84e+07 & 8.12 - 10.07 & 1/2 - 3/2 \\
6371.37 & -0.10 & 0.39 & SiII & 6371.36 & 6.80e+07 & 8.12 - 10.07 & 1/2 - 1/2 \\
\\                   
3986.76 & -0.06 & 0.20 & MgI & 3986.75 & 7.30e+06 & 4.35 - 7.45 & 1 - 2 \\
4057.51 & -0.07 & 0.20 & MgI & 4057.51 & 1.02e+07 & 4.35 - 7.40 & 1 - 2 \\
4703.01 & -0.06 & 0.22 & MgI & 4702.99 & 2.19e+07 & 4.35 - 6.98 & 1 - 2 \\
5172.69 & -0.46 & 0.33 & MgI & 5172.68 & 3.37e+07 & 2.71 - 5.11 & 1 - 1 \\
5183.60 & -0.50 & 0.32 & MgI & 5183.60 & 5.61e+07 & 2.72 - 5.11 & 2 - 1 \\
5528.41 & -0.06 & 0.24 & MgI & 5528.40 & 1.39e+07 & 4.35 - 6.59 & 1 - 2 \\
\\                   
3895.64 & -0.15 & 0.17 & FeI & 3895.66 & 9.39e+06 & 0.11 - 3.29 & 1 - 0 \\
3899.04 & -0.06 & 0.26 & FeI & 3899.03 & 7.60e+05 & 2.45 - 5.63 & 4 - 4 \\
3899.70 & -0.22 & 0.20 & FeI & 3899.71 & 2.58e+06 & 0.09 - 3.27 & 2 - 2 \\
3900.53 & -0.21 & 0.16 & FeI & 3900.51 & 7.90e+06 & 3.24 - 6.42 & 3 - 3 \\
3902.94 & -0.23 & 0.20 & FeI & 3902.95 & 2.14e+07 & 1.56 - 4.73 & 3 - 3 \\
3903.89 & -0.09 & 0.21 & FeI & 3903.90 & 7.61e+06 & 2.99 - 6.17 & 4 - 4 \\
3906.47 & -0.13 & 0.15 & FeI & 3906.48 & 8.32e+05 & 0.11 - 3.28 & 1 - 1 \\
3907.93 & -0.07 & 0.17 & FeI & 3907.93 & 6.67e+06 & 2.76 - 5.93 & 3 - 2 \\
3917.18 & -0.09 & 0.14 & FeI & 3917.18 & 4.34e+05 & 0.99 - 4.15 & 2 - 3 \\
3920.26 & -0.19 & 0.17 & FeI & 3920.26 & 2.60e+06 & 0.12 - 3.28 & 0 - 1 \\
3922.91 & -0.32 & 0.21 & FeI & 3922.91 & 1.08e+06 & 0.05 - 3.21 & 3 - 4 \\
3927.92 & -0.29 & 0.20 & FeI & 3927.92 & 2.60e+06 & 0.11 - 3.27 & 1 - 2 \\
3930.29 & -0.30 & 0.19 & FeI & 3930.30 & 1.99e+06 & 0.09 - 3.24 & 2 - 3 \\
3940.86 & -0.07 & 0.21 & FeI & 3940.88 & 1.20e+05 & 0.96 - 4.10 & 3 - 4 \\
3963.11 & -0.04 & 0.25 & FeI & 3963.10 & 1.50e+07 & 3.28 - 6.41 & 1 - 2 \\
3977.74 & -0.07 & 0.19 & FeI & 3977.74 & 6.41e+06 & 2.20 - 5.31 & 2 - 2 \\
3986.17 & -0.04 & 0.18 & FeI & 3986.17 & --- & --- - --- & 3 - 4 \\
3997.40 & -0.07 & 0.17 & FeI & 3997.39 & 1.26e+07 & 2.73 - 5.83 & 4 - 5 \\
3998.04 & -0.06 & 0.12 & FeI & 3998.05 & 5.70e+06 & 2.69 - 5.79 & 5 - 4 \\
4004.89 & -0.04 & 0.23 & FeI & 4004.83 & --- & --- - --- & 6 - 5 \\
4005.24 & -0.14 & 0.19 & FeI & 4005.24 & 2.04e+07 & 1.56 - 4.65 & 3 - 2 \\
4021.85 & -0.04 & 0.12 & FeI & 4021.87 & 8.55e+06 & 2.76 - 5.84 & 3 - 4 \\
4035.65 & -0.04 & 0.18 & FeI & 4035.59 & --- & --- - --- & 4 - 5 \\
4045.81 & -0.28 & 0.21 & FeI & 4045.81 & 8.62e+07 & 1.48 - 4.55 & 4 - 4 \\
4051.93 & -0.05 & 0.18 & FeI & 4051.91 & 3.30e+06 & 3.40 - 6.46 & 3 - 2 \\
4062.45 & -0.03 & 0.18 & FeI & 4062.44 & 1.85e+07 & 2.85 - 5.90 & 1 - 1 \\
4063.60 & -0.20 & 0.17 & FeI & 4063.59 & 6.65e+07 & 1.56 - 4.61 & 3 - 3 \\
4071.74 & -0.15 & 0.15 & FeI & 4071.74 & 7.64e+07 & 1.61 - 4.65 & 2 - 2 \\
4107.50 & -0.05 & 0.22 & FeI & 4107.49 & 1.74e+07 & 2.83 - 5.85 & 2 - 1 \\
4132.05 & -0.19 & 0.19 & FeI & 4132.06 & 1.18e+07 & 1.61 - 4.61 & 2 - 3 \\
4134.67 & -0.06 & 0.17 & FeI & 4134.68 & 1.25e+07 & 2.83 - 5.83 & 2 - 3 \\
4143.41 & -0.08 & 0.18 & FeI & 4143.41 & 2.70e+07 & 3.05 - 6.04 & 4 - 4 \\
4143.87 & -0.27 & 0.23 & FeI & 4143.87 & 1.33e+07 & 1.56 - 4.55 & 3 - 4 \\
4147.68 & -0.07 & 0.29 & FeI & 4147.67 & 4.36e+05 & 1.48 - 4.47 & 4 - 3 \\
4163.66 & -0.04 & 0.15 & FeI & 4163.68 & --- & 2.69 - 5.67 & 5 - 5 \\
4173.46 & -0.25 & 0.21 & FeI & 4173.47 & --- & 3.01 - 5.99 & 2 - 1 \\
4177.66 & -0.12 & 0.22 & FeI & 4177.59 & 3.71e+04 & 0.91 - 3.88 & 4 - 4 \\
4191.43 & -0.12 & 0.29 & FeI & 4191.43 & 2.73e+07 & 2.47 - 5.43 & 2 - 1 \\
4202.02 & -0.18 & 0.19 & FeI & 4202.03 & 8.22e+06 & 1.48 - 4.43 & 4 - 4 \\
4216.19 & -0.09 & 0.18 & FeI & 4216.18 & 1.84e+04 & 0.00 - 2.94 & 4 - 4 \\
4217.52 & -0.05 & 0.25 & FeI & 4217.55 & 2.46e+07 & 3.43 - 6.37 & 1 - 2 \\
4219.36 & -0.03 & 0.13 & FeI & 4219.36 & 2.88e+07 & 3.57 - 6.51 & 5 - 6 \\
4222.21 & -0.05 & 0.22 & FeI & 4222.21 & 5.76e+06 & 2.45 - 5.39 & 3 - 3 \\
4227.40 & -0.13 & 0.20 & FeI & 4227.43 & 5.29e+07 & 3.33 - 6.26 & 5 - 6 \\
4235.93 & -0.09 & 0.21 & FeI & 4235.94 & 1.88e+07 & 2.43 - 5.35 & 4 - 4 \\
4238.80 & -0.05 & 0.19 & FeI & 4238.81 & 2.41e+07 & 3.40 - 6.32 & 3 - 4 \\
4247.42 & -0.05 & 0.21 & FeI & 4247.43 & 1.94e+07 & 3.37 - 6.29 & 4 - 5 \\
4250.11 & -0.11 & 0.21 & FeI & 4250.12 & 2.07e+07 & 2.47 - 5.39 & 2 - 3 \\
4250.79 & -0.18 & 0.20 & FeI & 4250.79 & 1.02e+07 & 1.56 - 4.47 & 3 - 3 \\
4260.49 & -0.11 & 0.21 & FeI & 4260.47 & 3.99e+07 & 2.40 - 5.31 & 5 - 5 \\
4271.15 & -0.12 & 0.23 & FeI & 4271.15 & 1.82e+07 & 2.45 - 5.35 & 3 - 4 \\
4271.77 & -0.33 & 0.22 & FeI & 4271.76 & 2.28e+07 & 1.48 - 4.39 & 4 - 5 \\
4282.40 & -0.11 & 0.20 & FeI & 4282.40 & 1.21e+07 & 2.18 - 5.07 & 3 - 2 \\
4294.10 & -0.23 & 0.20 & FeI & 4294.12 & 3.12e+06 & 1.48 - 4.37 & 4 - 4 \\
4315.02 & -0.19 & 0.28 & FeI & 4315.08 & 7.76e+06 & 2.20 - 5.07 & 2 - 2 \\
4324.99 & -0.04 & 0.18 & FeI & 4324.95 & --- & 2.18 - 5.06 & 2 - 3 \\
4325.77 & -0.16 & 0.20 & FeI & 4325.76 & 5.16e+07 & 1.61 - 4.47 & 2 - 3 \\
4375.93 & -0.11 & 0.18 & FeI & 4375.93 & 2.95e+04 & 0.00 - 2.83 & 4 - 5 \\
4383.54 & -0.27 & 0.20 & FeI & 4383.54 & 5.00e+07 & 1.48 - 4.31 & 4 - 5 \\
4404.75 & -0.25 & 0.21 & FeI & 4404.75 & 2.75e+07 & 1.56 - 4.37 & 3 - 4 \\
4415.11 & -0.16 & 0.18 & FeI & 4415.12 & 1.19e+07 & 1.61 - 4.42 & 2 - 3 \\
4442.33 & -0.06 & 0.21 & FeI & 4442.34 & 3.76e+06 & 2.20 - 4.99 & 2 - 2 \\
4459.12 & -0.05 & 0.18 & FeI & 4459.12 & 2.52e+06 & 2.18 - 4.96 & 3 - 3 \\
4461.65 & -0.05 & 0.16 & FeI & 4461.65 & 2.95e+04 & 0.09 - 2.87 & 2 - 3 \\
4494.56 & -0.06 & 0.18 & FeI & 4494.56 & 3.45e+06 & 2.20 - 4.96 & 2 - 3 \\
4528.61 & -0.11 & 0.20 & FeI & 4528.61 & 5.44e+06 & 2.18 - 4.91 & 3 - 4 \\
4592.05 & -0.05 & 0.29 & FeI & 4592.65 & 1.61e+05 & 1.56 - 4.26 & 3 - 3 \\
4920.50 & -0.08 & 0.25 & FeI & 4920.50 & 3.58e+07 & 2.83 - 5.35 & 5 - 4 \\
4957.51 & -0.17 & 0.45 & FeI & 4957.60 & 4.22e+07 & 2.81 - 5.31 & 6 - 5 \\
5171.59 & -0.15 & 0.30 & FeI & 5171.60 & 4.46e+05 & 1.48 - 3.88 & 4 - 4 \\
5232.95 & -0.07 & 0.22 & FeI & 5232.94 & 1.94e+07 & 2.94 - 5.31 & 4 - 5 \\
5266.54 & -0.03 & 0.22 & FeI & 5266.56 & 1.10e+07 & 3.00 - 5.35 & 3 - 4 \\
5269.53 & -0.27 & 0.25 & FeI & 5269.54 & 1.27e+06 & 0.86 - 3.21 & 5 - 4 \\
5270.36 & -0.10 & 0.20 & FeI & 5270.36 & 3.67e+06 & 1.61 - 3.96 & 2 - 1 \\
5328.03 & -0.21 & 0.23 & FeI & 5328.04 & 1.15e+06 & 0.91 - 3.24 & 4 - 3 \\
5328.53 & -0.12 & 0.29 & FeI & 5328.53 & 4.74e+05 & 1.56 - 3.88 & 3 - 3 \\
5371.48 & -0.15 & 0.23 & FeI & 5371.49 & 1.05e+06 & 0.96 - 3.27 & 3 - 2 \\
5397.12 & -0.13 & 0.25 & FeI & 5397.13 & 2.58e+05 & 0.91 - 3.21 & 4 - 4 \\
5405.77 & -0.11 & 0.21 & FeI & 5405.77 & 1.09e+06 & 0.99 - 3.28 & 2 - 1 \\
5429.70 & -0.14 & 0.26 & FeI & 5429.70 & 4.27e+05 & 0.96 - 3.24 & 3 - 3 \\
5446.92 & -0.17 & 0.29 & FeI & 5446.87 & 2.50e+04 & 1.61 - 3.88 & 2 - 3 \\
5455.59 & -0.09 & 0.26 & FeI & 5455.61 & 6.05e+05 & 1.01 - 3.28 & 1 - 1 \\
3938.34 & -0.14 & 0.23 & FeII & 3938.97 & 8.40e+05 & 5.91 - 9.06 & 3/2 - 5/2 \\
4122.61 & -0.16 & 0.35 & FeII & 4122.67 & 3.30e+04 & 2.58 - 5.59 & 5/2 - 5/2 \\
4178.85 & -0.24 & 0.18 & FeII & 4178.86 & 1.72e+05 & 2.58 - 5.55 & 5/2 - 7/2 \\
4233.16 & -0.41 & 0.19 & FeII & 4233.17 & 7.22e+05 & 2.58 - 5.51 & 5/2 - 7/2 \\
4233.59 & -0.05 & 0.19 & FeII & 4233.17 & 7.22e+05 & 2.58 - 5.51 & 5/2 - 7/2 \\
5234.62 & -0.22 & 0.25 & FeII & 5234.63 & 2.5e+05 & 3.22 - 5.89 & 7/2 - 5/2 \\
4258.16 & -0.10 & 0.31 & FeII & 4258.15 & 3.10e+04 & 2.70 - 5.62 & 3/2 - 3/2 \\
4273.31 & -0.14 & 0.35 & FeII & 4273.33 & 9.10e+04 & 2.70 - 5.60 & 3/2 - 1/2 \\
4296.55 & -0.19 & 0.19 & FeII & 4296.57 & 7.00e+04 & 2.70 - 5.59 & 3/2 - 5/2 \\
4303.17 & -0.25 & 0.21 & FeII & 4303.18 & 2.20e+05 & 2.70 - 5.58 & 3/2 - 3/2 \\
4351.77 & -0.29 & 0.22 & FeII & 4351.77 & 4.86e+05 & 2.70 - 5.55 & 3/2 - 5/2 \\
4384.31 & -0.06 & 0.18 & FeII & 4384.32 & 7.20e+03 & 2.66 - 5.48 & 11/2 - 9/2 \\
4385.38 & -0.22 & 0.23 & FeII & 4385.39 & 4.50e+05 & 2.78 - 5.60 & 1/2 - 1/2 \\
4416.82 & -0.25 & 0.22 & FeII & 4416.83 & 2.10e+05 & 2.78 - 5.58 & 1/2 - 3/2 \\
4489.20 & -0.11 & 0.21 & FeII & 4489.18 & 5.90e+04 & 2.83 - 5.59 & 7/2 - 5/2 \\
4491.39 & -0.15 & 0.19 & FeII & 4491.40 & 1.89e+05 & 2.86 - 5.62 & 3/2 - 3/2 \\
4508.28 & -0.17 & 0.18 & FeII & 4508.29 & 7.30e+05 & 2.86 - 5.60 & 3/2 - 1/2 \\
4515.32 & -0.15 & 0.18 & FeII & 4515.34 & 2.37e+05 & 2.84 - 5.59 & 5/2 - 5/2 \\
4520.21 & -0.12 & 0.20 & FeII & 4520.22 & 9.80e+04 & 2.81 - 5.55 & 9/2 - 7/2 \\
4549.51 & -0.31 & 0.26 & FeII & 4549.47 & 1.00e+06 & 2.83 - 5.55 & 7/2 - 5/2 \\
4555.88 & -0.21 & 0.21 & FeII & 4555.89 & 2.26e+05 & 2.83 - 5.55 & 7/2 - 7/2 \\
4576.33 & -0.08 & 0.19 & FeII & 4576.34 & 6.40e+04 & 2.84 - 5.55 & 5/2 - 5/2 \\
4582.82 & -0.07 & 0.26 & FeII & 4582.83 & 3.44e+04 & 2.84 - 5.55 & 5/2 - 7/2 \\
4620.52 & -0.06 & 0.23 & FeII & 4620.52 & 2.53e+04 & 2.83 - 5.51 & 7/2 - 7/2 \\
4629.33 & -0.13 & 0.21 & FeII & 4629.34 & 1.72e+05 & 2.81 - 5.48 & 9/2 - 9/2 \\
4666.76 & -0.11 & 0.27 & FeII & 4666.76 & 1.30e+04 & 2.83 - 5.48 & 7/2 - 9/2 \\
4731.45 & -0.08 & 0.23 & FeII & 4731.45 & 2.80e+04 & 2.89 - 5.51 & 5/2 - 7/2 \\
4923.92 & -0.44 & 0.29 & FeII & 4923.92 & 4.28e+06 & 2.89 - 5.41 & 5/2 - 3/2 \\
5018.44 & -0.51 & 0.31 & FeII & 5018.43 & 2.00e+06 & 2.89 - 5.36 & 5/2 - 5/2 \\
5169.03 & -0.71 & 0.30 & FeII & 5169.03 & 4.22e+06 & 2.89 - 5.29 & 5/2 - 7/2 \\
5197.57 & -0.19 & 0.21 & FeII & 5197.58 & 5.40e+05 & 3.23 - 5.62 & 5/2 - 3/2 \\
5234.62 & -0.22 & 0.25 & FeII & 5234.62 & 2.50e+05 & 3.22 - 5.59 & 7/2 - 5/2 \\
5264.79 & -0.18 & 0.37 & FeII & 5264.81 & 3.52e+04 & 3.23 - 5.58 & 5/2 - 3/2 \\
5284.10 & -0.11 & 0.23 & FeII & 5284.11 & 1.90e+04 & 2.89 - 5.24 & 5/2 - 7/2 \\
5316.65 & -0.34 & 0.31 & FeII & 5316.62 & 3.89e+05 & 3.15 - 5.48 & 11/2 - 9/2 \\
5325.54 & -0.07 & 0.32 & FeII & 5325.55 & 8.00e+04 & 3.22 - 5.55 & 7/2 - 7/2 \\
5425.23 & -0.03 & 0.26 & FeII & 5425.26 & 9.20e+03 & 3.20 - 5.48 & 9/2 - 9/2 \\
5534.84 & -0.07 & 0.20 & FeII & 5534.85 & 3.00e+04 & 3.24 - 5.48 & 11/2 - 9/2 \\
5991.35 & -0.02 & 0.23 & FeII & 5991.38 & 4.20e+03 & 3.15 - 5.22 & 11/2 - 9/2 \\
6147.69 & -0.03 & 0.23 & FeII & 6147.73 & 1.30e+05 & 3.89 - 5.90 & 3/2 - 1/2 \\
6238.35 & -0.05 & 0.35 & FeII & 6238.38 & 7.50e+04 & 3.89 - 5.88 & 3/2 - 3/2 \\
6247.54 & -0.10 & 0.35 & FeII & 6247.56 & 1.60e+05 & 3.89 - 5.88 & 5/2 - 3/2 \\
6456.39 & -0.09 & 0.37 & FeII & 6456.38 & 1.70e+05 & 3.90 - 5.82 & 7/2 - 5/2 \\
6516.08 & -0.06 & 0.39 & FeII & 6516.05 & 8.30e+03 & 2.89 - 4.79 & 5/2 - 7/2 \\
\\                   
3913.47 & -0.23 & 0.19 & TiII & 3913.46 & 1.60e+07 & 1.12 - 4.28 & 7/2 - 7/2 \\
4028.34 & -0.08 & 0.21 & TiII & 4028.34 & 5.10e+06 & 1.89 - 4.97 & 9/2 - 7/2 \\
4053.84 & -0.05 & 0.18 & TiII & 4053.83 & 4.20e+06 & 1.89 - 4.95 & 7/2 - 5/2 \\
4290.23 & -0.23 & 0.24 & TiII & 4290.23 & 4.60e+06 & 1.16 - 4.05 & 3/2 - 5/2 \\
4300.04 & -0.30 & 0.25 & TiII & 4300.05 & 7.70e+06 & 1.18 - 4.06 & 5/2 - 7/2 \\
4301.91 & -0.18 & 0.25 & TiII & 4301.93 & 6.20e+06 & 1.16 - 4.04 & 1/2 - 3/2 \\
4307.88 & -0.27 & 0.22 & TiII & 4307.90 & 4.60e+06 & 1.16 - 4.04 & 3/2 - 3/2 \\
4312.86 & -0.16 & 0.27 & TiII & 4312.87 & 4.10e+06 & 1.18 - 4.05 & 5/2 - 5/2 \\
4367.63 & -0.04 & 0.19 & TiII & 4367.66 & 1.90e+06 & 2.59 - 5.43 & 7/2 - 9/2 \\
4395.02 & -0.18 & 0.16 & TiII & 4395.04 & 9.40e+06 & 1.08 - 3.90 & 5/2 - 7/2 \\
4399.76 & -0.09 & 0.22 & TiII & 4399.77 & 3.10e+06 & 1.24 - 4.05 & 3/2 - 5/2 \\
4417.72 & -0.08 & 0.18 & TiII & 4417.72 & 2.10e+06 & 1.16 - 3.97 & 3/2 - 5/2 \\
4443.80 & -0.10 & 0.17 & TiII & 4443.80 & 1.10e+07 & 1.08 - 3.87 & 3/2 - 5/2 \\
4450.49 & -0.10 & 0.22 & TiII & 4450.49 & 2.00e+06 & 1.08 - 3.87 & 5/2 - 5/2 \\
4468.49 & -0.13 & 0.20 & TiII & 4468.50 & 1.00e+07 & 1.13 - 3.90 & 9/2 - 7/2 \\
4501.27 & -0.12 & 0.20 & TiII & 4501.27 & 9.80e+06 & 1.12 - 3.87 & 7/2 - 5/2 \\
4563.77 & -0.07 & 0.18 & TiII & 4571.98 & 1.20e+07 & 1.57 - 4.28 & 9/2 - 7/2 \\
4571.97 & -0.09 & 0.18 & TiII & 4563.77 & 8.80e+06 & 1.22 - 3.94 & 1/2 - 3/2 \\
4589.92 & -0.07 & 0.34 & TiII & 4589.95 & 1.30e+06 & 1.24 - 3.94 & 3/2 - 3/2 \\
5185.96 & -0.10 & 0.29 & TiII & 5185.91 & 1.40e+06 & 1.89 - 4.28 & 7/2 - 7/2 \\
\\                   
3944.01 & -0.30 & 0.20 & AlI & 3944.01 & 4.93e+07 & 0.00 - 3.14 & 1/2 - 1/2 \\ 
3961.52 & -0.25 & 0.20 & AlI & 3961.52 & 9.80e+07 & 0.01 - 3.14 & 3/2 - 1/2 \\
\\                   
4025.14 & -0.07 & 0.18 & CrI & 4025.01 & --- & --- - --- & --- - --- \\
4077.72 & -0.33 & 0.18 & CrI & 4077.68 & --- & --- - --- & --- - --- \\
4254.33 & -0.22 & 0.23 & CrI & 4254.35 & 3.15e+07 & 0.00 - 2.91 & 3 - 4 \\
4274.80 & -0.20 & 0.23 & CrI & 4274.80 & 3.07e+07 & 0.00 - 2.90 & 3 - 3 \\
4289.72 & -0.16 & 0.21 & CrI & 4289.72 & 3.16e+07 & 0.00 - 2.89 & 3 - 2 \\
4351.77 & -0.29 & 0.22 & CrI & 4351.77 & 1.20e+07 & 1.03 - 3.88 & 4 - 5 \\
4541.51 & -0.12 & 0.24 & CrI & 4541.51 & 3.80e+06 & 3.08 - 5.81 & 3 - 3 \\
4012.38 & -0.12 & 0.17 & CrII & 4012.47 & --- & --- - --- & --- - --- \\
4558.64 & -0.08 & 0.21 & CrII & 4558.66 & 8.80e+06 & 4.07 - 6.79 & 9/2 - 7/2 \\
4588.19 & -0.07 & 0.22 & CrII & 4588.20 & 1.20e+07 & 4.07 - 6.77 & 7/2 - 5/2 \\
4616.64 & -0.03 & 0.24 & CrII & 4616.63 & 4.00e+06 & 4.07 - 6.76 & 3/2 - 3/2 \\
4824.11 & -0.08 & 0.22 & CrII & 4824.13 & 1.70e+06 & 3.87 - 6.44 & 9/2 - 9/2 \\
4848.19 & -0.06 & 0.25 & CrII & 4848.23 & 2.60e+06 & 3.86 - 6.42 & 7/2 - 7/2 \\
4876.45 & -0.06 & 0.26 & CrII & 4876.47 & 1.60e+06 & 3.86 - 6.41 & 7/2 - 5/2 \\
5208.47 & -0.10 & 0.32 & CrII & 5197.58 & 5.06e+07 & 0.94 - 3.32 & 2 -3 \\ 
\\                   
4030.73 & -0.18 & 0.27 & MnI & 4030.76 & 1.70e+07 & 0.00 - 3.08 & 5/2 - 7/2 \\
4033.06 & -0.17 & 0.23 & MnI & 4033.07 & 1.65e+07 & 0.00 - 3.07 & 5/2 - 5/2 \\
4034.46 & -0.13 & 0.23 & MnI & 4034.49 & 1.58e+07 & 0.00 - 3.07 & 5/2 - 3/2 \\
4110.99 & -0.06 & 0.18 & MnI & 4110.90 & --- & --- - --- & --- - --- \\
\hline
\end{longtable}

\section{The disk properties of EX Lupi revisited \label{diskmass}}

Starting with the SED from Sipos et al. (2009), we have fitted the spectral energy distribution (SED) using the RADMC code (Dullemond \& Dominik 2004).
The star is very variable in the optical and IR, and most of the data available are non-simultaneous, 
so the results have to be handled with care. Nevertheless, we can in this way constrain the basic
disk parameters, including the estimated
disk mass. Assuming that the disk is populated by grains with sizes 0.1-100 $\mu$m with a collisional distribution,
plus some optically thin population of small (0.1 to 2 $\mu$m) grains in the inner hole (to account for the
strong silicate feature), we can obtain reasonable fits to the SED with disk masses
as low as 1-3$\times$10$^{-3}$ M$_\odot$. The mass is highly dependent on the maximum grain size and on
the detailed disk structure, which is beyond the scope of this paper. 
Our estimate of the inner hole in the disk is roughly consistent with the values of Sipos et al. 2009 ($\sim$0.4-0.5 AU,
compared to their estimate of $\sim$0.3 AU), noting that the hole radius is also highly dependent on the
grain properties and structure of the inner disk rim. In particular, the hole size depends on the sizes of the small grains
within the hole and whether we assume them to be distributed all over the inner disk or concentrated near the rim.
Sipos et al. (2009) also demonstrated that a smoothly-curved rim is necessary to reproduce the IR flux.
We also find that the best fit to the quiescence SED is obtained assuming a stellar radius of 1.5 R$_\odot$,
slightly lower than the previous value of 1.6 R$_\odot$, although this depends on the quiescence data used,
given the small scale variability of the star.
From this brief analysis, we confirm the presence of a dust hole and also suggest that the disk mass does not
necessarily needs to be significantly larger than typical disk masses for similar CTTS, if we
assume a reasonable maximum grain size.

\section{Accretion rate estimates during quiescence and their effects on the radial velocities\label{acc-appendix}}

We estimate the accretion rate of EX~Lupi using the accretion-related emission lines.
Our FEROS/HARPS data are not flux-calibrated, and were often taken under non-photometric conditions. Given the 
variability of the object, it is not even possible to estimate the line flux as required to calculate
accretion rates (Alcal\'{a} et al. 2014), so we opt for using archival X-shooter data
from the ESO database.
X-shooter has three spectrograph arms (UVB, VIS, and NIR ), providing simultaneous wavelength coverage from 
$\sim$300 to $\sim$2480\,nm. EX~Lupi was observed as part of program 085.C-0764(A) 
on  May 4, 2010. This epoch is intermediate (within a month) between two of our FEROS observations. 
The slit widths used were 0$\farcs$5, 0$\farcs$5, and 
0$\farcs$4, for exposure times of  600\,s, 480\,s,  and 1200\,s, respectively, for the UVB, VIS, and NIR arms. We used 
the 1D spectra, which had been  calibrated for wavelength and flux from ESO Phase 3 Data  Release. With the relations 
between the line luminosity and accretion luminosity, we derive  the accretion luminosity of EX~Lupi  using 10 emission lines 
including He\,I~5876, 6678, 7065\,\AA, H$\alpha$, Ca\,II~8498, 8542, 8862\,\AA, Pa${\beta}$, Pa${\gamma}$, and Pa${\sigma}$
(Alcal\'{a} et al. 2014).  
The mass accretion rates can then be calculated from 
the inferred accretion luminosities using the following relation:
\begin{equation}
\dot{M}_{\rm acc}=\frac{L_{{\rm acc}}R_{\star}}{{\rm G}M_{\star} (1-\frac{R_{\star}}{R_{\rm in}})},
\end{equation}
\noindent where $R_{\rm in}$ denotes the truncation radius of the disk, which is taken to be 5\,$R_{\star}$ (Gullbring et al. 1998). 
G is the gravitational constant, $M_{\star}$ is the stellar mass, and $R_{\star}$ is the stellar radius. 
The accretion rates estimated from the 10  lines range from $2\times10^{-10}$ to $6\times10^{-10}$\,$M_{\odot}$\,yr$^{-1}$  
with an average of 4$\pm2\times10^{-10}$\,$M_{\odot}$\,yr$^{-1}$. 
In order to put these values in context with our HARPS/FEROS observations, we used the H$\alpha$ line
to infer whether the X-shooter spectrum was taken at a low or a high accretion rate, with respect to the typical
quiescence values. The H$\alpha$ 10\% velocity width in the X-shooter data is 320 km/s. The two closest FEROS spectra
(taken 10 days before and 18 days after the X-shooter data) have H$\alpha$ 10\% widths of 320 and 360 km/s, respectively.
These values are between low and intermediate, comparing the rest of FEROS/HARPS data analysed,
although variations in days/weeks are often seen within our collection. Comparing the strength of the Ca II BC of the lines,
the X-shooter spectrum has also an intermediate-strength BC (note the difference in resolving power), also stronger than the
two closest FEROS spectra (which have negligible Ca II BC). With this information, we consider that the measured
accretion rate of 4$\pm2\times10^{-10}$\,$M_{\odot}$\,yr$^{-1}$ is probably an intermediate value within the typical
variations of EX Lupi in quiescence.

\begin{figure*}
\centering
\begin{tabular}{cc}
\includegraphics[width=0.45\linewidth]{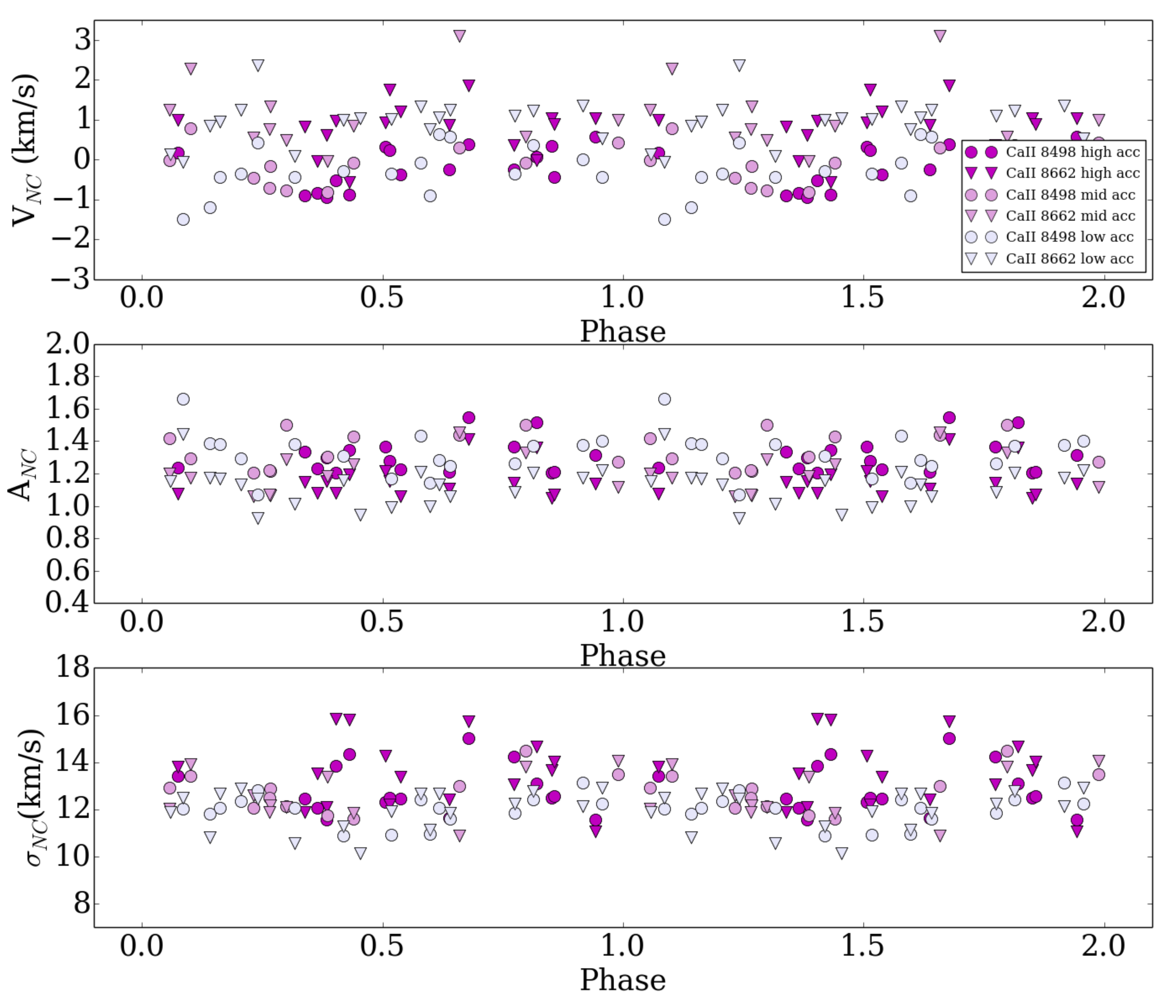} &
\includegraphics[width=0.45\linewidth]{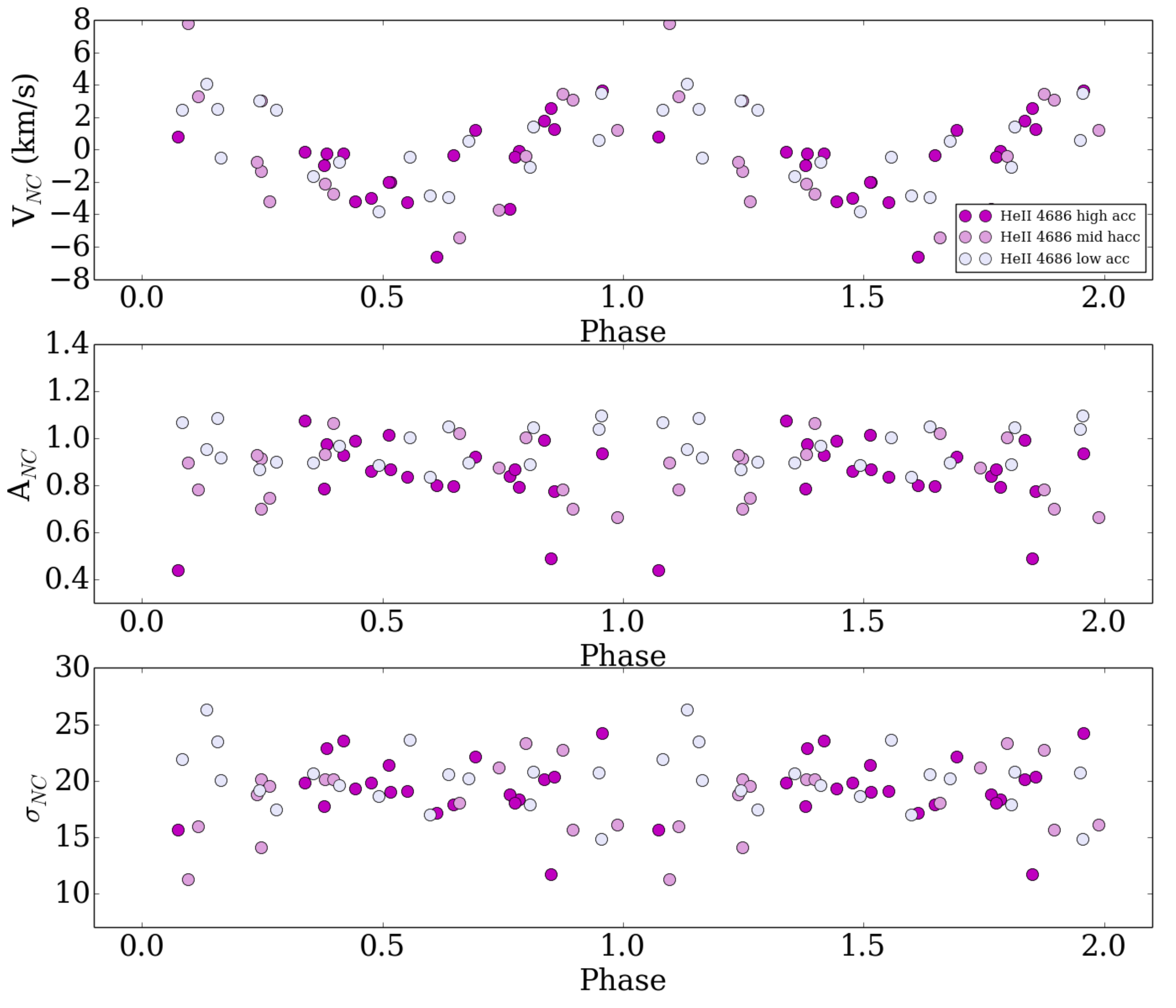} \\
\end{tabular}
\caption{Line fit parameters for the NC of the Ca II IR (left) and He II lines (right), colour-coded
according to their accretion rate (see text). The upper panel shows the radial velocity of the NC line (V$_{NC}$), the
middle panel shows the amplitude of the NC line (A$_{NC}$), and the bottom pannel
shows the Gaussian width ($\sigma_{NC}$). There is no evident modulation of the RV with the accretion rate, and 
only the line width for the Ca II NC seems to be higher for the higher accretion rates, although this could be 
also an effect of the presence of a stronger underlying BC. The signatures observed in the NC are thus very robust 
with respect to variations of the accretion rate within one order of magnitude. 
 \label{accphase-fig}}
\end{figure*}

Increases in the H$\alpha$ EW and velocity wings are usually interpreted as increases in the 
accretion rate (Natta et al. 2004). In EX Lupi, the BC of the Ca II IR triplet becomes stronger when 
the accretion rate increases. For the objects with
lower accretion rates, according to H$\alpha$, the BC of the Ca II IR triplet becomes undetectable.
We thus can use the strength of the BC of the Ca II IR lines to classify our data according to their
accretion levels, considering "high accretion"
(BC peak $\geq$1 over the normalised continuum levels), "intermediate accretion" (BC peak 0.2-1 over the normalised
continuum levels), and "low accretion" (BC peak $<$0.2 or undetectable). These values would roughly correspond
to accretion rates of a few times 10$^{-9}$M$_\odot$/yr, between 5$\times$10$^{-10}$-10$^{-9}$M$_\odot$/yr,
and below 5$\times$10$^{-10}$M$_\odot$/yr. This classification allowed us to examine to which extent
the NC is affected by variations in the accretion rate. Surprisingly, variations in the accretion rate within 
this range do not seem to affect the NC radial velocities nor amplitudes (Figure \ref{accphase-fig}). 
Only the width of the NC of the  Ca II IR triplet seems
to increase in some of the higher accretion spectra, but this could be also due to contamination by the stronger BC,
especially in the case where the BC is centred around zero velocity.
There is also no significant correlation between the phase and the strength of the accretion rate. In fact, 
for datasets taken on subsequent days, we usually observe that the accretion rate/strength of the features
is relatively constant, independently of the shape of the feature.

\section{Cross-correlation tables for the Ca II line fits \label{appendix-xcorCaII}}

The following tables include the result of the cross-correlation of the three Gaussian
fit for the Ca II 8498 and 8662\AA\ lines. For all the correlations, we perfomed
a Spearman rank test providing the correlation coefficient (r) and false-alarm probability (p)
for the various Gaussian parameters:
amplitude (A1, A2, A3), central wavelength (C1, C2, C3), and width (S1, S2, S3). 
The numbers indicate to which one of the three Gaussians involved in the fit corresponds:
narrow component (1), weakest broad component (2), strongest broad component (3).
For the autocorrelation of every line, we check the relation between all different
fit parameters to explore potential common dependencies between the three components
(Tables \ref{xcorgaussians1-table} and \ref{xcorgaussians2-table}). With the two-line
correlation, we explore whether the independent fit of the two Ca II 8498 and 8662\AA\ lines
provides consistent results, as both lines of the triplet are expected to be
similar (Table  \ref{xcorgaussians-table}).
The most significant autocorrelations appear between the width and amplitude of the lines,
suggesting that broader lines tend to be also stronger. This is consistent with 
an accretion scenario where an increased rate requires more material coming from more
distant regions in the disk and results in the heating of a larger part of the accretion structure.
The width of the NC could also be affected in case of increased accretion, although this could
be also due to contamination by the strong BC.
In the two-line cross-correlation, we observe that all the parameters are strongly correlated (as
expected for the two lines being nearly identical) except for the
center of the second Gaussian component. This is probably due to the fitting being degenerated:
the main BC component is often well reproduced by a single Gaussian, so the remaining Gaussian component
is often weak and poorly defined. Although the three Gaussian model offers very good fits of the
line profiles, we observe small deviations from the symmetric Gaussian profiles in the BC that suggest a
more complex situation.

\begin{table*}
\caption{Correlation coefficients for the Gaussian fits to the Ca II 8498 \AA lines.} 
\label{xcorgaussians1-table}
\begin{tabular}{l c c c c c c c c c}
\hline\hline
 & A1  &    C1     &    S1      &  A2        &  C2  & S2 & A3 & C3 & S3  \\
\hline
A1 &  &  0.12/0.45 & 0.24/0.12  & -0.08/0.60 & 0.31/0.04 & -0.15/0.32 & -0.30/0.05 &  0.31/0.04 & -0.35/0.02  \\
C1 &  &            & 0.37/0.01  & 0.11/0.48 & 0.21/0.16 & 0.26/0.08 & 0.20/0.20 & 0.05/0.76 & 0.01/0.96  \\
S1 &  &            &            & 0.41/0.005 & {\bf0.50/6e-4} & 0.11/0.47  & 0.28/0.06 & -0.08/0.61  &0.20/0.18   \\
A2 &  &            &            &            & {\bf0.56/6e-5}  & 0.32/0.03 & {\bf0.66/2e-5} & -0.38/0.01 & {\bf0.59/2e-5}  \\
C2 &  &            &            &            &            & 0.23/0.12 & 0.18/0.24 &  -0.13/0.37 & 0.14/0.35  \\
S2 &  &  	   &            &            &            &           & 0.24/0.11 & 0.14/0.35 &  -0.04/0.78 \\
A3 &  &            &            &            &            &           &           & {\bf-0.61/9e-6} & {\bf0.48/8e-4}  \\
C3 &  &   &   &  &  &  &  &  									& {\bf -0.55/8e-5}  \\
\hline
\end{tabular}
\tablefoot{Autocorrelations in the 3-Gaussian fits for the CaII 8498\AA\ line.
Spearman r correlation
coefficient and p probability uncorrelated. Strong correlation (false-alarm probability $<$5e-3)
are marked in bold.}
\end{table*}

\begin{table*}
\caption{Correlation coefficients for the Gaussian fits to the CaII 8662 \AA lines.} 
\label{xcorgaussians2-table}
\begin{tabular}{l c c c c c c c c c}
\hline\hline
     &    A1         &    C1       &    S1      &  A2 &  C2  & S2 & A3 & C3 & S3  \\
\hline
A1 &                 & -0.11/0.45  & 0.19/0.20  & 0.12/0.43  & 0.28/0.05 & -0.07/0.66 & -0.14/0.35 & 0.07/0.63 &  -0.14/0.36 \\
C1 &                 &             & -0.09/0.55 & -0.11/0.46 & 0.02/0.88 & -0.02/0.91 & 0.01/0.90 & -0.25/0.09 &  0.14/0.35 \\
S1 &                 &             &            &  {\bf 0.45/2e-3} & 0.21/0.16 & 0.40/0.007 & {\bf 0.48/9e-4} & -0.17/0.25  & {\bf 0.45/2e-3}  \\
A2 &    	     &   	   &   		&  		   & 0.40/6e-3 & 0.36/0.02  & {\bf 0.53/2e-4} & -0.19/0.22  & 0.37/0.01  \\
C2 &  		     &   	   &   		&  		   &  	       & 0.06/0.70  & -0.2/0.15 & 0.09/0.58 & -0.04/0.79  \\
S2 &  		     &   	   &  		&  		   &  		&  	    &  0.36/0.01 & -0.11/0.46 & 0.1/0.27  \\
A3 &  		     &   	   &   		&  		   &  		& 	    &  		 & {\bf-0.52/3e-4} & {\bf0.68/3e-7}  \\
C3 & 		    &   	   &   		&  		   &  		&  	    &  		 &  		   &  {\bf-0.63/4e-6} \\
\hline
\end{tabular}
\tablefoot{Autocorrelations in the 3-Gaussian fits for the 8662\AA\ line.
Strong correlation (false-alarm probability $<$5e-3)
are marked in bold.}
\end{table*}

\begin{table*}
\caption{Cross-correlation coefficients for the Gaussian fits to the CaII 8498 and 8662$\AA$ lines.} 
\label{xcorgaussians-table}
\begin{tabular}{l c }
\hline\hline
 Components  &    r/p  \\
\hline
A1$_{8498}$/A1$_{8662}$ & 0.86/4e-14    \\
C1$_{8498}$/C1$_{8662}$ & 0.66/7e-7    \\
S1$_{8498}$/S1$_{8662}$ & 0.63/3e-6   \\
A2$_{8498}$/A2$_{8662}$ & 0.88/2e-15  \\
C2$_{8498}$/C2$_{8662}$ & 0.18/0.23   \\
S2$_{8498}$/S2$_{8662}$ & 0.49/6e-4   \\
A3$_{8498}$/A3$_{8662}$ & 0.81/2e-11   \\
C3$_{8498}$/C3$_{8662}$ & 0.55/8e-5    \\
S3$_{8498}$/S3$_{8662}$ & 0.54/1e-4  \\
\hline
\end{tabular}
\tablefoot{Cross-correlations in the 3-Gaussian fits for the  CaII 8498 and 8662$\AA$ lines. 
For each pair, the correlation coefficent r and
the false-alarm probability p are given as r/p.}
\end{table*}

\section{The behaviour and properties of individual lines \label{appendix-lines}}

This appendix describes in detail the properties of the most prominent lines observed in EX Lupi.

\vskip 0.3truecm
\large \textit{{The Ca II lines}}

The two Ca II IR triplet lines that are covered by FEROS (at 8498 and 8662\AA, the line
at 8542\AA\ falls within the gap) behave in a very similar way
and have similar profiles. They have a very stable NC and a remarkably variable BC, which appears 
redshifted and blueshifted up to velocities over $\pm$200 km/s, and with a peak located up to $\pm$100 km/s. 
The presence of a strong BC depends on the accretion rate, becoming undetectable
at low accretion rates. The NC is always detected and does not change much (its
peak, once extracted from the BC, is always in the range of 1-1.5 over the normalised continuum).
The lines do not show any significant RV modulation, although there is a clear offset between
their velocity and the zero velocity of the photosphere. There is also a $<$1 km/s systematic
offset between the velocity of the 8498 and 8662\AA, probably caused by the differences in
the line optical depths (see main text).

\begin{figure*}
\centering
\includegraphics[width=0.7\linewidth]{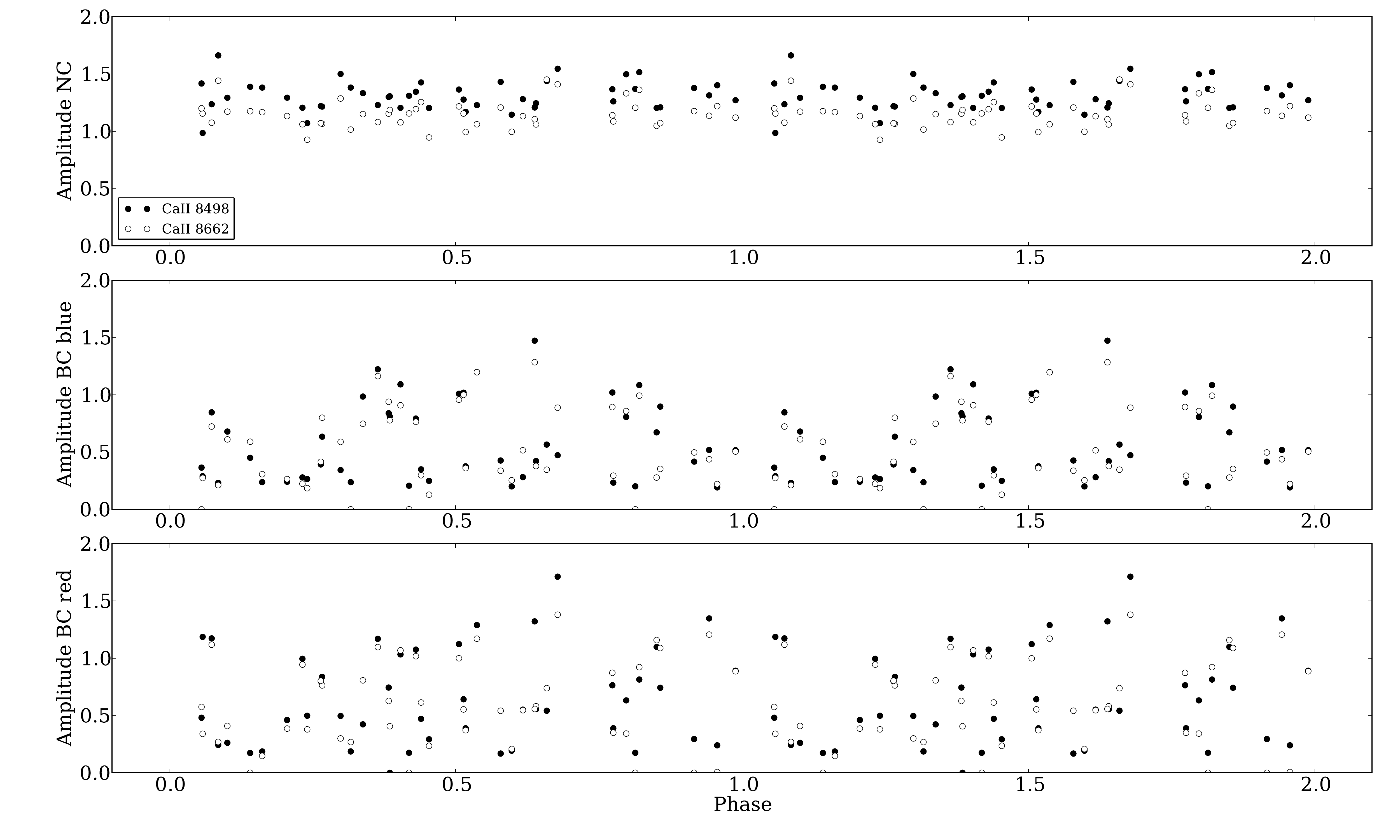}
\caption{Phase-folded curves (using the photospheric absorption line period and phase, which is nearly
identical to the He II 4686\AA\ one) for the amplitude
of the multi-Gaussian model used to fit the CaII IR lines. Top: narrow component. Middle: blueshifted
BC. Bottom: redshifted BC. There is a marginal amplitude modulation of the BC amplitude.
 \label{ampliCaIIIR}}
\end{figure*}

Regarding the amplitude of the BC, there is a low-significance modulation, especially at times of
high accretion/strong BC (Figure \ref{ampliCaIIIR}). The amplitude is maximal at 
zero velocity, which is the opposite behaviour that we see with high-excitation lines such as
He II. This is consistent with the Ca II IR lines (and, in particular, the BC) being produced in a less
embedded area of the accretion column. Therefore, we get more emission when the structure
is facing us (at zero velocity), while for lines that are produced more deeply in the accretion
column, the strongest emission is seen when we are looking sidewards to the structure.

In the IR, we also find the CaII 8248 and 8927\AA\ . Both lines are narrow, very weak and typically noisy, so
we cannot obtain much information. The  8927\AA\ line
is not detectable in spectra taken at low accretion.
The CaII H and K lines in the near-UV have a very different morphology compared to the IR lines.
The NC+BC structure is not so evident. The BC has a very strong blueshifted absorption, which
is not smooth, but also appears to have some structure. Some redshifted 
absorption could be also present on certain dates. The blueshifted absorption is clearly
correlated with the strength of the accretion rate and likely results from an accretion-powered
wind.

\begin{figure*}
\centering
\begin{tabular}{c}
\includegraphics[width=0.7\linewidth]{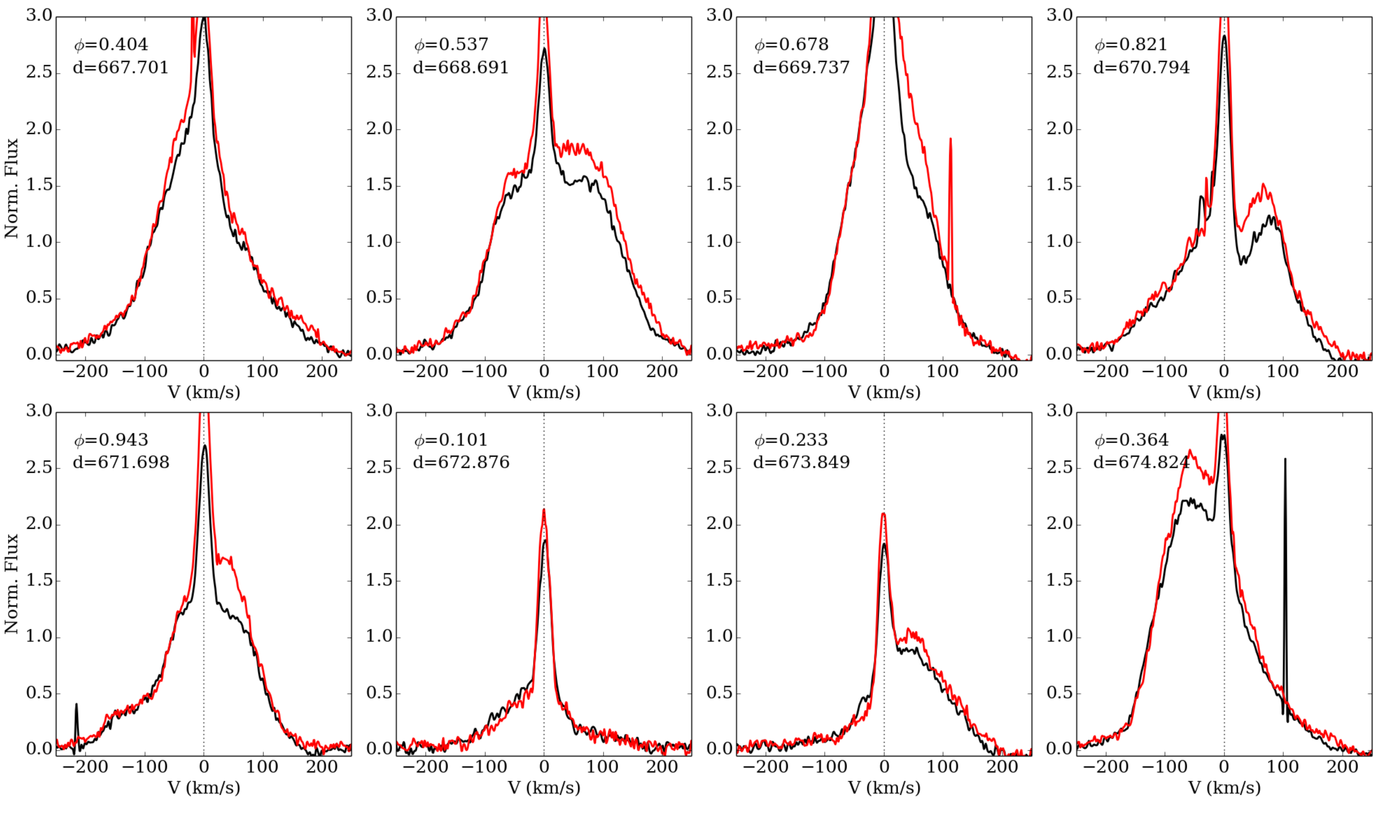} \\
\includegraphics[width=0.7\linewidth]{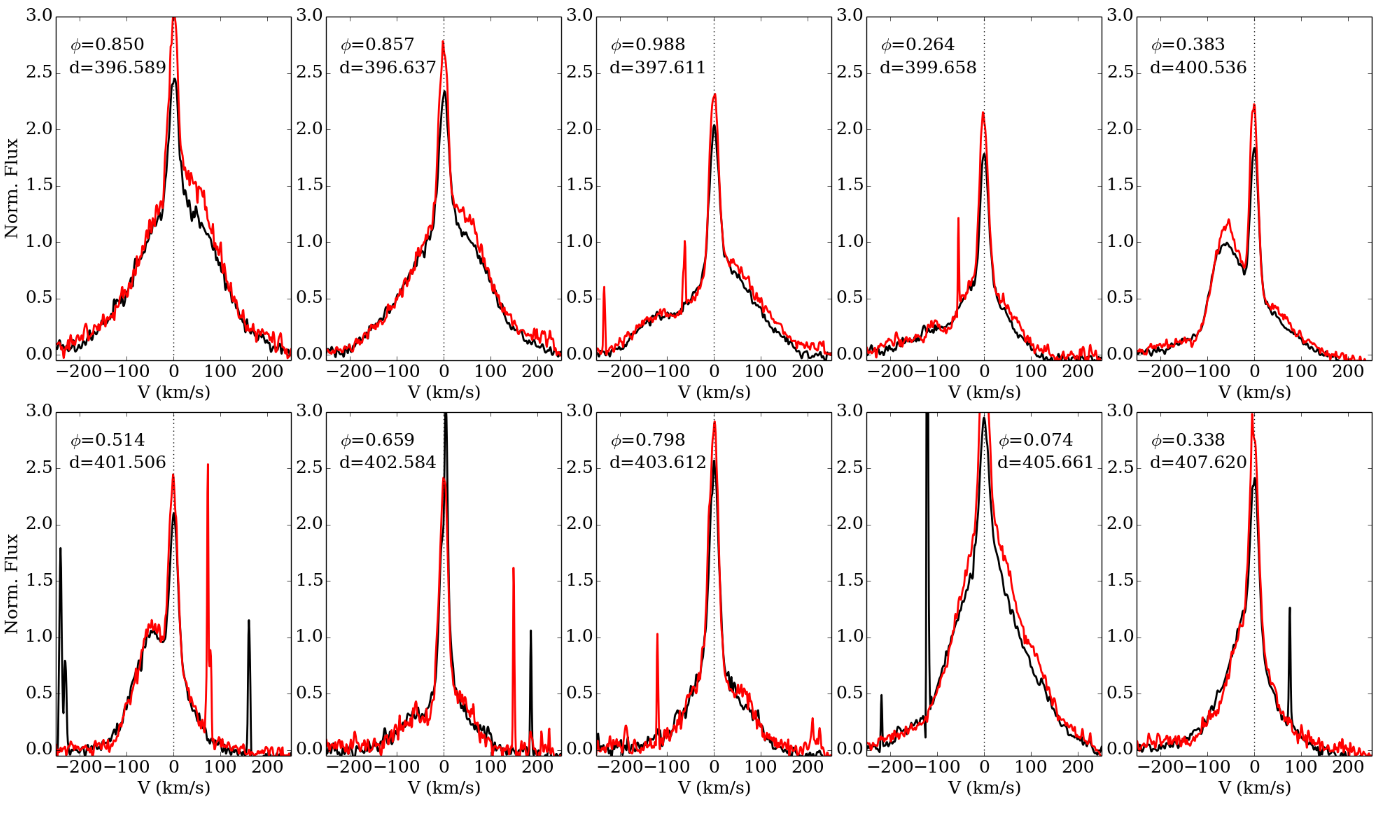} \\
\end{tabular}
\caption{Some examples of day-to-day modulation of the BC of the Ca II lines (8498\AA\ line in red,
8662\AA\ line in black). For each set, phase ($\phi$) and date (d=JD-2455000) are given.
 \label{BCphase-fig}}
\end{figure*}

\begin{figure*}
\centering
\begin{tabular}{cc}
\includegraphics[width=0.45\linewidth]{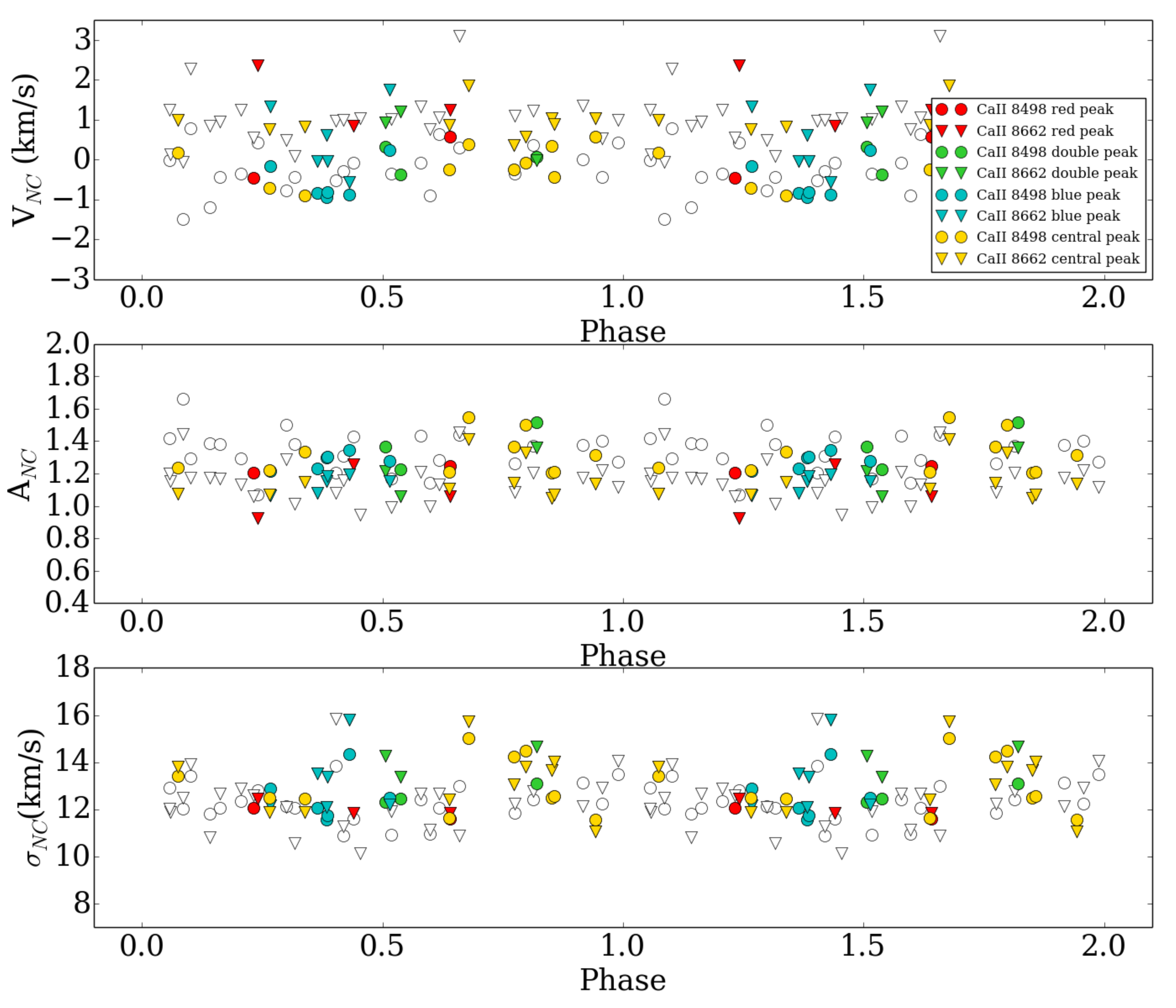} &
\includegraphics[width=0.45\linewidth]{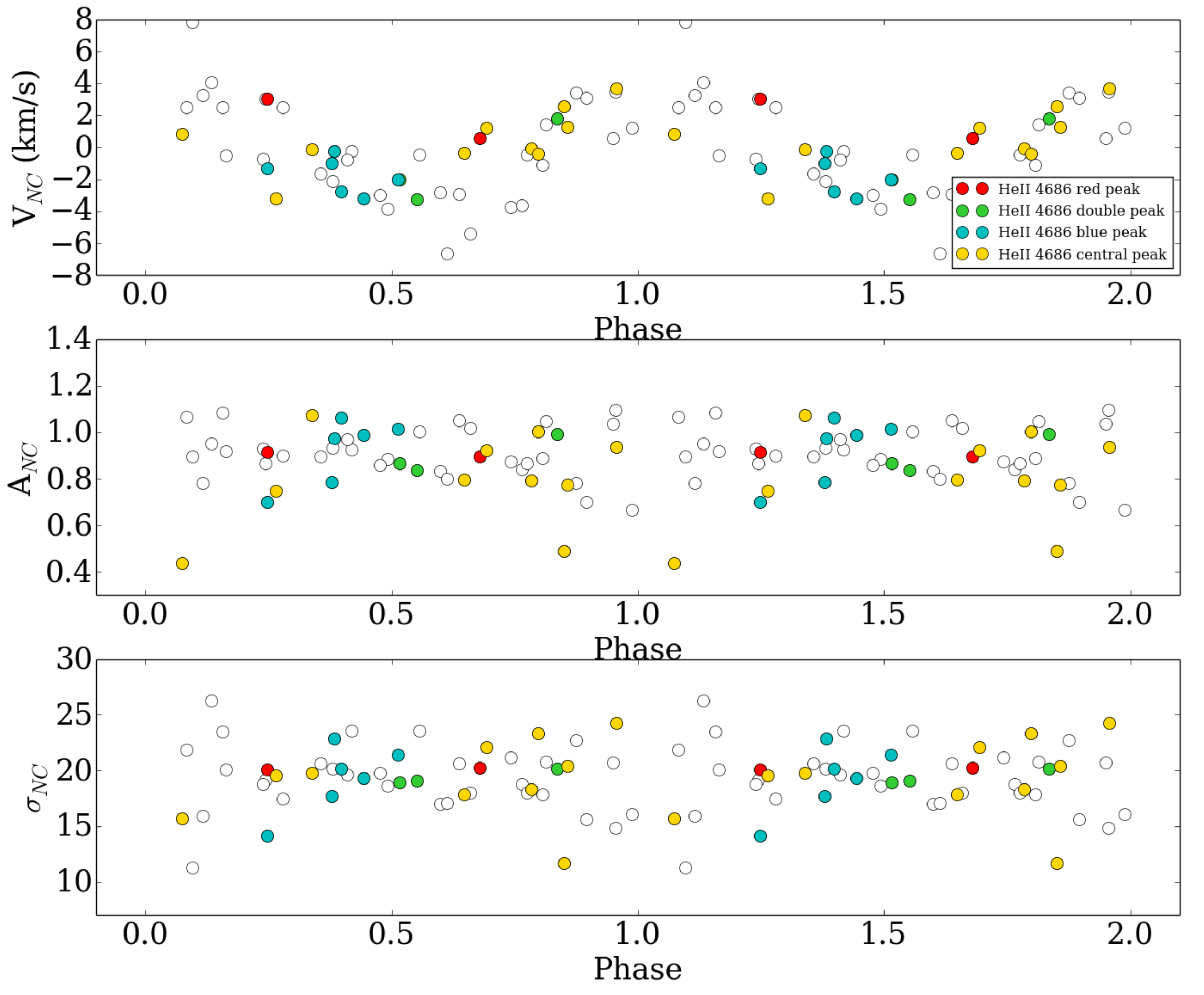} \\
\end{tabular}
\caption{ Shape of the BC (classified as "centred", "blueshifted", "redshifted", "double-peaked", see text)
vs phase. Only data points with strong BC and a clear peak are classified.}
 \label{peakphase-fig}
\end{figure*}

Examining the shape of the BC of the Ca II IR triplet vs the phase, 
we find that they are not completely independent. The rapid day-to-day variability observed during outburst (SA12)
with strong shifts of the BC peak is also manifest here when comparing observations taken on consecutive days
(Figure \ref{BCphase-fig}).
Selecting only the cases where the BC is strong ($\geq$0.3 over the continuum level) and has a simple shape,
we classified the shape of the peak as "centred" (if the velocity of the BC peak is in the range $\pm$20 km/s),
"redshifted" (for velocities $>$20 km/s), "blueshifted" (for velocities $<$-20 km/s, and "double-peak" (if the
BC has two peaks with height differences $<$0.2). For each case, we measured both the 8498 and 8662\AA\ lines, and
the final classification depends on the averaged result, plus a visual inspection to remove uncertain cases (e.g.,
those where the top of the BC is flat over a large velocity range). 
Most of the data points are 
consistent with either central or blueshifted BC, given that redshifted components tend to fail the strength
 criterion, being often weaker. Blueshifted BC appear at phases between 0.24-0.5, corresponding to
the phases for which the emission lines move from no-shift to blueshifted (Figure \ref{peakphase-fig}). 
A KS test shows a probability of
0.9\% that the phase correlation is random, which points to a significant correlation between the BC shape and
the NC line modulation, and thus to a common dynamical origin. If we assume that the BC is formed in the same
accretion structure where the NC comes from, but at a larger distance, the presence of blueshifted BC
at these phases would indicate a slightly trailing column, where the material is seen as coming to the observer
while the footprint of the accretion column where the NC originates moves from zero to maximum blueshifted velocity
with respect to the observer.

\vskip 0.3truecm
\large \textit{{The He I and He II lines}}

We detect several strong He I lines: at 4471, 5015, 5875, 6678, and 7065\AA. The He I lines
differ from the metallic lines in that they do not have a clear NC+BC structure, although this
structure was present in the outburst (SA12). They are typically broader than the NC of
the metallic lines. This was also observed in outburst (SA12) and has been noted in
other stars (Beristain et al. 1998), with the stronger 5875\AA\ line being up to $\sim$20-30 km/s, but not as 
broad as to be regarded as BC. In some cases, a weak BC (which can be blueshifted or redshifted)
is also seen. The sharp
cut of the 5875 and 6678\AA\ lines on the blue side could be a blueshifted
absorption, a possible wind signature. Unlike the Ca II lines, the He I lines become
sometimes very weak when the accretion rate drops.

The He I lines show the same kind of modulation as observed in the metallic lines.
The amplitude of the modulation is of the same order than the RV amplitude for the photospheric
absorption lines, although there is a clear offset between the zero velocities of both. This is
a similar effect to what Gahm et al. (2013) observed for the He I lines of RU Lupi.
We also observe a mild modulation of the amplitude, similar to what is observed for 
other metallic lines (Mg I, Fe I, Fe II), but with more noise, maybe because the line
fits are not so well defined due to the presence of blueshifted absorption components.
The potential blueshifted absorption component, which could correspond to a wind, does not show any clear RV variation. 

The He II line at 4686\AA\ is remarkably strong for a low-mass star. Like the
He I lines, it has only one component that is broader than the usual metallic NC. The line also
appears to have sometimes a redshifted tail (maybe due to an infalling, turbulent component), 
although it could also be caused by a nearby line
(there are many weak and strong lines in the blue and UV part of the spectrum). He II shows the
most remarkable and tight RV modulation pattern of all lines, 
considering that the line is relatively weak and that it is located
in a rather noisy part of the spectrum. Its periodogram shows a very sharp peak at 7.41d,
the same as the period observed for the photospheric absorption lines. There is no clear modulation
of the line amplitude, although an increase in the blue continuum during episodes of stronger accretion
may mask accretion-related flux variations.

\vskip 0.3truecm
\large \textit{{The Fe I and Fe II lines}}

The Fe II lines are among the strongest lines after the typical H, He, and Ca II ones.
In particular the multiplet 42 is strong in emission, with good S/N lines at 4923 and 5018 \AA\ . 
The lines show the NC+BC structure, although in this case the BC is much weaker than for
Ca II. The BC shows the same kind of modulation as observed for Ca II. The NC
reveals a clear RV modulation with an amplitude comparable to that of the photospheric
absorption line, but clearly off-phase. There is also a small modulation of the amplitude, of
the same type as described for He II (smaller amplitude at zero velocity) and thus opposite to
the amplitude variations of the Ca II IR BC.

The same result is found in other Fe II and Fe I transitions. Other strong iron lines include Fe II at 5362,
5316 \AA\ (this one is very strong and may have sometimes a weak blueshifted absorption), 4549, 4383, 4385 \AA\,
and Fe I lines at 5269, 5270 \AA. Fe I 5371.49 \AA\ has a nearby unidentified line at about 5369.95 \AA, both
of which are weak. The Fe I lines in the 3700-4000 \AA\ region are usually noisy and weak. 
In general, the behaviour of Fe II and Fe I lines is similar. Their BC show bulk velocity shifts that
are consistent with those of Ca II.
Their NC have clear RV modulations, although some of the lines are too weak or too noisy to find
a clear pattern in them. Different lines have different zero-point RV and amplitudes.

\vskip 0.3truecm
\large \textit{{The Mg I multiplet at 5167, 5172, 5183\AA}}

Several Mg I lines are also detected in the quiescence spectra, among them the multiplet
at 5167, 5172, and 5183\AA. The lines are strong and mostly narrow, with BC that are usually
faint and noisy but become stronger when the BC of Ca II IR gets stronger. They show similar
velocity patterns as the Ca II and Fe I, Fe II lines. The NC shows the usual RV modulation observed
in Fe I and Fe II, opposite to the photospheric lines RV, and the same variations in amplitude vs phase,
with the lines getting stronger at the maximum blueshifted and redshifted velocity.
As for the Ca II lines, there is also a small offset in velocity betwen the 5167\AA\ line and the
other two, although in this case contamination by a nearby strong Fe I line cannot be excluded.

\vskip 0.3truecm
\large \textit{{The O I 7774 triplet and 8446\AA\ line}}

The O I triplet around 7774 \AA\ and 8446 \AA\ are also very strong in EX Lupi.
The 8446 \AA\ complex, being a blended triplet itself, has a complex structure, very extended, and
rather weak or undetectable at low accretion rates. It does not have a clear NC, although it
is narrower in the centre and has very extended BC wings (up to $\pm$150 km/s) that
changes from blueshifted to redshifted following the patterns observed for Ca II. The line could have sometimes a weak
redshifted absorption, although due to the low S/N and the redshifted emission of the BC, it is not
evident. Due to the complexity of the line, it is not possible to study its RV.

The OI 7774 triplet is much stronger and always detected. The 7772 \AA\ line is very strong and narrow and easy to fit, giving
very good results. Having a typical width around 14-16 km/s, it also shows a BC when the accretion
rate increases. The behaviour of the BC is hard to establish due to the presence of the other triplet
lines. Fitting the two weaker lines is often a problem due to the contamination by the BC of the
strong 7772 \AA\ line. The 7772 \AA\ line displays the same kind of RV modulation as the rest of
metallic lines, and has the same weak modulation of the line amplitude. Its width could be also
modulated, becoming thinner when the amplitude decreases, but this is very uncertain due to the
complexity of the continuum around the triplet where the BC of all three lines are blended.

\vskip 0.3truecm
\large \textit{{Other lines}}

As mentioned in SA12, EX Lupi in quiescence shows more than 200 emission lines, of which at present
we have only investigated the strongest ones. Other lines include neutral elements (such as Mn I and
Cr I), which are weaker than the mentioned Fe I and Mg I lines, and other ionized species (mostly
Si II and Ti II). Table \ref{narrow-table} contains the list of strong, unblended lines identified in the spectra.

The Cr I 4254 and 4274\AA\ lines are relatively strong and narrow ($\sim$8-10 km/s), 
although they are partially affected by
photospheric absorption lines. Despite being noisier, they show the same RV
and amplitude modulation as the other metallic lines. 

The Si II 6347 and 6371\AA\ lines are also relatively strong. The 6347\AA\ line shows sometimes
evidence for a BC with the usual velocity pattern, although the lines become nearly undetectable
when the accretion rate decreases. Both lines show the RV modulation observed for the rest of 
metallic lines, although the amplitude of the modulation is slightly lower than observed
for the photospheric absorption lines.

The Ti II lines at 4300, 4302, and 4307\AA\ are also detected, despite the poor S/N of the
spectra in this area. They are narrow (8-12 km/s), although one of the spectra shows a hint of a BC.
Despite the high noise level, they show the usual RV modulation with a peak-to-peak amplitude about 6km/s,
and the usual sinusoidal modulation, although in blue and UV lines the amplitude
can be easily distorted by poor continuum determination when the S/N is low.

Other lines such as Mn I at 4030\AA\ are very faint, so any attempt to fit and extract them is noise-dominated.

\end{appendix}

\end{document}